\pdfoutput=1
\def\forcepreprint{1}
\documentclass[12pt,authoryear]{elsarticle}
\usepackage[OT1]{fontenc}
\newtheorem{proposition}{Proposition}
\newtheorem{lemma}{Lemma}
\newtheorem{corollary}{Corollary}
\newtheorem{remark}{Remark}
\usepackage{graphicx} 
\usepackage{tikz}
\usetikzlibrary{arrows.meta,positioning}
\usepackage{epstopdf}
\usepackage[applemac]{inputenc}
\usepackage{amsmath}
\usepackage{amssymb}
\usepackage{tcolorbox}
\usepackage{framed}

\usepackage[ruled,vlined]{algorithm2e}
\usepackage{bm}
\usepackage{natbib}
\usepackage{float}
\usepackage{pbox}
\usepackage{multirow}
\usepackage{array}
\usepackage{lscape}
\usepackage{color}
\usepackage{subcaption}
\usepackage[section]{placeins}
\usepackage{natbib}
\usepackage{tabularx}
\usepackage{multirow}
\usepackage{booktabs}
 \usepackage{longtable}
\usepackage{array}
\usepackage{multirow}
\usepackage{wrapfig}
 \usepackage{float}
\usepackage{colortbl}
 \usepackage{pdflscape}
\usepackage{tabu}
\usepackage{threeparttable}
\usepackage{threeparttablex}
\usepackage[normalem]{ulem}
 \usepackage{makecell}
\usepackage{xcolor}
\usepackage[toc,page]{appendix} 
\usepackage{geometry}
\usepackage{enumitem}
\setlist[enumerate]{label*=\arabic*.} 

\newlength\titlepageskip

\newif\ifblind \blindfalse
\ifdefined\forceblind \blindtrue \fi
\newif\ifpreprint \preprintfalse
\ifdefined\forcepreprint \preprinttrue \blindfalse \fi
\ifpreprint \usepackage{url} \fi

\begin{document}
\makeatletter
\def\ps@pprintTitle{%
  \let\@oddhead\@empty
  \let\@evenhead\@empty
  \let\@oddfoot\@empty
  \let\@evenfoot\@empty
}
\ifpreprint
\def\ps@pprintTitle{%
  \let\@oddhead\@empty
  \let\@evenhead\@empty
  \def\@oddfoot{\footnotesize\itshape Preprint submitted for publication
    (November 2026). This version has not been peer reviewed.\hfil}%
  \let\@evenfoot\@oddfoot
}
\fi
\makeatother
\def\appendixpage{\vspace*{8cm} 
\begin{center} 
\Huge\textbf{Annexes} 
\end{center} 
} 
\def\appendixname{}%

\begin{frontmatter}
\title{Flexible nonlinear spatial autoregressive models under spatial confounding: a gradient boosting approach with closed-form estimation} 
\ifblind\else
\author{Ghislain Geniaux}
\address{ghislain.geniaux@inrae.fr, UR Ecodeveloppement, INRAE, Avignon, France}
\fi
\begin{abstract}
This paper develops a boosting-based framework for estimating
nonlinear spatial autoregressive models (SAR and SEM). Functional
gradient descent boosting with spline base learners provides
automatic selection of covariates and functional forms, while the
spatial parameter is estimated by an extension of the closed-form,
determinant-free estimator (CFE) of \citet{smirnov2020}: we show that
it remains consistent, with the same asymptotic variance formula, for
any additive estimator admitting a smoothing-matrix representation,
and we prove that the underlying quasi-score is Neyman-orthogonal to
the nonlinear nuisance, which makes the flexible, regularized first
stage safe.

This flexibility, however, exposes the estimator to a spatial
confounding problem that must be solved for the framework to deliver
structural estimates. The geoadditive smooth of the coordinates,
$s(u,v)$, a default ingredient of flexible spatial specifications,
competes with the autoregressive term for the same low-frequency
spatial variation, and purely predictive tuning resolves the
competition in favour of the surface, biasing the spatial parameter
--- the geoadditive trap. We therefore equip the estimator with
targeted safeguards: a spectral identification diagnostic, a spectral
penalty on the surface, spatial cross-fitting, stopping iterations
selected by spatial cross-validation, and heteroskedasticity- and
spatial-correlation-robust sandwich variances.

Monte Carlo experiments, including a factorial design that generates
the trap, show that the safeguarded estimators restore test size and
confidence-interval coverage while matching maximum likelihood
accuracy at a fraction of its computational cost. An empirical
application to the Lucas County, Ohio, housing market illustrates the
workflow.
\end{abstract}

\end{frontmatter}

\break

\break 
\section{Introduction}

Spatial autoregressive (SAR) models constitute a central framework in
spatial econometrics for capturing spatial dependence across
observations \citep{ord1975estimation,anselin1988spatial}.
They allow researchers to quantify spatial spillovers, that is,
how outcomes observed at one location are influenced by outcomes
or characteristics observed at neighboring locations.

In their standard formulation, spatial autoregressive models assume
linear relationships between explanatory variables and the outcome
variable. Yet in many empirical applications these relationships are
inherently nonlinear, and ignoring the nonlinearity generates spatial
autocorrelation in regression residuals even when the true
data-generating process involves no spatial dependence
\citep{mcmillen2003spatial}, leading classical spatial econometric
tests \citep{Anselin199677,elhorst2010applied} to over-diagnose
spillovers.

Combining flexible nonlinear regression with spatial autoregressive
structures raises three interconnected challenges that have not been
jointly addressed in a unified framework. We take them in turn,
together with the strand of literature that each has generated.

The first challenge is \emph{computational and theoretical}.
Likelihood-based estimation requires evaluating the spatial
log-determinant, which becomes computationally demanding for large
datasets. A rich literature has introduced nonlinear components into
spatial autoregressive models within this likelihood or GMM
tradition: semiparametric additive SAR models estimated with splines
\citep{gress2004using,basile2004semi}, partially linear models
estimated by profile quasi-maximum likelihood \citep{su2010profile},
semiparametric GMM estimators \citep{su2012semiparametric}, pairwise
difference estimators \citep{zhang2013pairwise}, flexible
semiparametric specifications \citep{malikov2017semiparametric},
and, more recently, neural-network based SAR models
\citep{xiao2023nonlinear}, high-dimensional variable selection
\citep{zhang2024variable}, robust procedures \citep{tang2024robust},
and sieve-based varying-coefficient models \citep{lu2025smooth}.
These estimators remain computationally demanding at scale, and most
offer no automatic selection of covariates or functional forms. A
separate strand of the literature, developed for the \emph{linear}
SAR model and independently of any nonlinear concern, attacks the
computational bottleneck itself: \citet{smirnov2001} showed that
the structure of SAR models allows the repeated evaluation of the
log-determinant to be avoided, the indirect-inference estimator of
\citet{kyriacou2017} inverts the analytic binding function of the
(inconsistent) OLS estimator in pure SAR regression, and the
Closed-Form Estimator (CFE) of \citet{smirnov2020} solves a
quadratic quasi-score equation, yielding an analytic estimator of
the spatial parameter (Section~\ref{sec:cfe} reviews this family in
detail). But replacing the parametric regression function by a
flexible additive estimator may affect the validity of the moment
conditions underlying such methods, and --- to the best of our
knowledge --- no existing semiparametric spatial estimator provides
a formal proof that determinant-free estimation remains valid when
the regression function is estimated nonparametrically. Establishing
that proof is the first object of this paper.

The second challenge is \emph{structural identification}. A
recurring difficulty in spatial regression is the potential
confounding between spatial dependence and spatially structured
covariates. As \citet{debarsy2025identification} document in a
recent comprehensive review, standard specification-search
strategies relegate spillovers to a side-effect of the data-driven
chosen model --- which is not an appropriate identification strategy
--- and geographically-based interaction matrices are at best a
mismeasurement of the true connectivity structure. The nonlinear
setting sharpens the difficulty: spatial autocorrelation in
residuals is often the imprint of incorrect functional forms rather
than of genuine spillovers \citep{mcmillen2003spatial}, and when
explanatory variables are strongly spatially autocorrelated --- as
is common with geographic proximity variables such as distances to
points of interest --- a flexible specification may capture spatial
signal that could alternatively be attributed to the spatial lag.
Covariate selection, which is routine in flexible modelling, runs
straight through this challenge: a selection mechanism rewarded for
predictive fit will retain spatially structured candidates precisely
when they absorb spillover signal.

The third challenge concerns \emph{spatial confounding within the
specification itself}, and is the extreme case of the selection
problem just described. A bivariate geographic smoother $s(u,v)$ is
a default ingredient of flexible spatial specifications, and for
good reason: omitted spatially structured covariates are the rule
rather than the exception. Yet when covariates themselves carry
geography, the surface and the autoregressive term compete for the
same low-frequency spatial variation, and the identification of the
spatial parameter can collapse entirely --- the \emph{geoadditive
trap}. The difficulty is structural: both allocations of that
variation fit the data equally well, so no purely predictive
criterion --- cross-validation of the fit, REML selection of the
smoothing parameters --- can arbitrate between them, and predictive
tuning of the complexity dials (boosting iterations, degrees of
freedom of the surface) systematically resolves the competition in
favour of the surface. This is the spatial-econometric face of the
\emph{spatial confounding} studied in mixed models
\citep{reich2006effects,hodges2010adding,hanks2015restricted,
page2017estimation}; the remedy developed there, restricted spatial
regression, corrects the smoother rather than the estimating
equation, and over-correction turns out to be as damaging as the
trap itself.

This paper provides a unified framework that addresses the three
challenges jointly. Functional gradient descent boosting is the
estimation vehicle: a simple modification of the gradients and loss
functions in the Gaussian case extends component-wise boosting to
SAR and SEM specifications formulated as additive models, providing
automatic selection of covariates and functional forms
\citep{buhlmann2011statistics} while keeping spatial dependence
interpretable as residual spillover effects, and maintaining full
flexibility in the choice of base learners. Within this framework,
the contribution of the paper is threefold.

First, we establish the theoretical validity of the Closed-Form
Estimator for penalized spline additive models --- the central
theoretical result of the paper. In the fixed-basis regime in which
the estimators are implemented --- all smooth terms built on spline
bases of fixed dimension, the ``dictionary'' of the boosting
literature --- the fitted smoothers of the three auxiliary
regressions collapse onto a common spline projection, so that the
CFE estimator is consistent and asymptotically normal, inheriting
the asymptotics of the linear estimator of \citet{smirnov2020}
applied to the spline design, under standard moment conditions and
an identification condition generically satisfied when $W$ is not
collinear with the space spanned by $f(X)$. The closed-form variance
is inherited together with its documented conservativeness, which
grows with the strength of spatial dependence. Crucially, the
spatial autoregressive structure introduces \emph{no additional
regularity requirement} on the nonlinear components beyond their
representability in the spline basis. For component-wise gradient
boosting --- the Boosting Semi-Parametric Additive (BSPA) estimators,
the boosting-fitted counterpart of the penalized GAM first stage
--- the formal results extend conditionally on the selection path,
with a stopping iteration growing logarithmically in $n$; Monte
Carlo evidence consistently shows spline-based and boosting-based
variants to be indistinguishable, and the unconditional treatment of
the selection path is left to future research
(Section~\ref{sec:conclusion}).

Second, we recast the identification question in the language of
semiparametric efficiency. The CFE moment is a partialling-out score
in the sense of \citet{robinson1988}, and we prove that it is
Neyman-orthogonal \citep{chernozhukov2018}: its Gateaux derivative
with respect to each first-stage regression function vanishes at the
truth, so that first-order estimation errors of the flexible first
stage do not move the moment --- the property that makes
regularized, data-driven nonlinear components safe. Orthogonality
protects against estimation error; it cannot manufacture
identification, and the impossibility described in the third
challenge means that the remedies must act on the score and on the
identification side rather than on predictive fit. The framework is
equipped accordingly. An observable spectral diagnostic $\pi_k$
measures, before any structural estimate is computed, the overlap
between the fitted smoother and the dominant eigenvectors of $W$. A
spectral penalty removes the redundant low-frequency flexibility of
the surface while keeping it in the model; it is tuned by the
stability of $\hat\rho$ along the penalty path, never by a
predictive criterion. Spatial cross-fitting eliminates the
finite-sample own-observation bias of the quasi-score. The stopping
iteration of the boosting estimators is selected by spatial-block
cross-validation, and inference rests on heteroskedasticity- and
spatial-correlation-robust (HC0/SHAC) sandwich variances; for SEM
specifications near the boundary, a switching criterion to full ML
restores reliable identification and inference.

Third, we provide a factorial Monte Carlo design (E3) that
generates the geoadditive trap by varying independently where
spatial structure enters the true covariates, what contaminates the
candidate set, and how the fitted model is specified. The design
quantifies the identification collapse (bias on $\hat\rho$, test
size, confidence-interval coverage), traces the contribution of each
safeguard, and shows that the safeguarded estimators restore size
and coverage while leaving the fit of the nonlinear components
essentially untouched --- whereas a deliberately over-penalized arm
of restricted-spatial-regression type illustrates the cost of
correcting the smoother instead of the score.

The proposed methods are evaluated through three Monte Carlo
designs. The baseline design (E1) covers two data-generating processes
(DGPs; SAR and SEM), four levels of spatial dependence, three sample sizes, and two
signal-to-noise ratios, with the stopping iteration of the boosting
arms selected by spatial-block cross-validation. Design E2 assesses
robustness to unobserved spatial heterogeneity in the disturbances.
Design E3 is the factorial design described above.

The approach is illustrated on the Lucas County (Ohio) house-price
data, a classical hedonic benchmark of spatial econometrics. The
application exercises the complete workflow on real data: a
candidate set including the geoadditive surface and distance
variables, a menu of spatial weight matrices, the spectral
diagnostic and penalty, cross-fitting, and sandwich-based inference
on the spatial parameter.

All estimators introduced in this article are implemented in the
\textit{spboost} R package.

The remainder of the paper is organized as follows.
Section~\ref{sec:model} presents the boosting-based framework,
introduces the Closed-Form Estimator, establishes its asymptotic
validity in the nonlinear additive setting, and develops the
orthogonality result together with the safeguards against the
geoadditive trap. Section~\ref{sec:simulations} reports Monte Carlo
evidence. Section~\ref{sec:empirical} presents the empirical
application to the Lucas County housing data.
Section~\ref{sec:conclusion} concludes.

\section{A boosting framework for nonlinear spatial autoregressive
models}
\label{sec:model}

This section develops the estimation framework and the theoretical
foundations of the proposed nonlinear spatial autoregressive model.
The key idea is that determinant-free estimation of the spatial
parameter remains valid even when the regression function is estimated
nonparametrically through additive models. The proposed framework combines
three elements that are usually treated
separately in the literature: (i) additive nonlinear regression,
(ii) a spatial autoregressive structure, and (iii) determinant-free
estimation of the spatial parameter.

Rather than estimating these components independently, the proposed
framework integrates them within a single estimation procedure. The
additive regression function is estimated using functional gradient
boosting, while the spatial autoregressive parameter is obtained from a
closed-form quasi-score equation that remains valid in the nonlinear
setting. This joint formulation leads to a computationally efficient
estimator for nonlinear spatial autoregressive models.

\subsection{Functional gradient descent boosting with additive models}

Statistical boosting introduced by \citet{friedman2000additive}
provides a natural bridge between machine learning algorithms and
interpretable statistical models through generalized additive models
(GAMs) \citep{hastie1990generalized}.
Gradient boosting methods construct flexible regression functions by
sequentially combining simple base learners using a functional gradient
descent procedure \citep{friedman2001greedy,buhlmann2011statistics}.
These methods handle high-dimensional data efficiently while allowing
flexible nonlinear modeling.

Consider a Gaussian additive regression model in which a continuous
outcome $y$ depends on a set of explanatory variables $X$,
\[
y = f(X)+\varepsilon = \beta_0 + \sum_{j=1}^p h_j(X_j) + \varepsilon,
\]
where $h_j(\cdot)$ denotes an unknown smooth function of covariate
$X_j$. Given a loss function $l(\cdot)$ and a class of base learners
$h(\cdot)$, functional gradient descent boosting estimates the
regression function by minimizing the expected loss
\[
\hat f(\cdot) = \arg\min_{f(\cdot)} E_{Y,X}[l(Y,f(X))].
\]
At each iteration $t$, the algorithm fits a base learner to the
negative gradient of the loss evaluated at the current estimate
$f^{t-1}(X)$. For the squared error loss, the negative gradient
corresponds to the current residual vector. The fitted function is
updated according to
\[
\hat f^{m+1}(X) = \hat f^{m}(X) + \alpha \hat h_{j^*}(X_{j^*}),
\]
where $0<\alpha<1$ is a learning rate controlling the step size.

In component-wise boosting \citep{buhlmann2003boosting}, base learners
are specified for each covariate separately to maintain an
interpretable additive structure. Regularization is achieved through
shrinkage and early stopping: the algorithm is run for a finite
number of iterations $m_{\mathrm{stop}}$, which acts as the main
complexity parameter of the fit --- few iterations yield a heavily
shrunken fit, while letting the algorithm run approaches the
unpenalized solution. Because nothing in the gradient descent itself
tells the algorithm when to stop, $m_{\mathrm{stop}}$ must be fixed
beforehand on the basis of an external criterion --- an information
criterion or, more commonly, cross-validation (CV). The choice of that
criterion is far from innocuous in a spatial setting;
Section~\ref{sec:mstop} is devoted to it. An important property is that
boosting-based additive models allow simultaneous estimation of nonlinear functional relationships and 
automatic variable selection, provided that the candidate base learners 
do not induce observational equivalence with the spatial structure.
This property is particularly valuable in spatial econometric
applications, where the number of candidate covariates often
exceeds what can be theoretically motivated, and where including
irrelevant spatially structured variables risks confounding genuine
spillover effects with covariate-driven spatial patterns.

\subsection{Extending boosting to spatial autoregressive models}

We consider three nonlinear spatial autoregressive specifications:
the Spatial AutoRegressive model (SAR), the Spatial Error Model
(SEM), and their combination (SARAR):
\begin{eqnarray}
y &=& \rho Wy + f(X) + \varepsilon \qquad \text{(SAR)} \\
y &=& f(X) + (I-\lambda W)^{-1}\varepsilon \qquad \text{(SEM)} \\
y &=& \rho Wy + f(X) + (I-\lambda W)^{-1}\varepsilon \qquad
\text{(SARAR)},
\end{eqnarray}
where $y$ is an $n$-vector of observations, $X$ denotes the matrix
of explanatory variables, $f(X)=\beta_0+\sum_j h_j(X_j)$ is an
additive regression function with intercept $\beta_0$, and $W$ is
a row-standardized spatial weights matrix. The intercept is
estimated jointly with the smooth components; following standard
practice for GAM and boosting estimators, all covariates are
centred prior to estimation so that $\beta_0$ corresponds to the
global mean of the response \citep{wood2011fast}.

The set of candidate base learners that defines $f(X)$ deserves one
remark at this stage. Beyond univariate smooths of the covariates,
applied work routinely adds a bivariate smooth of the geographic
coordinates, $s(u,v)$ --- a thin-plate or tensor-product surface
\citep{wood2003thin,wood2017generalized} --- to absorb unobserved
spatially structured heterogeneity. This \emph{geoadditive}
component is a default reflex of practitioners both in models
without any explicit spatial dependence, where it is at the centre
of the spatial-confounding debate
\citep{hodges2010adding,reich2006effects,paciorek2010}, and in
semiparametric spatial econometrics, where it coexists with an
autoregressive term \citep{basile2004semi,basile2014modeling}. In
the SAR specification above, however, $s(u,v)$ and $\rho Wy$
compete for the same low-frequency spatial variation of $y$: their
coexistence raises an identification question that is central to
this paper and is analysed, together with its remedies, in
Section~\ref{sec:remedies}.

When the spatial parameter is unknown, its estimation differs across
specifications. In the SAR model, where $\rho$ enters multiplicatively
through the spatial lag $Wy$, estimation can be based on the profile
log-likelihood
\begin{equation}
\label{eqL}
\ln L(\rho)=C+\ln |I-\rho W|
-\frac{n}{2}\ln
\left(
\frac{(y-\rho Wy-f(X))'(y-\rho Wy-f(X))}{n}
\right),
\end{equation}
where $C=-(n/2)\ln(2\pi)-(n/2)$. In the SEM model, $\lambda$ governs
the spatial structure of the disturbance term through $(I-\lambda W)^{-1}$,
and a symmetric profile likelihood can be derived by premultiplying
the model by $(I-\lambda W)$. The SARAR case combines both. In all
specifications, the Jacobian term $\ln|I-\rho W|$ depends only on
the spatial parameter and not on $f(X)$. Consequently, conditional on
a fixed value of the spatial parameter, estimation of $f(X)$ proceeds
by minimizing the residual sum of squares of the spatially transformed
dependent variable, which corresponds to redefining the gradient and
loss functions of the boosting algorithm as summarized in
Table~\ref{tabgrad}.

\begin{table}[htp]
\scriptsize
\caption{Gradients and quadratic loss functions conditional on spatial
parameters}
\begin{center}
\begin{tabular}{c|c|c|c|c}
& Aspatial & SAR & SEM & SARAR \\
\hline
Gradient & $(y-f)$ & $(y-\rho Wy-f)$ & $(I-\lambda W)(y-f)$ &
$(I-\lambda W)(y-\rho Wy-f)$ \\
Loss Function & $\sum(y-f)^2$ & $\sum(y-\rho Wy-f)^2$ &
$\sum(((I-\lambda W)(y-f))^2)$ &
$\sum(((I-\lambda W)(y-\rho Wy-f))^2)$ \\
\hline
\end{tabular}
\end{center}
\label{tabgrad}
\end{table}

The SARAR specification combines the SAR and SEM gradients shown in
Table~\ref{tabgrad} and is handled within the same algorithmic
framework. However, joint estimation of $\rho$ and $\lambda$ in the
SARAR model introduces cross-parameter dependencies that are absent
from the marginal SAR and SEM cases, and the finite-sample properties
of the SARAR estimator are not separately investigated here.
Establishing those properties --- in particular the identification
conditions under joint determinant-free estimation --- is left for
future work.

Although the likelihood-based framework provides a natural way to
combine boosting with spatial autoregressive models, it remains
computationally demanding because the profile likelihood requires
repeated evaluations of the spatial log-determinant. We therefore
now turn to an alternative determinant-free estimation strategy.

\subsection{Determinant-free estimation of the spatial parameter}
\label{sec:cfe}

A crucial question is whether determinant-free spatial estimation
remains valid when the regression function is estimated
nonparametrically. If the additive estimator altered the structure
of the residual projection operator, the quasi-score equation
underlying the CFE approach would no longer hold.
The following section shows that this difficulty does not arise.

Estimating spatial autoregressive models typically requires evaluating
the log-determinant of matrices of the form $(I-\rho W)$ or
$(I-\lambda W)$, which becomes computationally demanding for large
samples. To avoid this difficulty we rely on the Closed-Form Estimator
(CFE) proposed by \citet{smirnov2020}, which replaces the true
likelihood score by a quasi-score whose expectation vanishes
asymptotically \citep{gourieroux1984}.

The CFE belongs to a broader effort to remove the computational
bottleneck of likelihood-based spatial estimation. A first strand
accelerates the log-determinant itself, reducing maximum likelihood
to the evaluation of a scalar function of $\rho$: eigenvalue-based
computation \citep{ord1975estimation}, sparse matrix factorisation
with vectorised grid evaluation of the profile likelihood
\citep{pace1997sparse,pace1997quick}, stochastic Monte Carlo trace
estimation
\citep{barry1999}, Chebyshev approximation \citep{pace2004chebyshev},
and the characteristic polynomial approach of \citet{smirnov2001};
see \citet{bivand2013} for a systematic comparison of these methods.
A second strand removes numerical optimisation altogether through
closed-form estimation: the indirect inference estimator of
\citet{kyriacou2017} inverts the analytic binding function of the
(inconsistent) OLS estimator in pure SAR regression, and the CFE of
\citet{smirnov2020} solves a quadratic quasi-score equation. The CFE
is distinctive within this family in requiring neither the
log-determinant nor any functional of $(I-\rho W)^{-1}$, but only
the spatial lags $Wy$ and $W^2y$ together with auxiliary residuals;
this is precisely what allows the regression stage to be replaced by
an arbitrary additive or boosting-based nonparametric estimator,
which is the route taken in this paper. The CFE moment also has a
structural property that, to our knowledge, has not been noted
before: being built from the residuals of regressions of $y$, $Wy$
and $W^2y$ on the covariates, it is a partialling-out score in the
sense of \citet{robinson1988}, hence a Neyman-orthogonal moment for
the partially linear SAR model in the sense of the double machine
learning literature \citep{chernozhukov2018}. First-order estimation
errors in the flexible first stage do not move the moment
(Appendix~\ref{app:orthogonality}); the residual finite-sample
channel --- the own-observation correlation between smoother
estimation error and residuals --- is removed by computing the
auxiliary residuals out-of-fold, a cross-fitted variant of the
estimator evaluated in Section~\ref{sec:simulations}.

In the linear SAR model $y=\rho Wy + X\beta + \varepsilon$, the
spatial parameter is obtained from the quadratic equation
\[
\rho=\frac{b-\sqrt D}{2a}, \qquad D=b^2-4ac,
\]
where
\[
a=e_1'e_2,\quad b=e_0'e_2+e_1'e_1,\quad c=e_0'e_1,
\]
and $e_0$, $e_1$, $e_2$ are auxiliary residuals from regressions
involving $y$, $Wy$, and $W^2y$. For the SEM model
$y=X\beta+u$, $u=\lambda Wu+\varepsilon$, a similar closed-form
solution is obtained using two auxiliary regressions involving $y$
and $Wy$, with coefficients constructed through the spatial
weighting matrix $W$:
$a_\lambda=e_1'We_1$, $b_\lambda=e_0'We_1+e_1'We_0$,
$c_\lambda=e_0'We_0$. To handle practical departures from the ideal linear-asymptotic setting---such as finite-sample effects and the use of boosting-regularized residuals---we implement a robust root-selection rule based on a residual variance proxy (see Appendix~\ref{appendix_cfe_linear} for details).

The CFE also provides closed-form asymptotic variance estimators.
For the SAR model,
\[
\hat\sigma^2(\rho)=\frac{\hat\sigma^2}{b-2\hat\rho a},
\]
and for the SEM model,
\[
\hat\sigma^2(\lambda)=\frac{\hat\sigma^2}{b_\lambda-2\hat\lambda
a_\lambda},
\]
where $\hat\sigma^2=\hat e'\hat e/n$.

In nonlinear spatial models, the auxiliary residuals are constructed
from additive regressions:
\[
e_0=y-\sum \hat h_0(x),\quad
e_1=Wy-\sum \hat h_1(x),\quad
e_2=W^2y-\sum \hat h_2(x).
\]
These regressions are estimated on a common spline basis of
fixed dimension, either by generalized additive models with
penalized splines or by boosting-based additive models,
yielding the penalized (\texttt{GAM\_SAR\_CFE},
\texttt{GAM\_SEM\_CFE}) and boosted (\texttt{BSPA\_SAR\_CFE},
\texttt{BSPA\_SEM\_CFE}) estimators, which adaptively select between the exact solution and fallback approximations depending on the numerical stability of the quadratic discriminant.

When the discriminant $D$ becomes negative, the exact CFE solution
does not admit a real root. In such cases, the framework
automatically triggers a hierarchy of
determinant-free fallback approximations to maintain numerical 
robustness. The ratio approximation $\tilde\rho_R={e_0'e_1}/({e_0'e_2+e_1'e_1})$
and the Wald-type instrumental-variable (IV) estimator
$\tilde\rho_{IV}={e_1'e_0}/{e_1'e_1}$ serve
as simplified representations of the CFE orthogonality condition 
(see Appendix~\ref{appendix_cfe_linear} for their geometric derivation). For the SEM model, the
weighted projection $\tilde\lambda_W={e_1'We_0}/{e_1'We_1}$
provides the corresponding fallback.

\begin{figure}[htbp]
\begin{center}
\begin{tikzpicture}[
box/.style={
draw,
rounded corners,
align=center,
minimum width=5.2cm,
minimum height=1.2cm
},
smallbox/.style={
draw,
rounded corners,
align=center,
minimum width=4.8cm,
minimum height=1.1cm
},
arrow/.style={->,thick},
aparrow/.style={->,thick,dashed}
]
\node[box] (root) {
Generalized CFE framework\\
determinant-free estimation of spatial parameters
};
\node[box,below=1.6cm of root, xshift=-4cm] (sar0) {
GAM\_SAR\_CFE\\
BSPA\_SAR\_CFE

};
\node[box,below=1.6cm of root, xshift=4cm] (sem0) {
GAM\_SEM\_CFE\\
BSPA\_SEM\_CFE
};
\node[box,below=1cm of sem0,minimum height=1.7cm] (sem1) {
SEM exact CFE estimator\\[5pt]
$\hat{\lambda}_{CFE}=\frac{b-\sqrt{b^2-4ac}}{2a}$,\quad
$\begin{smallmatrix}a=e_1'We_1\\ b=e_0'We_1+e_1'We_0\\ c=e_0'We_0\end{smallmatrix}$
};
\node[box,anchor=south,minimum height=1.7cm] (sar1) at (sar0.south |- sem1.south) {
SAR exact CFE estimator\\[5pt]
$\hat{\rho}_{CFE}=\frac{b-\sqrt{b^2-4ac}}{2a}$
};
\node[smallbox,below=of sar1] (sar2) {
SAR ratio approximation\\
$\tilde{\rho}_R=\frac{e_0'e_1}{e_0'e_2+e_1'e_1}$
};
\node[smallbox,below=of sar2] (sar3) {
SAR Wald / IV estimator\\
$\tilde{\rho}_{IV}=\frac{e_1'e_0}{e_1'e_1}$
};
\node[smallbox,below=of sem1] (sem2) {
SEM weighted projection estimator\\
$\tilde{\lambda}_W=\frac{e_1'We_0}{e_1'We_1}$
};
\draw[arrow] (root) -- node[left, xshift=-3mm]{SAR} (sar0);
\draw[arrow] (root) -- node[right, xshift=3mm]{SEM} (sem0);
\draw[arrow] (sar0) -- node[right]{} (sar1);
\draw[arrow] (sem0) -- node[right]{} (sem1);
\draw[aparrow] (sar1) -- node[right, align=left]
  {fallback ($b^2-4ac<0$ or $a\approx 0$)} (sar2);
\draw[aparrow] (sar2) -- node[right, align=left]
  {further fallback ($b\approx 0$)} (sar3);
\draw[aparrow] (sem1) -- node[right, align=left]
  {fallback ($b^2-4ac<0$):\\ minimises $Q_W(\lambda)=(e_0-\lambda e_1)'W(e_0-\lambda e_1)$} (sem2);
\end{tikzpicture}
\caption{Hierarchy of CFE estimators. Solid arrows lead to the exact
closed-form solutions of the SAR and SEM quadratic moment conditions;
dashed arrows lead to the approximate and simplified estimators used
as numerical fallbacks. The hierarchy applies to the boosted
(\texttt{BSPA}) and penalized (\texttt{GAM}) variants alike.}
\label{fig:cfe_hierarchy_branches}
\end{center}
\end{figure}
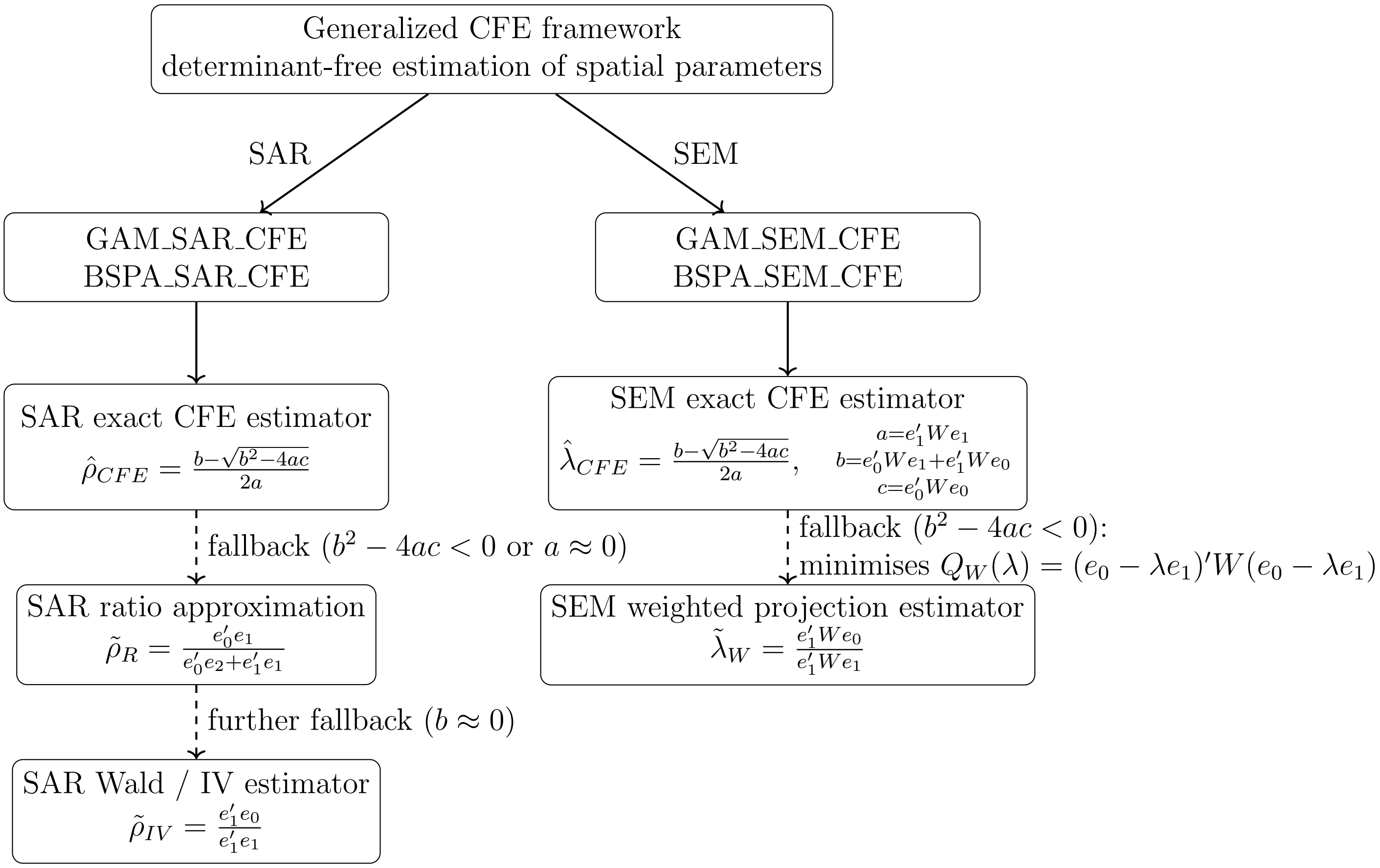

A word on nomenclature. Estimator names read
\texttt{<first stage>\_<spatial model>\_<spatial estimator>}: the
first tag identifies the first-stage fitting algorithm (\texttt{BSPA}
for component-wise boosting, \texttt{GAM} for penalized splines), the
second the spatial specification (\texttt{SAR}, \texttt{SEM},
\texttt{SARAR}), and the third the estimator of the spatial parameter
(\texttt{ML} or \texttt{CFE}). The \texttt{BSPA} and \texttt{GAM}
first stages fit one and the same additive specification --- by
component-wise boosting with automatic selection in the first case,
by penalized likelihood in the second --- and we accordingly speak of
the \emph{boosted} and \emph{penalized} variants of an estimator in
the text. Figure~\ref{fig:cfe_hierarchy_branches}
summarizes this hierarchy. In practice the fallbacks matter only
for SEM: across all Monte Carlo replications the SAR discriminant
remained positive, whereas the weighted projection activates
systematically as $\hat\lambda$ approaches the boundary
($\hat\lambda > 0.85$), compensating for the near-singularity of
$(I-\lambda W)$ at the price of overly conservative standard
errors, as documented in Appendix~\ref{appendix_additional_mc}.

\subsection{Quasi-linear decomposition and asymptotic properties}
\label{sec:asymptotic_nonlinear}

A central theoretical question is whether the CFE estimator remains
valid when the regression function $f(X)$ is estimated
nonparametrically. In principle, replacing the true regression function
by a flexible additive estimator could distort the quasi-score equation
used by the CFE method.

We show that this concern does not arise under standard regularity
conditions: the estimation error of the additive component does not
affect the asymptotic behaviour of the CFE estimator. Consequently, the
spatial parameter remains consistently estimable and its variance
formula coincides asymptotically with the linear case.

The key theoretical insight is that once the nonlinear additive
component has been estimated, the spatial autoregressive parameter
can be identified from the residual projection operator associated
with the additive estimator. This leads to a quasi-linear representation of
the CFE estimating equation that remains valid even when the regression
function is estimated nonparametrically through additive or
boosting-based methods.

The CFE derivation relies on the quasi-linear decomposition of the
residual process. The boosting algorithm operates on the residual
vector $e(\rho)=y-\rho Wy-f(X)$, which admits the first-order
approximation
\[
e(\rho) \approx e_0 - \rho e_1,
\]
where $e_0=My$ and $e_1=MWy$, with $M=(I-S)$ the residual projection
operator associated with the additive estimator. This residualization corresponds
to projecting the model onto the orthogonal complement of the
additive component, similarly to a Frisch--Waugh--Lovell decomposition.
This decomposition holds exactly when the
smoothing operator is linear and fixed, which is how the estimator is
implemented: the auxiliary residuals are computed once, with smoothing
parameters held fixed across $\rho$.
Appendix~\ref{appendix_formal_bound} bounds the additional error that
would arise if the smoothing parameter were re-optimised along $\rho$:
that error is controlled by the observable quantity
$|\hat\mu(\rho)-\hat\mu(0)|$
(Proposition~\ref{prop:bound}, Appendix~\ref{appendix_formal_bound}).
The asymptotic results of Appendix~\ref{appendix_asymptotic_cfe} are
established directly for the fixed-smoother estimator and do not rely
on that bound.

The spectral separation condition~(HSS) of
Appendix~\ref{appendix_formal_bound} provides a complementary,
empirically verifiable diagnostic of the overlap between the nonlinear
base learners and the dominant eigenvectors of $W$; it is not used in
the formal proofs, and large overlaps flag configurations in which the
identification condition deserves empirical scrutiny.

This quasi-linear decomposition is formally identical to the
structure used in the linear derivation of the CFE estimator.
Consequently, the quadratic equation used in the CFE approach
remains a valid first-order approximation in nonlinear additive
models.

The proposed framework also reflects a modelling hierarchy between
nonlinear covariate effects and residual spatial dependence. By
allowing $f(X)$ to absorb spatially structured variation that can
be explained by observed covariates, the spatial autoregressive
parameter is interpreted conservatively as capturing residual
spillover effects. This modelling hierarchy is not innocuous: if
the true data-generating process contains strong genuine spatial
diffusion, an excessively flexible nonlinear component may attenuate
the estimate of $\rho$. The boosting regularization through weak
base learners, small learning rate, and early stopping limits the
complexity of $f(X)$ and reduces this risk.

We establish the following asymptotic properties of the CFE estimator
in the nonlinear additive framework. Full statements and proofs are provided in
Appendix~\ref{appendix_asymptotic_cfe}, while the main intuition
behind the assumptions is discussed here.

Assumption~(A1) requires spatial regularity of $W$: the matrix is
row-standardised with zero diagonal, its eigenvalues lie in the
closed unit disc with the unit eigenvalue (associated with the
constant eigenvector) simple, its row and column sums are uniformly
bounded, and the spatial multiplier $(I-\rho W)^{-1}$ remains bounded
uniformly over the admissible parameter space. This is the standard
stability condition for spatial autoregressive processes
\citep{anselin1988spatial,kelejian2001} and is satisfied by
row-normalised weight matrices with a bounded number of neighbours.
The equal-weight scheme studied by \citet{kelejian2002} and
\citet{baltagi2006} satisfies (A1) but is excluded by the
identification condition~(A3), as discussed in
Appendix~\ref{appendix_asymptotic_cfe}.

Assumption~(A2) states the fixed-basis regime in which the
estimators are implemented: each smooth term uses a spline basis of
fixed size --- the default behaviour of standard penalized-spline and
component-wise boosting implementations --- and the
additive components are representable in that basis. Flexible
additive fits are then estimated at the parametric rate; the effect
of a small approximation error, and the connection with the
nonparametric sieve regime in which the basis grows with $n$, are
quantified in Appendix~\ref{appendix_asymptotic_cfe}
(Remarks~\ref{rem:approx} and~\ref{rem:sieve}).
The smooth components used in the Monte Carlo experiments ---
$f_1(x_1)=x_1-0.5$ (linear), the quasi-V-shaped $f_2$ and the sinusoidal
$f_3$ --- have a measured basis approximation error of at most
$3\times10^{-2}$ at the default basis sizes
(Remark~\ref{rem:approx}). Crucially, (A2) does not depend on the
spatial structure of the model --- the spatial autoregressive
framework introduces no additional requirement on $f$ beyond its
representability in the basis.

Assumption~(A3) is a standard identification condition: the
population counterpart of the CFE estimating equation has a unique
solution at $\rho_0$. It is analogous to the rank condition in GMM
and is generically satisfied when $W$ is not collinear with the space
spanned by $f(X)$. The formal statement requires
$n^{-1}(e_1-\rho_0 e_2)^\top(e_1-\rho_0 e_2)\xrightarrow{p}Q_0>0$;
see Appendix~\ref{appendix_asymptotic_cfe}.

Assumption~(A4) collects standard moment conditions --- bounded
fourth moments of the errors and bounded spatial lags of $y$ ---
that are required to control the stochastic terms in the asymptotic
expansion and are mild in any spatial autoregressive framework with
light-tailed disturbances.

The key difficulty in the nonlinear SAR framework is that the
additive estimator depends on the spatial parameter through the
transformed response $(I-\rho W)y$. As a consequence, the smoothing
operator varies with $\rho$, which could in principle affect the
asymptotic properties of the CFE estimator.

The following propositions summarize the main theoretical result of
the paper: the dependence of the smoothing operator on the spatial
parameter is asymptotically negligible, so that determinant-free
estimation retains the same asymptotic properties as in the linear
case.

\begin{proposition}[Consistency, Appendix~\ref{appendix_asymptotic_cfe}]
\label{prop:consistency_main}
Under \emph{(A1)--(A4)},
$\hat\rho_{CFE}\xrightarrow{p}\rho_0$.
\end{proposition}

\begin{proposition}[Asymptotic normality and conservative variance,
Appendix~\ref{appendix_asymptotic_cfe}]
\label{prop:variance_main}
Under \emph{(A1)--(A5)},
$\sqrt{n}\,(\hat\rho_{CFE}-\rho_0)$ has the same limiting normal
distribution as the linear closed-form estimator of
\citet{smirnov2020} applied to the regressor matrix $\mathbf{B}$,
and $\hat\sigma^2(\rho)=\hat\sigma^2/(b-2\hat\rho_{CFE}\,a)$
converges in probability to the associated quasi-score information
bound. As documented in the linear case, this bound over-estimates
the realized sampling variance, increasingly so as $\rho_0$ grows,
so that the resulting confidence intervals are conservative
(\citealp{smirnov2020}, Tables~3 and~6;
Appendix~\ref{appendix_additional_mc}).
\end{proposition}

\begin{remark}
The asymptotic theory matches the implementation: three separate
penalized-spline regressions, each with its own GCV- or
REML-selected smoothing parameters. The key structural fact
(Lemma~\ref{lem:collapse}, Appendix~\ref{appendix_asymptotic_cfe})
is that, on a fixed basis, all three fitted smoothers collapse
onto the same orthogonal projection at rate $O_p(n^{-1})$, so the
nonlinear CFE inherits the asymptotics of the linear CFE of
\citet{smirnov2020} with regressor matrix $\mathbf{B}$. No
nonparametric convergence-rate condition is required at this level;
the sieve regime, in which such conditions reappear, is discussed
in Remark~\ref{rem:sieve} of
Appendix~\ref{appendix_asymptotic_cfe}.
\end{remark}

\begin{remark}[Scope of Propositions~\ref{prop:consistency_main}
and~\ref{prop:variance_main}]
\label{rem:scope}
Propositions~\ref{prop:consistency_main}
and~\ref{prop:variance_main} are established for additive
estimators admitting a fixed smoothing matrix representation
$\hat f(X) = S_\mu y$, which covers penalized spline GAMs under
Generalized Cross-Validation (GCV) or Restricted Maximum Likelihood (REML) smoothing parameter selection
\citep{wood2011fast}. The formal argument relies on the
differentiability of $S_\mu$ with respect to $\mu$ and the
implicit function theorem applied to the GCV criterion
(Appendix~\ref{appendix_formal_bound}); these tools are not
directly available for component-wise gradient boosting, which
does not produce a fixed smoothing matrix.

For the boosted estimators \texttt{BSPA\_SAR\_CFE} and
\texttt{BSPA\_SEM\_CFE}, the formal proof covers the
case where the auxiliary regressions $e_0$, $e_1$, $e_2$ are
estimated by penalized spline GAMs. When they are estimated by
component-wise boosting, Corollary~\ref{cor:boosting} in
Appendix~\ref{appendix_asymptotic_cfe} extends the formal results
\emph{conditionally on the selection path}: on a fixed basis,
the boosting operator converges geometrically to the same spline
projection \citep{buhlmann2003boosting}, so that a stopping
iteration growing logarithmically in $n$ suffices. The
unconditional treatment of the data-dependent selection path and of
cross-validated number of iterations (\texttt{mstop}) remains open, and empirical evidence
provides the complementary support: the Monte Carlo experiments of
Section~\ref{sec:simulations} systematically show that
\texttt{BSPA\_SAR\_CFE} and \texttt{GAM\_SAR\_CFE} yield
virtually identical bias, RMSE, and coverage rates across all
configurations, providing direct finite-sample evidence that
the asymptotic properties of Propositions~\ref{prop:consistency_main}
and~\ref{prop:variance_main} extend to the boosting variants.
Removing the conditioning on the realised selection path and
treating cross-validated \texttt{mstop} formally is a direction
for future research (see Section~\ref{sec:conclusion}).
\end{remark}

The finite-sample performance of the variance formula is assessed
empirically through coverage rates reported in
Appendix~\ref{appendix_asymptotic_cfe}.

\subsection{Orthogonality, remedies, and inference}
\label{sec:remedies}

The structural property identified in Section~\ref{sec:cfe} --- the
CFE moment is a partialling-out score in the sense of
\citet{robinson1988} --- organises everything the estimator can and
cannot absorb. Writing the quasi-score as a functional of the three
first-stage regression functions, its Gateaux derivative with
respect to each of them vanishes at the truth
(Proposition~\ref{prop:orthogonality},
Appendix~\ref{app:orthogonality}): the moment is Neyman-orthogonal
in the sense of \citet{chernozhukov2018}. Two features of this
result are specific to the spatial setting. First, the endogeneity
of $Wy$ is immaterial for orthogonality, because the nuisances are
conditional expectations given the \emph{own} covariates $x_i$ only.
Second, independence of the innovations suffices;
heteroskedasticity is allowed for consistency. Orthogonality is
what makes a flexible, regularized first stage safe: first-order
estimation errors in $\hat h_0,\hat h_1,\hat h_2$ do not move the
moment.

Orthogonality protects against estimation \emph{error}; it cannot
manufacture identification. The distinction matters most for the
geoadditive component $s(u,v)$ introduced in
Section~\ref{sec:model}. A practitioner has every reason to want it
in the candidate set: omitted spatially structured covariates are
ubiquitous in applied work, and the surface is the standard
device for absorbing them. Yet when covariates themselves carry
geography --- spatially autocorrelated regressors, distance
variables --- the combination of a \emph{free} $s(u,v)$ with the
autoregressive term degrades $\hat\rho$ and invalidates inference
on it: this is the \emph{geoadditive trap}, the spatial-econometric
face of the spatial confounding documented in models without
autoregressive terms \citep{hodges2010adding,paciorek2010}. Simply
banning the surface from the candidate set is not an acceptable
answer; the trap splits into two separate channels, each with a
remedy that keeps $s(u,v)$ in the model.

\paragraph{Population channel: identification redundancy}
When the candidate base learners span the dominant spatial eigenvectors of $W$,
the surface term and the autoregressive term $\rho Wy$ compete for
the same low-frequency variation and the identification condition
(A3) degenerates. The overlap is observable: the spectral projection
coefficient $\pi_k(\hat\mu)=\boldsymbol{\phi}_k^\top
\mathbf{S}_{\hat\mu}\boldsymbol{\phi}_k$
(Appendix~\ref{appendix_formal_bound}, Figure~\ref{fig:pi_k})
measures the fraction of the $k$-th eigenvector of $W$ absorbed by
the smoother. The remedy acts on the surface smoother itself: we
introduce a \emph{spectral penalty} that augments the usual
roughness penalty of $s(u,v)$ with a second penalty
$\mu_s\,\|\boldsymbol{\Phi}_{K_s}^\top\mathbf{B}
\boldsymbol{\theta}\|^2$ on the projection of the fitted surface
onto the $K_s$ dominant eigenvectors of $W$ (default $K_s=100$;
basis \texttt{spx} in the implementation). The penalty is defined
on the penalized-spline representation of the surface and therefore
concerns the penalized estimators, where $\mu_s$ enters as an
ordinary second smoothing penalty, fully compatible with the
fixed-basis theory of Section~\ref{sec:asymptotic_nonlinear}. For
the boosted estimators, the complexity of the surface
base learner is governed instead by the stopping iteration, whose
identification-aware selection is the subject of
Section~\ref{sec:mstop}.
Importantly, $\mu_s$ cannot be tuned by a prediction-targeted
criterion: REML is structurally indifferent to how low-frequency
variation is allocated between $s(u,v)$ and $\rho Wy$, since both
allocations fit $y$ equally well
(Section~\ref{sec:E3p_results}). We therefore evaluate $\hat\rho$
along a fixed $\mu_s$-grid and select the plateau of the
trajectory.

\paragraph{Finite-sample channel: own-observation correlation}
Even with the population overlap removed, a finite-sample term
survives that orthogonality does not cover: the same observations are
used to fit the smoothers and to evaluate their residuals, so
$e_{k,i}$ retains a mechanical correlation with the estimation error
of the smoother at observation $i$ --- $\widehat m_1$ depends on
$\varepsilon_i$ through $Wy$ --- which biases the quadratic forms
$a,b,c$ of the quasi-score (Remark~\ref{rem:crossfit}). The remedy
is \emph{cross-fitting}: the
auxiliary residuals $e_0,e_1,e_2$ are computed out-of-fold, using
spatially scattered folds (small k-means cells of the coordinates
allocated round-robin to $K$ folds) so that each fold remains
representative of the whole domain, and the interpretable final fit
is re-estimated on the full sample. This is the standard double
machine-learning device \citep{chernozhukov2018} adapted to spatial
data.

\paragraph{Correct the score, not the smoother}
The two remedies share a principle: they repair the \emph{moment}
--- its identification content and its sampling noise --- while
leaving the covariate smooths untouched. The opposite strategy,
restricting the covariate smooths so that nothing spatial survives
in $f(X)$, is a pathology rather than a cure: it expels genuinely
spatial covariate signal into the autoregressive term and produces
spurious $\hat\rho>0$ with severe anticonservatism, the mechanism
documented for restricted spatial regression
\citep{hodges2010adding,hanks2015restricted,hughes2013,keller2020}.
Section~\ref{sec:E3p_results} reproduces both failure directions
within a single design and shows that the combination of the
spectral penalty and cross-fitting restores oracle-like behaviour,
whereas the over-restricted variant rejects a true $\rho=0$ in
100\% of replications. The diagnostic $\bar\pi$ (the average of
$\pi_k$ over the dominant eigenvectors) is used accordingly: it
flags the specifications in which identification deserves scrutiny
and verifies that the spectral penalty is operating, while the
decision itself rests on the stability of $\hat\rho$ along the
$\mu_s$ path.

\paragraph{Inference}
The closed-form variance
$\hat\sigma^2(\rho)=\hat\sigma^2/(b-2\hat\rho a)$ of
Proposition~\ref{prop:variance_main} is valid but conservative, and
targets homoskedastic innovations. We complement it with
heteroskedasticity-robust sandwich variants obtained by the delta
method on the quasi-score: with moment contributions
$m_i=(e_0-\hat\rho e_1)_i(e_1-\hat\rho e_2)_i$ and denominator
$d=b-2\hat\rho a$,
\[
\widehat{\operatorname{var}}_{\mathrm{HC0}}(\hat\rho)
=\frac{\sum_i m_i^2}{d^2},
\qquad
\widehat{\operatorname{var}}_{\mathrm{SHAC}}(\hat\rho)
=\frac{\sum_i m_i^2+\sum_{i\neq j}\kappa_{ij}\,m_i m_j}{d^2},
\]
where $\kappa_{ij}$ selects pairs within graph distance two in $W$,
in the spirit of the spatial HAC estimator of \citet{kelejian2007}.
The cross-terms matter: HC0 ignores the neighbourhood correlation
induced by $W\boldsymbol{\varepsilon}$ and under-covers, while SHAC
is close to nominal under strong heteroskedasticity
(Remark~\ref{rem:sandwich}). Both variants combine with
cross-fitting.

\subsection{Selection of the stopping iteration \texttt{mstop}}
\label{sec:mstop}

This subsection concerns the boosted estimators only;
the penalized variants involve no
stopping iteration, their complexity being governed by the
smoothing parameters. For component-wise boosting, the stopping
iteration \texttt{mstop} controls the complexity of the nonlinear
component $\hat f(X)$: it plays the role that the smoothing
parameter $\mu$ plays for penalized splines and the spectral
parameter $\mu_s$ of Section~\ref{sec:remedies} plays for the
geographic surface. The same
structural caveat therefore applies: a purely prediction-targeted
criterion is indifferent to how low-frequency variation is
allocated between $\rho Wy$ and the smooth terms, and cannot detect
identification loss. The dedicated sub-experiment of
Section~\ref{sec:E3p_results} confirms this mechanism: under a free
geographic smoother, random $K$-fold cross-validation on the global
RMSE of $Y$ inflates \texttt{mstop} (median around 290) and erodes
the robustness that moderate stopping provides, whereas
\emph{spatial-block} $K$-fold cross-validation --- the folds are $K$
k-means clusters of the coordinates, each cluster held out once
(leave-one-cluster-out) --- selects markedly
smaller values (median 130--168) and preserves the size of the test
of $\rho=0$. The mechanism is leakage: with random folds, the
neighbours of a held-out observation sit in the training folds, so
a surface overfitted to the spatial noise still predicts well and
the criterion rewards it; holding out whole clusters removes this
complicity. Spatial blocking does perforate the neighbourhood
graph, since the held-out observations lose their neighbours from
the training set --- a legitimate concern when the cross-validated
fit itself is the deliverable, as in predictive calibration of $W$;
here the folds only select a complexity level, and the model is
then re-estimated on the full sample.

Accordingly, this spatial-block protocol is the
recommended \texttt{mstop} protocol whenever the candidate set
includes geographic smoothers or spatially structured covariates,
and it is the protocol used for all boosted arms of the Monte
Carlo tables of Section~\ref{sec:simulations}. In trap-free
settings the two protocols select similar values, so little is lost
by adopting the spatial protocol as default. Default search grids
use an upper bound of the order of $n$ for SAR specifications and
$2n$ for SEM
specifications; the larger SEM bound reflects the finer nonlinear
fit required to estimate spatial error dependence. The ceiling
should be generous: in well-specified designs the criterion is flat
beyond the optimum and may select values near the bound without
harm to $\hat\rho$.

The asymptotic results of Section~\ref{sec:asymptotic_nonlinear}
should be interpreted as
conditional on a regularization path satisfying the standard
nonparametric convergence rate, or, in the penalized-spline case, on a
smoothing parameter selected by a smooth criterion such as GCV or REML.
For component-wise boosting, cross-validation of \texttt{mstop}
introduces a discrete data-dependent model-selection step that is not
covered by the formal proof. Its use in \texttt{BSPA\_SAR\_CFE} is
therefore justified empirically by the Monte Carlo evidence and by the
known asymptotic connection between small-step $L_2$ boosting and
penalized spline regularization.

\paragraph{SAR estimators}
For \texttt{BSPA\_SAR\_CFE} and \texttt{BSPA\_SAR\_ML} in trap-free
configurations --- no geographic smoother among the candidates and
weakly structured covariates --- the choice of protocol is benign:
random and spatial-block cross-validation select similar
\texttt{mstop} values, and estimation is insensitive to moderate
departures from the optimum, so a fixed \texttt{mstop} of a few
hundred iterations is an acceptable computational shortcut. As soon
as the candidate set includes $s(u,v)$ or distance-type covariates,
only the spatial-block protocol is reliable
(Section~\ref{sec:E3p_results}).

\paragraph{SEM estimators}
The situation is more complex for SEM estimators, because a second
difficulty --- specific to the SEM filter and unrelated to the trap
--- arises near the boundary of the parameter space. In trap-free
configurations, random cross-validation is reliable for
\texttt{BSPA\_SEM\_ML} across the full range of $\lambda$, and for
\texttt{BSPA\_SEM\_CFE} as long as $\hat\lambda \leq 0.9$. When
$\hat\lambda > 0.9$, however,
cross-validation on the global RMSE of $Y$ becomes unreliable:
the selected \texttt{mstop} systematically reaches the upper
bound of the search grid, accompanied by an upward drift of
$\hat\lambda$ toward 1 and a substantial increase in bias. This
pathological behaviour reflects the near-singularity of
$(I - \lambda W)$ near the boundary of the parameter space,
which allows $\hat f(X)$ to absorb residual spatial structure,
misleading the cross-validation criterion.

To address this issue, \texttt{BSPA\_SEM\_CFE} implements a
switching rule: when the estimated spatial parameter satisfies
$\hat\lambda > 0.9$, the cross-validation criterion is
automatically switched from the global RMSE of $Y$ to the RMSE
of the spatially filtered residuals
$(I - \hat\lambda W)(Y - \hat f(X))$, which is less sensitive
to spatial absorption near the boundary. (This $0.9$ threshold
governs the cross-validation criterion only; it is distinct from
the $\approx 0.85$ level at which the weighted-projection fallback
of Section~\ref{sec:cfe} starts to activate.)
While this filtered criterion provides more stable \texttt{mstop} selection 
near the boundary, it remains a second-best adjustment for the inherent 
ill-conditioning of the SEM filter; results at $\hat\lambda > 0.9$ should 
therefore be treated as indicative rather than definitive. A similar instability affects
\texttt{GAM\_SEM\_CFE}, where the inherent flexibility of the
penalised spline representation interferes with the identification
of $\lambda$ at high spatial dependence levels, independently of
any \texttt{mstop} choice. For this reason, we recommend
\texttt{BSPA\_SEM\_ML} with random cross-validation as the
preferred SEM estimator in high spatial dependence settings.
\texttt{BSPA\_SEM\_CFE} with the switching rule provides an
acceptable alternative when maximum likelihood estimation is
computationally prohibitive, but results at $\lambda=0.9$ should
still be interpreted with caution.

The empirical consequences of these selection rules are documented
in Section~\ref{sec:simulations}: for the SAR case, the dedicated
\texttt{mstop} sub-experiment of the E3 design
(Section~\ref{sec:E3p_results}) compares fixed, random-CV and
spatial-block-CV protocols on both the control and the adversarial
covariate configurations; for the SEM case, the E1 experiments
document the boundary behaviour and the switching rule.

\subsection{Estimation algorithm}

Algorithm~\ref{alg:cfe} summarizes the complete estimation procedure. 
To ensure memory efficiency for very large samples (e.g., $n = 250,000$), 
the \texttt{spboost} package leverages \texttt{C++} via \texttt{Rcpp} 
and \texttt{RcppEigen} for all operations involving the spatial 
weights matrix $W$. This sparse implementation allows 
the three auxiliary boosting regressions and the subsequent 
CFE quadratic solve to be executed on standard hardware without 
memory overflow.
The spatial parameter is obtained directly from closed-form
expressions based on auxiliary residuals, eliminating the need to
repeatedly evaluate spatial log-determinants.

\begin{algorithm}[H]
\resizebox{\linewidth}{!}{%
\small
\begin{minipage}{\linewidth}
\caption{Determinant-free SAR/SEM estimation via the Closed-Form
Estimator (CFE), as implemented in the R package \textit{spboost}}
\label{alg:cfe}
\SetAlgoLined

Choose the first-stage \texttt{method} for the auxiliary regressions:
\texttt{gam} (penalized spline GAM, \texttt{mgcv}) or
\texttt{gamboost} (component-wise gradient boosting with P-splines,
\texttt{mboost}), with its regularization: smoothing parameter $\mu$
for \texttt{gam}; learning rate $\alpha$ and stopping iteration
$m_{stop}$ (spatial-block $K$-fold CV recommended) for
\texttt{gamboost}.

\BlankLine
\emph{Optional safeguards (Section~\ref{sec:remedies}).} If a
geographic smoother $s(u,v)$ is a candidate, use its spectrally
penalized version (\texttt{spx}, $\mu_s$ chosen on a grid by the
plateau rule) and monitor $\pi_k$; compute the auxiliary residuals
out-of-fold on spatially scattered folds (\texttt{crossfit}).

\BlankLine
\textbf{SAR case}\;
Regress $y$, $Wy$ and $W^2y$ on $X$ with \texttt{method}; keep the
residuals $e_0$, $e_1$, $e_2$\;
Compute $a=e_1'e_2$, $b=e_0'e_2+e_1'e_1$, $c=e_0'e_1$ and
$D=b^2-4ac$\;
\eIf{$D \ge 0$ and $a\not\approx 0$}{
exact solution $\hat\rho_{CFE}=\frac{b-\sqrt D}{2a}$\;
}{
fallback $\tilde\rho_R=c/b$\quad(then $\tilde\rho_{IV}=c/(e_1'e_1)$
if $b\approx 0$)\;
}
\BlankLine
\textbf{SEM case}\;
Regress $y$ and $Wy$ on $X$ with \texttt{method}; keep the residuals
$e_0$, $e_1$\;
Compute $a=e_1'We_1$, $b=e_0'We_1+e_1'We_0$, $c=e_0'We_0$ and
$D=b^2-4ac$\;
\eIf{$D \ge 0$}{
exact solution $\hat\lambda_{CFE}=\frac{b-\sqrt{D}}{2a}$\;
}{
fallback: weighted projection
$\tilde\lambda_W=\frac{e_1'We_0}{e_1'We_1}$\;
}
\BlankLine
Estimate the final model on the spatially filtered response
using \texttt{method} with the estimated spatial parameter\;
Return the spatial parameter with its variance estimates ---
closed-form $\hat\sigma^2/(b-2\hat\rho a)$, \texttt{HC0} and
\texttt{SHAC} sandwiches --- and the spectral diagnostic $\pi_k$.

\end{minipage}
}
\end{algorithm}

\section{Finite-sample properties}
\label{sec:simulations}
\subsection{Monte Carlo simulation design}

To evaluate the finite-sample properties of the proposed estimators, we conduct a series of Monte Carlo experiments based on synthetic spatial datasets. Two types of spatial autoregressive DGPs are considered: the spatial autoregressive model (SAR) and the spatial error model (SEM). Four levels of spatial dependence are examined, ranging from no spatial autocorrelation to very strong dependence, namely $\rho \in \{0,0.2,0.6,0.9\}$ for the SAR model and $\lambda \in \{0,0.2,0.6,0.9\}$ for the SEM model.

The nonlinear SAR and SEM specifications are defined as follows:

\begin{eqnarray*}
Y_i &=& \rho W Y + \beta_0 + \sum_{j=1}^{3} f_j(x_{ij}) + \epsilon_i \qquad \text{(nonlinear SAR)}, \\
Y_i &=& \beta_0 + \sum_{j=1}^{3} f_j(x_{ij}) + (I-\lambda W)^{-1}\epsilon_i \qquad \text{(nonlinear SEM)}.
\end{eqnarray*}

Spatial locations are randomly drawn from the unit square $[0,1]^2$. The spatial weights matrix $W$ is row-normalized and constructed using the four nearest neighbours ($k=4$). The error term $\epsilon_i$ follows an independent normal distribution. Its variance is calibrated to obtain signal-to-noise ratios (SNR) equal to either $0.7$ or $0.9$, following the approach proposed by \citet{pace2012spatial}.

Three explanatory variables are generated independently from a uniform distribution over $[0,1]$. Two nonlinear functional forms are used: the second variable follows a quasi-V-shaped $C^\infty$ function, while the third follows a sinusoidal function. The first variable remains linear in order to assess the behavior of nonlinear estimators when some covariate effects are correctly specified as linear. 

All explanatory variables are centered prior to estimation so that the intercept corresponds to the global mean of the response variable. This normalization is particularly important for GAM and boosting estimators, whose smooth components are estimated relative to an offset representation.

The functional forms are defined as

\begin{eqnarray*}
f_1(x_1) &=& x_1- 0.5, \\
f_2(x_2) &=& \sqrt{3\bigl[(2x_2-1)^2+\delta^2\bigr]}-\delta\sqrt{3}, \quad \delta=0.01, \\
f_3(x_3) &=& \frac{\sin(2\pi x_3) + 1}{2} .
\end{eqnarray*}

The function $f_2$ provides a quasi-V-shaped profile that is $C^\infty$:
the smoothing parameter $\delta=0.01$ rounds the vertex at $x_2=\tfrac{1}{2}$
while keeping the maximum deviation from a strict V below $0.02$ in absolute
value. By symmetry, $f_2'(\tfrac{1}{2})=0$ and $f_2\in C^\infty([0,1])$,
satisfying assumption~(A2) with $s=\infty$.

\paragraph{Experiment E1: baseline Monte Carlo experiments}

In a first set of experiments (E1), we evaluate all estimators considered in this paper across the parameter configurations described above. The number of observations varies across $n \in \{1000,2000,5000\}$ and the signal-to-noise ratio takes values $SNR \in \{0.7,0.9\}$. For the boosting-based estimators, the stopping iteration is selected in every replication by spatial-block cross-validation with a ceiling of 2000 iterations, following the protocol of Section~\ref{sec:mstop}; the plain \textit{gamboost} benchmark uses spatially scattered folds.

\paragraph{Experiment E2: unobserved spatial heterogeneity in the errors}

In a second experiment (E2), we investigate the robustness of the
estimators to unobserved spatial heterogeneity in the error term,
$\tilde\epsilon_i = \epsilon_i + 0.5\,
\sqrt{|u_i - 0.35| + |v_i - 0.65|}$, a deterministic spatial shift
of the disturbances that no term of the fitted model represents.

\paragraph{Experiment E3: a factorial design for selection and
spatial confounding}

The last experiment merges the former selection and misspecification
scenarios into a single factorial design that varies independently
the three ingredients of the geoadditive identification trap
(Section~\ref{sec:E3p_results}): \emph{where} spatial structure
enters the true covariates, \emph{what} contaminates the candidate
set, and \emph{how} the regression is specified.
\begin{itemize}
\item \textbf{Axis S --- spatial structure of the true covariates.}
S0: none (i.i.d.\ covariates, as in E1);
S1: $x_1$ is spatially autocorrelated,
$\tilde x_1=(I-0.7W)^{-1}x_1$;
S2: $x_2$ is a deterministic function of the coordinates,
$\tilde x_2=\sqrt{(u_i-0.2)^2+(v_i-0.2)^2}$.
\item \textbf{Axis C --- contamination of the candidate set.}
C0: the true covariates only; C1: two irrelevant covariates
correlated with the true ones ($0.2$ with $x_1$, $0.7$ with $x_2$)
plus i.i.d.\ noise; C2: three normalised Euclidean distances to
fixed points ($d_1,d_2,d_3$) plus noise --- irrelevant candidates
that \emph{look like} spatial autocorrelation.
\item \textbf{Axis F --- specification of the fitted model.}
F0: oracle (true covariates only);
F1: full specification (all candidates);
F2: F1 plus a \emph{free} bivariate geographic smooth $s(u,v)$;
F2p: F1 plus the spectrally \emph{penalized} $s(u,v)$ of
Section~\ref{sec:remedies} (basis \texttt{spx},
$k_{\mathrm{spec}}=100$ dominant eigenvectors of $W$), with $\mu_s$
selected by REML or fixed on the grid
$\mu_s\in\{10^{-2},\dots,10^{4}\}$;
F2c and F2pc: as F2 and F2p with cross-fitting (suffix c:
out-of-fold residuals in the CFE quasi-score,
Remark~\ref{rem:crossfit}; the cross-fitted oracle F0c serves as
benchmark);
F3: F1 plus a deliberately \emph{over-penalized} $s(u,v)$ (spectral
penalty applied to the covariate smooths as well, S2 cells only),
included as a cautionary benchmark of restricted-spatial-regression
type.
\end{itemize}
Under C0 the full specification coincides with the oracle, which leaves
33 distinct cells, each crossed with
$\rho_0\in\{0,0.2,0.6,0.9\}$; $n=1000$, SNR$=0.7$, $nr=1000$
replications with common random numbers across cells and arms. The
estimators are \texttt{GAM\_SAR\_CFE} (all F arms) and
\texttt{BSPA\_SAR\_CFE} (F0--F2, default \texttt{mstop}$=100$). Two
companion sub-experiments complete the design: a \texttt{mstop}
sub-experiment on the control (S0$\times$C0) and adversarial
(S2$\times$C2) cells ($nr=250$) comparing fixed
\texttt{mstop}$\in\{100,400,1000\}$ with cross-validated values
under random and spatial-block folds; and a sample-size check
replicating both cells at $n=5000$ ($nr=250$).

For each fit we record the bias and RMSE of $\hat\rho$, the
empirical size of the 5\% $z$-test of $H_0:\rho=0$ and the coverage
of the 95\% confidence interval (closed-form variance), the RMSE of
the nonlinear components, the spectral projection diagnostic
$\bar\pi$ (mean of $\pi_k$ over the 25 dominant eigenvectors of
$W$, Figure~\ref{fig:pi_k}), and the boosting selection shares;
component biases are omitted because centring constraints keep them
near zero.
\ifpreprint
Replication-level results, job scripts and a frozen design document are
available in the replication repository
\url{https://github.com/ggeniaux/replication_spboost_geniaux_2026}
(folder \path{replication_scripts/e3prime}).
\else
Per-replication outputs, job scripts and a frozen design
document are provided with the reproduction material
(\texttt{inst/scripts/e3prime}).
\fi

Table~\ref{MCdesignE3_E4} summarises the simulation design across all experiments; the list of estimators and R packages is provided in Table~\ref{MCdesign_estimator} (Appendix~\ref{appendix_additional_mc}).

\begin{table}[htp]
\caption{Monte Carlo design}
\label{MCdesignE3_E4}
\begin{center}
\resizebox{0.95\linewidth}{!}{%
\begin{tabular}{|l|c|c|c|}
\hline
 & E1 & E2 & E3 \\
\hline
DGP & SAR, SEM & SAR & SAR \\
\hline
$\rho_0$ (or $\lambda_0$) & \multicolumn{3}{c|}{$\{0,\,0.2,\,0.6,\,0.9\}$} \\
\hline
$\epsilon_i$ & $\mathcal N(0,\sigma^2)$ i.i.d.
 & $\epsilon_i + 0.5\sqrt{|u_i-0.35|+|v_i-0.65|}$
 & $\mathcal N(0,\sigma^2)$ i.i.d. \\
\hline
$n$ & $\{1000,\,2000,\,5000\}$ & 2000 & 1000 (check at 5000) \\
\hline
SNR & $\{0.7,\,0.9\}$ & 0.7 & 0.7 \\
\hline
Replications $nr$ & \multicolumn{3}{c|}{1000 (sub-experiments: 250)} \\
\hline
Locations, $W$ & \multicolumn{3}{c|}{uniform on $[0,1]^2$;
row-normalised 4-nearest-neighbour $W$} \\
\hline
Nonlinear components & \multicolumn{3}{c|}{$f_1(x)=x-0.5$;\quad
$f_2(x)=\sqrt{3[(2x-1)^2+\delta^2]}-\delta\sqrt3,\ \delta=0.01$;\quad
$f_3(x)=(\sin(2\pi x)+1)/2$} \\
\hline
Spatial structure of & none & none
 & S0 none / S1 $\tilde x_1=(I-0.7W)^{-1}x_1$ \\
true covariates (S) & & & / S2 $\tilde x_2=$ distance to $(0.2,0.2)$ \\
\hline
Candidate contamination (C) & none & none
 & C0 none / C1 correlated (0.2, 0.7) $+$ noise \\
 & & & / C2 distances $d_1$--$d_3$ $+$ noise \\
\hline
Specification arms (F) & oracle & oracle
 & oracle / full / free $s(u,v)$ / spectral penalty \\
 & & & (REML or fixed $\mu_s$) / cross-fitted / over-penalized \\
\hline
Number of cells & 48 & 4 & $33\times4$ ($+$ \texttt{mstop} and $n$
sub-experiments) \\
\hline
\end{tabular}}
\end{center}
\end{table}

\subsection{Monte Carlo results}

\subsubsection{Monte Carlo results for E1 experiments}

The E1 Monte Carlo experiments assess the finite-sample behaviour of the estimators under controlled conditions. The objective is to examine how estimation accuracy evolves with the sample size and the signal-to-noise ratio while the data-generating process remains fixed. This design allows a direct comparison of the proposed semiparametric spatial estimators with standard benchmarks under varying degrees of spatial dependence.

Detailed results for a representative configuration ($n=1000$, $\mathrm{SNR}=0.9$) are reported in Tables~\ref{tab:SAR_E1_1000_09} and \ref{tab:SEM_E1_1000_09} for SAR and SEM designs, respectively. Synthetic tables (Tables~\ref{tab:SAR_E1_synt} and \ref{tab:SEM_E1_synth}) summarise the results across sample sizes and signal-to-noise ratios in order to highlight the main finite-sample patterns.

Because the smooth components are estimated under a centering constraint, their average bias is theoretically close to zero. Consequently, the tables report both bias and Root Mean Squared Error (RMSE) for the spatial parameter and the intercept term $f_0$, while only RMSE values are reported for the nonlinear components $f_1$, $f_2$, and $f_3$.

\paragraph{SAR design}

Table~\ref{tab:SAR_E1_1000_09} reports the results for the nonlinear SAR data-generating process. When $\rho=0$, the non-spatial nonlinear benchmarks (\textit{gam} and \textit{gamboost}) estimate the smooth components accurately. However, their performance deteriorates rapidly as spatial dependence increases. For example, the RMSE of $f_1$ obtained with \textit{gam} increases from 0.0104 when $\rho=0$ to 0.2082 when $\rho=0.9$. This reflects the well-known effect that ignoring spatial dependence induces spatially correlated residuals and distorts the estimation of nonlinear components.

The linear SAR benchmark (\textit{lagsarlm}) exhibits a qualitatively different type of misspecification. Because the true covariates $x_1$, $x_2$, $x_3$ are i.i.d., their nonlinear effects --- though entirely omitted from the linear specification --- do not induce spatial autocorrelation in the residuals. As a consequence, \textit{lagsarlm} achieves nearly unbiased estimates of $\rho$ across all values (bias below 0.003 in Table~\ref{tab:SAR_E1_1000_09}). Its failure is of a different nature: the nonlinear components are simply not modelled, so the RMSE for $f_2$ and $f_3$ stays at 0.50 and 0.22 respectively --- essentially the norm of their nonlinear parts --- at every value of $\rho$, while the linear $f_1$ is, by construction, correctly captured. This result highlights an important distinction: with spatially unstructured covariates, a linear SAR model can reliably identify the spatial parameter while leaving the nonlinear signal entirely unexplained.

By contrast, the proposed semiparametric SAR estimators display stable behaviour across all values of $\rho$. Both the boosted estimators (\texttt{BSPA\_SAR\_ML}, \texttt{BSPA\_SAR\_CFE}) and the penalized estimators fitted on the known specification (\texttt{GAM\_SAR\_ML}, \texttt{GAM\_SAR\_CFE}) simultaneously achieve small bias and RMSE for the spatial parameter while maintaining moderate RMSE levels for the nonlinear components. An important feature of these estimators is that, unlike the non-spatial benchmarks, the RMSEs of $f_1$, $f_2$, and $f_3$ remain stable as $\rho$ increases. In Table~\ref{tab:SAR_E1_1000_09}, \texttt{BSPA\_SAR\_CFE} yields essentially constant RMSE values for $f_1$ of approximately 0.0153 across all values of $\rho$ (from 0 to 0.9), whereas the corresponding values for \textit{gam} rise from 0.0104 to 0.2082. The near-constant RMSE profile of \texttt{BSPA\_SAR\_CFE} reflects the clean separation achieved by the spatial filtering step: once $\rho WY$ is properly accounted for, the quality of the nonlinear component estimates is essentially independent of the strength of spatial dependence.

Among the semiparametric variants, \texttt{BSPA\_SAR\_ML} and \texttt{BSPA\_SAR\_CFE} are virtually indistinguishable (bias difference below 0.001 at all $\rho$), confirming that the determinant-free step introduces no meaningful efficiency loss. The synthetic results in Table~\ref{tab:SAR_E1_synt} show the expected convergence pattern across sample sizes and SNR levels.

\paragraph{SEM design}

Table~\ref{tab:SEM_E1_1000_09} reports the corresponding SEM results. Spatial error dependence enters through the disturbance rather than the dependent variable, so misspecification of non-spatial estimators causes efficiency losses rather than the strong structural distortions observed under SAR. The semiparametric SEM estimators remain well-behaved at low to moderate $\lambda$: \texttt{BSPA\_SEM\_CFE} matches or slightly outperforms \texttt{BSPA\_SEM\_ML} in that range. The picture reverses at $\lambda=0.9$, where the ill-conditioning of $(I-\lambda W)^{-1}$ causes the CFE variant to deteriorate (RMSE$(\hat\lambda)=0.0395$ versus $0.0210$ for ML, bias $=-0.0329$), consistent with the recommendation in Section~\ref{sec:mstop} to prefer full ML near the boundary.

The synthetic SEM results (Table~\ref{tab:SEM_E1_synth}) confirm this pattern: \texttt{GAM\_SEM\_CFE} is competitive at weak dependence but loses accuracy markedly as $\lambda$ approaches 1 (RMSE of $f_1$ rising from 0.0123 to 0.1267 at $n=1{,}000$, SNR=0.9). The SEM case is therefore treated as secondary in the remainder of the paper, which focuses on the SAR specification most relevant for the empirical application.

\begin{table}[!h]
\centering
\caption{Results for experiment E1, nonlinear SAR, SNR$=0.9$, $n=1000$. Monte Carlo results over 1000 replications; \texttt{mstop} of the boosting arms selected by spatial-block cross-validation (median reported).}
\label{tab:SAR_E1_1000_09}
\centering
\resizebox{0.85\linewidth}{!}{%
\begin{tabular}[t]{llcccccccc}
\toprule
\multicolumn{2}{c}{ } & \multicolumn{2}{c}{Spatial parameter} & \multicolumn{2}{c}{Intercept} & \multicolumn{3}{c}{Nonlinear components} & \multicolumn{1}{c}{Boosting} \\
\cmidrule(l{3pt}r{3pt}){3-4} \cmidrule(l{3pt}r{3pt}){5-6} \cmidrule(l{3pt}r{3pt}){7-9} \cmidrule(l{3pt}r{3pt}){10-10}
\multicolumn{2}{c}{ } & \multicolumn{2}{c}{$\rho$} & \multicolumn{2}{c}{$f_0$} & \multicolumn{1}{c}{$f_1$} & \multicolumn{1}{c}{$f_2$} & \multicolumn{1}{c}{$f_3$} & \multicolumn{1}{c}{$m_{stop}$} \\
Estimator &   & Bias & RMSE & Bias & RMSE & RMSE & RMSE & RMSE & Median\\
\midrule
 & 0 & - & - & 0.0001 & 0.0070 & 0.0104 & 0.0260 & 0.0184 & -\\
 & 0.2 & - & - & 0.0001 & 0.0090 & 0.0117 & 0.0274 & 0.0192 & -\\
 & 0.6 & - & - & -0.0002 & 0.0236 & 0.0405 & 0.0710 & 0.0501 & -\\
\multirow{-4}{*}{\raggedright\arraybackslash gam} & 0.9 & - & - & -0.0024 & 0.1528 & 0.2082 & 0.3410 & 0.2499 & -\\
\cmidrule{1-10}
 & 0 & - & - & 0.0001 & 0.0070 & 0.0153 & 0.0266 & 0.0169 & 1190\\
 & 0.2 & - & - & 0.0001 & 0.0090 & 0.0166 & 0.0282 & 0.0181 & 1105\\
 & 0.6 & - & - & -0.0002 & 0.0236 & 0.0431 & 0.0728 & 0.0509 & 420\\
\multirow{-4}{*}{\raggedright\arraybackslash gamboost} & 0.9 & - & - & -0.0024 & 0.1528 & 0.2016 & 0.3441 & 0.2488 & 180\\
\cmidrule{1-10}
 & 0 & -0.0021 & 0.0428 & -0.0202 & 0.0640 & 0.0187 & 0.4990 & 0.2219 & -\\
 & 0.2 & -0.0023 & 0.0385 & -0.0202 & 0.0640 & 0.0187 & 0.4990 & 0.2219 & -\\
 & 0.6 & -0.0022 & 0.0247 & -0.0202 & 0.0640 & 0.0187 & 0.4990 & 0.2220 & -\\
\multirow{-4}{*}{\raggedright\arraybackslash lagsarlm} & 0.9 & -0.0012 & 0.0084 & -0.0202 & 0.0640 & 0.0187 & 0.4990 & 0.2220 & -\\
\cmidrule{1-10}
 & 0 & -0.0010 & 0.0198 & 0.0002 & 0.0070 & 0.0153 & 0.0267 & 0.0169 & 1162\\
 & 0.2 & -0.0009 & 0.0179 & 0.0001 & 0.0070 & 0.0153 & 0.0267 & 0.0169 & 1167\\
 & 0.6 & -0.0007 & 0.0115 & 0.0001 & 0.0070 & 0.0153 & 0.0267 & 0.0170 & 1162\\
\multirow{-4}{*}{\raggedright\arraybackslash BSPA\_SAR\_ML} & 0.9 & -0.0003 & 0.0038 & 0.0001 & 0.0071 & 0.0152 & 0.0267 & 0.0169 & 1180\\
\cmidrule{1-10}
 & 0 & -0.0009 & 0.0199 & 0.0001 & 0.0070 & 0.0153 & 0.0267 & 0.0169 & 1165\\
 & 0.2 & -0.0008 & 0.0179 & 0.0001 & 0.0070 & 0.0153 & 0.0267 & 0.0169 & 1168\\
 & 0.6 & -0.0006 & 0.0121 & 0.0001 & 0.0070 & 0.0153 & 0.0267 & 0.0170 & 1164\\
\multirow{-4}{*}{\raggedright\arraybackslash BSPA\_SAR\_CFE} & 0.9 & -0.0003 & 0.0049 & 0.0001 & 0.0071 & 0.0152 & 0.0267 & 0.0170 & 1179\\
\cmidrule{1-10}
 & 0 & -0.0010 & 0.0199 & 0.0001 & 0.0070 & 0.0104 & 0.0260 & 0.0184 & -\\
 & 0.2 & -0.0009 & 0.0180 & 0.0001 & 0.0070 & 0.0105 & 0.0260 & 0.0184 & -\\
 & 0.6 & -0.0007 & 0.0115 & 0.0001 & 0.0070 & 0.0104 & 0.0260 & 0.0184 & -\\
\multirow{-4}{*}{\raggedright\arraybackslash GAM\_SAR\_ML} & 0.9 & -0.0004 & 0.0039 & 0.0001 & 0.0071 & 0.0104 & 0.0260 & 0.0184 & -\\
\cmidrule{1-10}
 & 0 & -0.0009 & 0.0196 & 0.0001 & 0.0070 & 0.0103 & 0.0260 & 0.0184 & -\\
 & 0.2 & -0.0024 & 0.0179 & 0.0001 & 0.0070 & 0.0103 & 0.0260 & 0.0184 & -\\
 & 0.6 & -0.0041 & 0.0129 & 0.0001 & 0.0071 & 0.0104 & 0.0261 & 0.0184 & -\\
\multirow{-4}{*}{\raggedright\arraybackslash GAM\_SAR\_CFE} & 0.9 & -0.0032 & 0.0073 & 0.0001 & 0.0074 & 0.0104 & 0.0261 & 0.0184 & -\\
\bottomrule
\end{tabular}}
\end{table}

\begin{table}[!h]
\centering
\caption{Results for experiment E1, nonlinear SEM, SNR$=0.9$, $n=1000$. Monte Carlo results over 1000 replications; \texttt{mstop} of the boosting arms selected by spatial-block cross-validation (median reported).}
\label{tab:SEM_E1_1000_09}
\centering
\resizebox{0.85\linewidth}{!}{%
\begin{tabular}[t]{llcccccccc}
\toprule
\multicolumn{2}{c}{ } & \multicolumn{2}{c}{Spatial parameter} & \multicolumn{2}{c}{Intercept} & \multicolumn{3}{c}{Nonlinear components} & \multicolumn{1}{c}{Boosting} \\
\cmidrule(l{3pt}r{3pt}){3-4} \cmidrule(l{3pt}r{3pt}){5-6} \cmidrule(l{3pt}r{3pt}){7-9} \cmidrule(l{3pt}r{3pt}){10-10}
\multicolumn{2}{c}{ } & \multicolumn{2}{c}{$\lambda$} & \multicolumn{2}{c}{$f_0$} & \multicolumn{1}{c}{$f_1$} & \multicolumn{1}{c}{$f_2$} & \multicolumn{1}{c}{$f_3$} & \multicolumn{1}{c}{$m_{stop}$} \\
Estimator &   & Bias & RMSE & Bias & RMSE & RMSE & RMSE & RMSE & Median\\
\midrule
 & 0 & - & - & 0.0001 & 0.0070 & 0.0104 & 0.0260 & 0.0184 & -\\
 & 0.2 & - & - & 0.0002 & 0.0087 & 0.0103 & 0.0260 & 0.0183 & -\\
 & 0.6 & - & - & 0.0002 & 0.0148 & 0.0101 & 0.0260 & 0.0183 & -\\
\multirow{-4}{*}{\raggedright\arraybackslash gam} & 0.9 & - & - & 0.0002 & 0.0282 & 0.0101 & 0.0260 & 0.0181 & -\\
\cmidrule{1-10}
 & 0 & - & - & 0.0001 & 0.0070 & 0.0153 & 0.0266 & 0.0169 & 1190\\
 & 0.2 & - & - & 0.0002 & 0.0087 & 0.0153 & 0.0266 & 0.0169 & 1190\\
 & 0.6 & - & - & 0.0002 & 0.0148 & 0.0151 & 0.0267 & 0.0168 & 1170\\
\multirow{-4}{*}{\raggedright\arraybackslash gamboost} & 0.9 & - & - & 0.0002 & 0.0282 & 0.0150 & 0.0267 & 0.0167 & 1185\\
\cmidrule{1-10}
 & 0 & -0.0021 & 0.0482 & -0.0201 & 0.0641 & 0.0187 & 0.4990 & 0.2219 & -\\
 & 0.2 & -0.1693 & 0.1758 & -0.0200 & 0.0642 & 0.0186 & 0.4990 & 0.2219 & -\\
 & 0.6 & -0.4720 & 0.4743 & -0.0198 & 0.0649 & 0.0184 & 0.4990 & 0.2219 & -\\
\multirow{-4}{*}{\raggedright\arraybackslash errorsarlm} & 0.9 & -0.6708 & 0.6725 & -0.0193 & 0.0680 & 0.0181 & 0.4990 & 0.2219 & -\\
\cmidrule{1-10}
 & 0 & -0.0031 & 0.0460 & 0.0001 & 0.0070 & 0.0152 & 0.0267 & 0.0169 & 1160\\
 & 0.2 & -0.0080 & 0.0422 & 0.0002 & 0.0089 & 0.0151 & 0.0264 & 0.0167 & 1210\\
 & 0.6 & -0.0106 & 0.0289 & 0.0007 & 0.0187 & 0.0126 & 0.0230 & 0.0140 & 1704\\
\multirow{-4}{*}{\raggedright\arraybackslash BSPA\_SEM\_ML} & 0.9 & -0.0181 & 0.0210 & 0.0023 & 0.0671 & 0.0052 & 0.0162 & 0.0068 & 2000\\
\cmidrule{1-10}
 & 0 & -0.0025 & 0.0456 & 0.0002 & 0.0070 & 0.0152 & 0.0267 & 0.0169 & 1162\\
 & 0.2 & -0.0063 & 0.0416 & 0.0002 & 0.0089 & 0.0151 & 0.0264 & 0.0167 & 1220\\
 & 0.6 & -0.0124 & 0.0304 & 0.0007 & 0.0187 & 0.0126 & 0.0230 & 0.0140 & 1698\\
\multirow{-4}{*}{\raggedright\arraybackslash BSPA\_SEM\_CFE} & 0.9 & -0.0329 & 0.0395 & 0.0022 & 0.0606 & 0.0053 & 0.0163 & 0.0069 & 2000\\
\cmidrule{1-10}
 & 0 & -0.0025 & 0.0456 & 0.0001 & 0.0071 & 0.0123 & 0.0340 & 0.0234 & -\\
 & 0.2 & -0.0063 & 0.0416 & 0.0002 & 0.0072 & 0.0308 & 0.0963 & 0.0633 & -\\
 & 0.6 & -0.0121 & 0.0303 & 0.0003 & 0.0077 & 0.0865 & 0.2495 & 0.1569 & -\\
\multirow{-4}{*}{\raggedright\arraybackslash GAM\_SEM\_CFE} & 0.9 & -0.0317 & 0.0393 & 0.0003 & 0.0079 & 0.1267 & 0.3331 & 0.2081 & -\\
\bottomrule
\end{tabular}}
\end{table}
 
\paragraph{Confidence intervals: reliable for SAR, fragile near the SEM boundary}

Empirical coverage rates of nominal 95\,\% confidence intervals
$\hat\rho_{CFE}\pm 1.96\,\hat\sigma(\rho)$ are reported in
Appendix~\ref{appendix_additional_mc}. For the SAR design, coverage
is at or above the nominal level across all values of $\rho$ and sample
sizes. The variance ratio
$\hat\sigma^2_{MC}/\overline{\hat\sigma^2(\hat\rho)}$ falls
substantially below~1 for large $\rho$ (0.25--0.33 at $\rho=0.9$),
indicating that the analytical formula over-estimates sampling
variance under strong autocorrelation, which explains the
conservative coverage (0.991--1.000) in that regime. For the SEM
design, the same nominal-to-conservative pattern holds up to
$\lambda=0.6$, but the boundary case $\lambda=0.9$ exposes the
fragility of the closed-form variance: the denominator $b-2\hat\lambda a$
becomes negative in a sizeable share of replications --- the regime in
which the \texttt{cfe\_weighted\_projection} fallback is activated ---
so that the mean closed-form variance can itself turn negative
(negative ratios in Table~\ref{tab:coverage_SEM}) and coverage falls
to 0.90--0.94 once replications with an invalid variance are counted
as non-covering. Practitioners should not rely on the closed-form CFE
confidence intervals for the SEM model when $\hat\lambda$
exceeds~0.85, and should switch to the full ML variance there, in line
with the recommendation of Section~\ref{sec:mstop}.

\paragraph{Computational scalability}

Table~\ref{comptime} reports single-core computation times at \texttt{mstop}$=150$ (lower bounds for boosting variants, since CV multiplies fitting time by $K$). The determinant-free formulation removes the main bottleneck of likelihood-based spatial estimation: \texttt{GAM\_SAR\_CFE} completes in 0.063~s at $n=1{,}000$ and 0.90~s at $n=250{,}000$, against 4.3~s and 55.7~s for \texttt{lagsarlm} (factor of~$\approx 60$) with the advantage of a more flexible nonlinear model. \texttt{BSPA\_SAR\_CFE}, which additionally performs variable selection, is itself faster than \texttt{lagsarlm} at $n=250{,}000$ (19.7~s vs 55.7~s). Above $n=12{,}000$, \textit{spboost} package switches from \texttt{gam} to \texttt{bam} (\textit{Big Additive Models}, the large-$n$ backend of \texttt{mgcv} that exploits discretisation of covariates and sparse matrix arithmetic to reduce complexity from $O(n^2)$ to $O(n)$), making \texttt{GAM\_SAR\_CFE} the most efficient estimator at large $n$; \texttt{BSPA\_SAR\_CFE} provides the best trade-off when variable selection is also required. For very large samples where boosting-based specification search becomes costly, we recommend a two-step workflow: run \texttt{BSPA\_SAR\_CFE} on one or several subsamples to identify the relevant covariates and functional forms, then re-estimate the resulting fixed specification on the full dataset using \texttt{GAM\_SAR\_CFE}, which benefits from the \texttt{bam} backend and remains the fastest option at scale.

\begin{table}
\caption{Average computation time in seconds, single-core execution (MacBook Pro, Apple M1 Max, 64 GB RAM, 10 performance cores + 2 efficiency cores)}
\centering
\resizebox{0.9\linewidth}{!}{
\begin{tabular}[t]{rrrrrrrrr}
\toprule
n & k & gamboost & gam & lagsarlm & BSPA\_SAR\_ML & GAM\_SAR\_ML & BSPA\_SAR\_CFE & GAM\_SAR\_CFE\\
\midrule
1000 & 3 & 0.157 & 0.013 & 4.310 & 0.730 & 0.176 & 0.189 & 0.063\\
5000 & 3 & 0.289 & 0.021 & 0.631 & 1.620 & 0.367 & 0.425 & 0.098\\
10000 & 3 & 0.209 & 0.036 & 1.148 & 2.603 & 0.770 & 0.762 & 0.159\\
50000 & 3 & 1.279 & 0.056 & 6.450 & 14.042 & 4.715 & 4.006 & 0.245\\
250000 & 3 & 5.959 & 0.261 & 55.715 & 81.492 & 24.794 & 19.650 & 0.903\\
\midrule
1000 & 10 & 0.129 & 0.285 & 4.343 & 2.065 & 5.302 & 0.561 & 1.578\\
5000 & 10 & 0.522 & 0.504 & 1.373 & 5.468 & 6.192 & 1.324 & 1.711\\
10000 & 10 & 0.742 & 0.350 & 2.511 & 8.641 & 5.497 & 2.467 & 1.685\\
50000 & 10 & 3.228 & 0.149 & 15.946 & 38.761 & 20.354 & 12.624 & 0.614\\
\midrule
1000 & 20 & 0.278 & 2.725 & 4.380 & 5.044 & 91.057 & 1.062 & 13.847\\
5000 & 20 & 0.791 & 2.714 & 2.880 & 11.752 & 42.920 & 2.771 & 11.541\\
10000 & 20 & 1.369 & 7.416 & 6.101 & 15.028 & 121.911 & 5.161 & 22.801\\
50000 & 20 & 5.827 & 0.345 & 39.053 & 71.428 & 114.824 & 21.428 & 1.428\\
\bottomrule
\end{tabular}}
\label{comptime}
\end{table}
 
\subsubsection{Monte Carlo results for the E2 experiment}

The E2 experiment extends the E1 benchmark by introducing unobserved
spatial heterogeneity into the disturbances, while keeping the sample
size fixed at $n=2000$ and the signal-to-noise ratio at
$\mathrm{SNR}=0.7$: the error becomes
$\tilde\epsilon_i=\epsilon_i+0.5\sqrt{|u_i-0.35|+|v_i-0.65|}$, a
deterministic spatial shift that no term of the fitted models
represents. (Spatial structure entering through the covariates is
studied in the E3 factorial below, where it interacts with the
candidate set and the specification.)

\begin{table}[htbp]
\centering
\caption{Results for experiment E2 (unobserved spatial heterogeneity in the disturbances), nonlinear SAR, $n=2000$, SNR$=0.7$, 1000 replications; \texttt{mstop} by spatial-block cross-validation.}
\label{tab:SAR_E2}
\resizebox{0.8\linewidth}{!}{%
\begin{tabular}{llcccccccc}
\toprule
\multicolumn{2}{c}{ } & \multicolumn{2}{c}{$\rho$} & \multicolumn{2}{c}{$f_0$} & \multicolumn{1}{c}{$f_1$} & \multicolumn{1}{c}{$f_2$} & \multicolumn{1}{c}{$f_3$} & Coverage\\
Estimator &   & Bias & RMSE & Bias & RMSE & RMSE & RMSE & RMSE & 95\% CI\\
\midrule
 & 0 & 0.0266 & 0.0411 & -0.0227 & 0.0543 & 0.0156 & 0.4994 & 0.2218 & 0.823\\
 & 0.2 & 0.0259 & 0.0382 & -0.0227 & 0.0540 & 0.0155 & 0.4994 & 0.2218 & 0.817\\
 & 0.6 & 0.0205 & 0.0270 & -0.0226 & 0.0528 & 0.0152 & 0.4994 & 0.2217 & 0.741\\
\multirow{-4}{*}{\raggedright\arraybackslash lagsarlm} & 0.9 & 0.0097 & 0.0112 & -0.0224 & 0.0498 & 0.0143 & 0.4994 & 0.2217 & 0.526\\
\cmidrule{1-10}
 & 0 & 0.0360 & 0.0426 & 0.0002 & 0.0095 & 0.0199 & 0.0337 & 0.0222 & 0.602\\
 & 0.2 & 0.0326 & 0.0384 & 0.0002 & 0.0093 & 0.0197 & 0.0334 & 0.0220 & 0.630\\
 & 0.6 & 0.0227 & 0.0261 & 0.0002 & 0.0087 & 0.0188 & 0.0321 & 0.0210 & 0.745\\
\multirow{-4}{*}{\raggedright\arraybackslash BSPA\_SAR\_CFE} & 0.9 & 0.0103 & 0.0113 & 0.0002 & 0.0067 & 0.0156 & 0.0275 & 0.0174 & 0.927\\
\cmidrule{1-10}
 & 0 & 0.0359 & 0.0424 & 0.0002 & 0.0095 & 0.0136 & 0.0320 & 0.0243 & 0.606\\
 & 0.2 & 0.0320 & 0.0378 & 0.0002 & 0.0093 & 0.0135 & 0.0318 & 0.0241 & 0.637\\
 & 0.6 & 0.0213 & 0.0249 & 0.0002 & 0.0087 & 0.0127 & 0.0307 & 0.0230 & 0.780\\
\multirow{-4}{*}{\raggedright\arraybackslash GAM\_SAR\_CFE} & 0.9 & 0.0091 & 0.0104 & 0.0002 & 0.0068 & 0.0105 & 0.0268 & 0.0192 & 0.946\\
\bottomrule
\end{tabular}}
\end{table}
 
Table~\ref{tab:SAR_E2} shows a consistent pattern across estimators.
The unobserved heterogeneity is partially attributed to the
autoregressive term, producing a modest positive bias on $\hat\rho$
for all three estimators --- largest when the true dependence is
weak and vanishing as $\rho_0$ grows --- while the nonlinear
components are estimated with RMSE values close to the E1 benchmark.
The linear \textit{lagsarlm} benchmark shares the same bias on
$\hat\rho$ but, as in E1, leaves the nonlinear signal unexplained.
The practical reading is that a smooth unobserved spatial shift in
the disturbances contaminates $\hat\rho$ moderately and
symmetrically across linear and semiparametric estimators: it is a
misspecification of the error model, not a weakness specific to the
flexible first stage.

\subsubsection{Monte Carlo results for the E3 factorial}
\label{sec:E3p_results}

\begin{figure}[htbp]
\centering
\includegraphics[width=\linewidth]{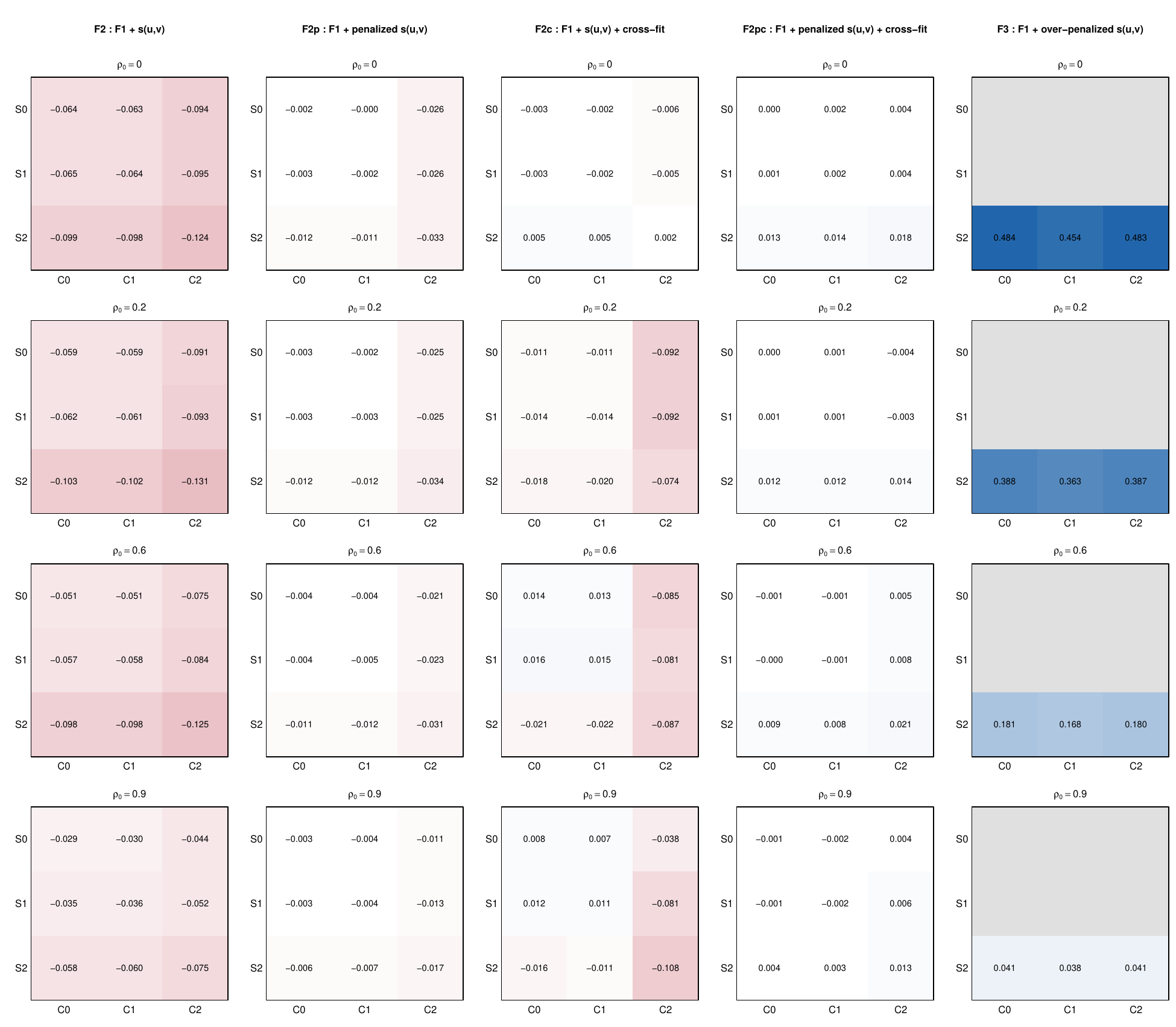}
\caption{Experiment E3: mean bias of $\hat\rho$ by cell
(within each panel, rows: axis S; columns: axis C) for five fitted
models of axis F (panel columns: F2, F2p, F2c, F2pc, F3) and four
values of $\rho_0$ (panel rows), \texttt{GAM\_SAR\_CFE}, 1000
replications per cell. Red: downward bias (the smoother absorbs the
autoregressive signal); blue: upward bias (true covariate signal
pushed into $\hat\rho$); grey: arm not defined in the cell.}
\label{fig:E3p_bias}
\end{figure}

\paragraph{The trap is graded, and geography --- not autocorrelation
--- drives it}
Figure~\ref{fig:E3p_bias} maps the mean bias of $\hat\rho$ over the
factorial, and Table~\ref{tab:E3p_main} details the control cell
(S0$\times$C0) and the adversarial cell (S2$\times$C2). Three facts
stand out. First, the free geographic smooth biases $\hat\rho$
downward in \emph{every} cell, including the benchmark setting with
i.i.d.\ covariates and no contamination: at S0$\times$C0 the bias of
the F2 arm ranges from $-0.051$ ($\rho_0=0.6$) to $-0.064$
($\rho_0=0$), against $-0.002$ to $-0.004$ for the oracle. Second,
the two axes combine monotonically: distance-type candidates (C2)
add $-0.02$ to $-0.03$, a geographic true covariate (S2) another
$-0.03$ to $-0.04$, up to $-0.131$ at S2$\times$C2 for
$\rho_0=0.2$. Third, a merely \emph{spatially autocorrelated} covariate is
almost harmless: S1 is indistinguishable from S0 throughout (bias
differences below $0.004$). The danger is not spatial correlation in
the regressors --- the case most discussed in the spatial
confounding literature \citep{paciorek2010,hughes2013,keller2020}
--- but geography entering the regression disguised as a covariate
or as a candidate. The bias attenuates at $\rho_0=0.9$ (down to
$-0.058$ at S2$\times$C0) without vanishing: a strong autoregressive
signal is harder, but not impossible, to absorb.

\begin{table}[htbp]
\centering
\caption{Experiment E3: the geoadditive trap and its remedies in the
control cell (S0$\times$C0) and the adversarial cell (S2$\times$C2).
Monte Carlo results over 1000 replications, $n=1000$, SNR$=0.7$.
Size: empirical rejection rate of the 5\% $z$-test of $H_0:\rho=0$
($\rho_0=0$ block); coverage: empirical coverage of the 95\%
confidence interval ($\rho_0=0.6$ block), both based on the closed-form
variance. $\bar\pi$: mean spectral projection over the 25 dominant
eigenvectors of $W$ (Appendix~\ref{app:orthogonality} and
Figure~\ref{fig:pi_k}). Fitted-model codes as in axis F of the design;
the over-penalized arm F3 runs in S2 cells only.}
\label{tab:E3p_main}
\resizebox{0.9\linewidth}{!}{%
\begin{tabular}{llccccc}
\toprule
Cell & Fitted model & Bias & RMSE & Size / Cov. & RMSE $f_2$ & $\bar\pi$\\
\midrule
\multicolumn{7}{l}{\textbf{$\rho_0 = 0$ --- size of the test of $\rho=0$}}\\
S0$\times$C0 & F0 (oracle) & -0.002 & 0.031 & 0.050 & 0.042 & 0.04\\
 & F2 (F1 + $s(u,v)$) & -0.064 & 0.074 & 0.440 & 0.042 & 0.15\\
 & F2p, REML $\mu_s$ & -0.061 & 0.070 & 0.397 & 0.042 & 0.14\\
 & F2p, $\mu_s=10^4$ & -0.002 & 0.031 & 0.051 & 0.042 & 0.04\\
 & F2c (F2 + cross-fit) & -0.003 & 0.031 & 0.067 & 0.042 & 0.11\\
 & F2pc (F2p + cross-fit) & 0.000 & 0.031 & 0.054 & 0.042 & 0.04\\
 & F2, boosted (\texttt{BSPA\_SAR\_CFE}) & -0.012 & 0.034 & 0.069 & -- & --\\
\addlinespace[0.6ex]
S2$\times$C2 & F0 (oracle) & -0.011 & 0.040 & 0.070 & 0.044 & 0.14\\
 & F1 (full specification) & -0.033 & 0.052 & 0.157 & 0.224 & 0.21\\
 & F2 (F1 + $s(u,v)$) & -0.124 & 0.134 & 0.798 & 0.595 & 0.32\\
 & F2p, REML $\mu_s$ & -0.113 & 0.123 & 0.733 & 0.207 & 0.30\\
 & F2p, $\mu_s=10^4$ & -0.033 & 0.052 & 0.157 & 0.223 & 0.21\\
 & F2c (F2 + cross-fit) & 0.002 & 0.036 & 0.114 & 0.437 & 0.25\\
 & F2pc (F2p + cross-fit) & 0.018 & 0.044 & 0.124 & 0.235 & 0.20\\
 & F3 (over-penalized) & 0.483 & 0.483 & 1.000 & 0.453 & 0.03\\
 & F2, boosted (\texttt{BSPA\_SAR\_CFE}) & 0.006 & 0.039 & 0.054 & -- & --\\
\addlinespace[0.6ex]
\midrule
\multicolumn{7}{l}{\textbf{$\rho_0 = 0.6$ --- coverage of the 95\% CI}}\\
S0$\times$C0 & F0 (oracle) & -0.004 & 0.020 & 0.988 & 0.042 & 0.04\\
 & F2 (F1 + $s(u,v)$) & -0.051 & 0.057 & 0.578 & 0.042 & 0.20\\
 & F2p, REML $\mu_s$ & -0.050 & 0.056 & 0.597 & 0.042 & 0.18\\
 & F2p, $\mu_s=10^4$ & -0.004 & 0.020 & 0.988 & 0.042 & 0.04\\
 & F2c (F2 + cross-fit) & 0.014 & 0.030 & 0.849 & 0.042 & 0.11\\
 & F2pc (F2p + cross-fit) & -0.001 & 0.019 & 0.988 & 0.042 & 0.04\\
 & F2, boosted (\texttt{BSPA\_SAR\_CFE}) & -0.002 & 0.020 & 0.991 & -- & --\\
\addlinespace[0.6ex]
S2$\times$C2 & F0 (oracle) & -0.011 & 0.029 & 0.980 & 0.067 & 0.13\\
 & F1 (full specification) & -0.031 & 0.043 & 0.885 & 0.218 & 0.20\\
 & F2 (F1 + $s(u,v)$) & -0.125 & 0.131 & 0.108 & 1.219 & 0.41\\
 & F2p, REML $\mu_s$ & -0.122 & 0.129 & 0.135 & 0.177 & 0.37\\
 & F2p, $\mu_s=10^4$ & -0.031 & 0.043 & 0.885 & 0.219 & 0.21\\
 & F2c (F2 + cross-fit) & -0.087 & 0.172 & 0.282 & 1.996 & 0.37\\
 & F2pc (F2p + cross-fit) & 0.021 & 0.037 & 0.885 & 0.246 & 0.19\\
 & F3 (over-penalized) & 0.180 & 0.181 & 0.000 & 0.453 & 0.03\\
 & F2, boosted (\texttt{BSPA\_SAR\_CFE}) & -0.003 & 0.027 & 0.982 & -- & --\\
\addlinespace[0.6ex]
\bottomrule
\end{tabular}}
\end{table}
 
\begin{figure}[htbp]
\centering
\includegraphics[width=0.85\linewidth]{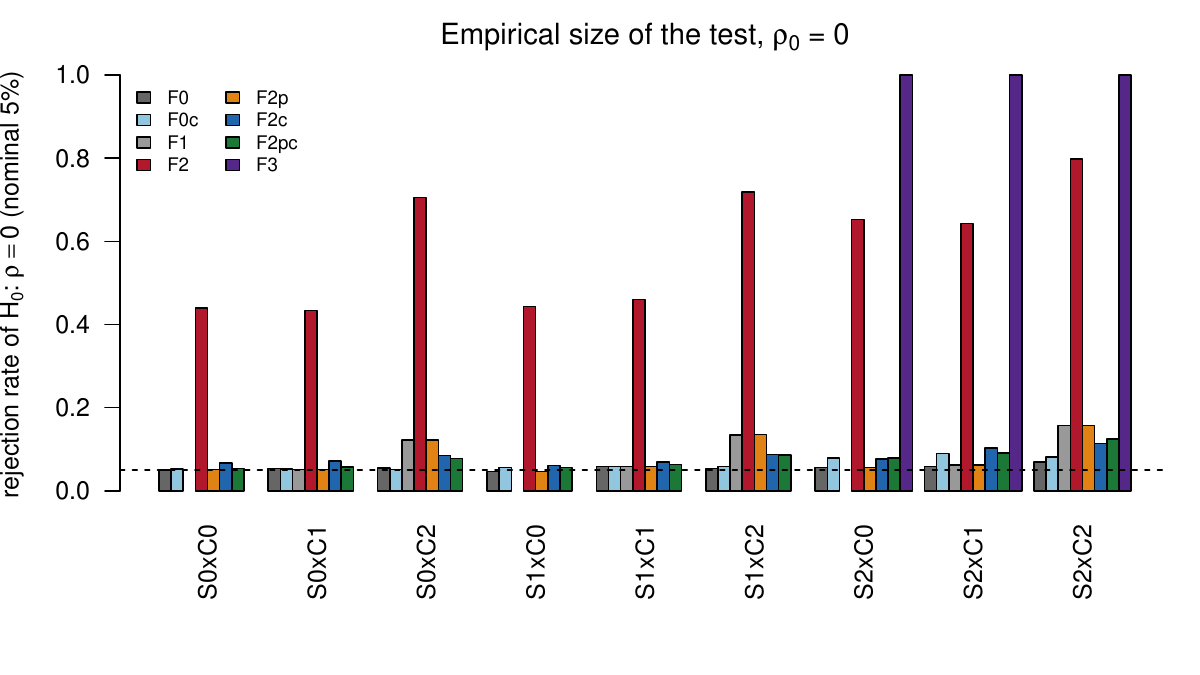}
\caption{Experiment E3: empirical rejection rate of the 5\%
$z$-test of $H_0:\rho=0$ when $\rho_0=0$, by cell and fitted model
of axis F (closed-form variance, 1000 replications; F0c is the
cross-fitted oracle). The dashed line is the nominal level.}
\label{fig:E3p_size}
\end{figure}

\paragraph{Testing $\rho=0$ under a free geoadditive specification
is invalid}
Figure~\ref{fig:E3p_size} reports the empirical size of the $z$-test
of $H_0:\rho=0$. Under the free smooth (F2), the test rejects in
43--72\% of replications at a 5\% nominal level --- 44\% already in
the healthy S0$\times$C0 cell: a free geoadditive specification
``detects'' spatial autocorrelation about one time in two when there
is none, in the direction of a spurious \emph{negative} $\hat\rho$.
The over-penalized arm F3 fails in the opposite direction: by
penalizing the covariate smooths as well, it expels the genuinely
spatial signal of $f_2(\tilde x_2)$ into the autoregressive term,
producing $\hat\rho\approx0.48$ when $\rho_0=0$ and a 100\%
rejection rate, while driving the spectral diagnostic to
oracle-like values ($\bar\pi\approx0.03$). This is the
anticonservatism documented for restricted spatial regression
\citep{hughes2013,keller2020} reproduced inside a single design,
and it shows that $\bar\pi$ is a diagnostic, not an objective
function. The two remedies below restore the size to 5--9\% in
every cell.

\begin{figure}[htbp]
\centering
\includegraphics[width=0.95\linewidth]{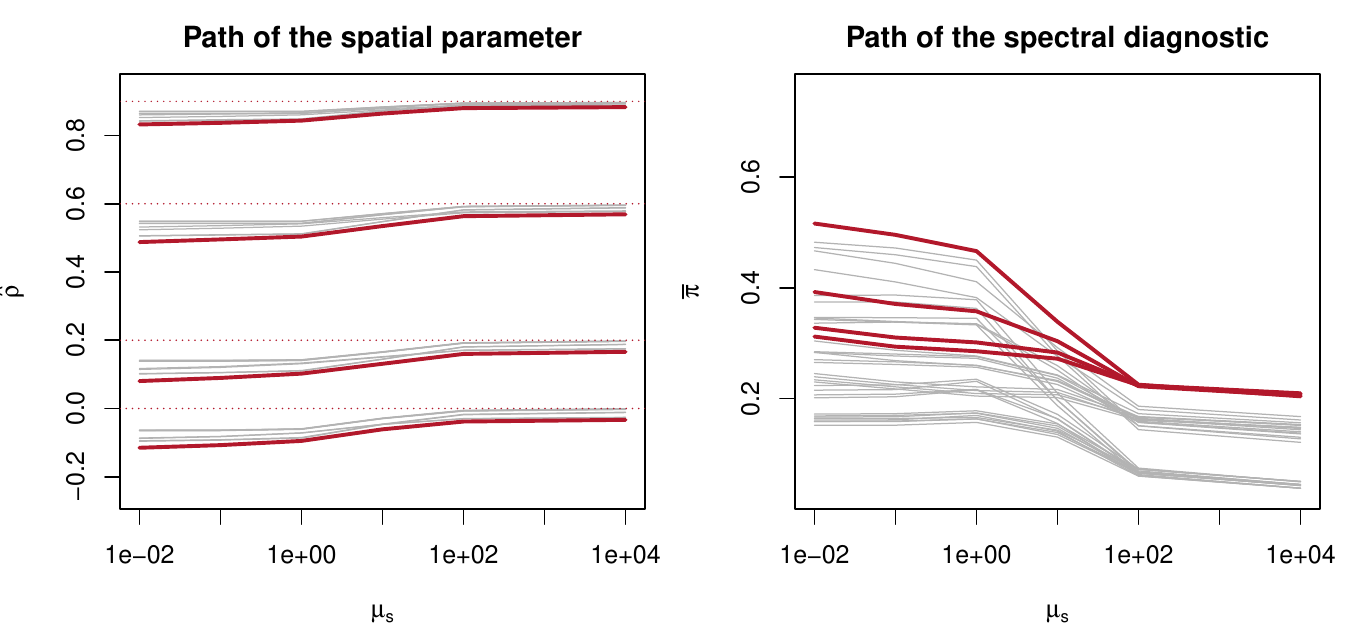}
\caption{Experiment E3: paths of $\hat\rho$ (left) and of the
spectral diagnostic $\bar\pi$ (right) along the spectral penalty
grid $\mu_s\in\{10^{-2},\dots,10^4\}$ (cell means; the adversarial
cell S2$\times$C2 in bold red, its true $\rho_0$ dotted).}
\label{fig:E3p_path}
\end{figure}

\paragraph{Two channels, two remedies, and their combination}
Section~\ref{sec:remedies} decomposes the trap into a population
channel (the free surface is redundant with $\rho W y$) and a
finite-sample channel (own-observation correlation between smoother
error and quasi-score residuals, formalized in
Appendix~\ref{app:orthogonality}). The factorial confirms that each
channel has its own remedy and that neither suffices alone.
The spectral penalty with $\mu_s$ selected by REML is inoperative
(in Table~\ref{tab:E3p_main}, the F2p rows with REML $\mu_s$ match F2
at S0$\times$C0 and, at S2$\times$C2, remove barely a tenth of the
bias, leaving the size at 73\%): a prediction criterion is
structurally indifferent to the allocation of spatial signal between
$s(u,v)$ and $\rho Wy$. With $\mu_s$ fixed at the upper end of the
grid, the penalty closes the surface channel: the F2p arm recovers
oracle-level bias everywhere except in the C2 column, where a
residual $-0.02$ to $-0.03$ flows through the \emph{free} smooths of
the distance candidates (and translates into a 12--14\% size).
Figure~\ref{fig:E3p_path} traces $\hat\rho$ and $\bar\pi$ along the
$\mu_s$ grid: the recovery is monotone and reaches a plateau,
suggesting a simple identification-oriented selection rule --- take
the smallest $\mu_s$ on the plateau --- in place of the invalid
REML choice.
Cross-fitting alone (F2c) closes the finite-sample channel: it
restores the size of the test in every cell (6--9\%, including C2)
and removes the bias at $\rho_0=0$ entirely ($-0.001$ at
S2$\times$C2). But under C2 concurvity it degrades the
\emph{estimation} of a nonzero $\rho_0$ (bias $-0.074$ to $-0.108$
for $\rho_0\geq0.2$, with doubled RMSE) and it is devastating for
the interpretable fit: the RMSE of $\hat f_2$ at S2$\times$C2
reaches 1.996, against 1.219 for the free smooth and 0.067 for the
oracle --- the out-of-fold score no longer disciplines the
within-sample concurvity between $s(\tilde x_2)$ and $s(u,v)$.
The combination (F2pc) is the only arm that closes both channels:
$|$bias$|\leq0.021$ in all 36 cells, empirical size between 5\% and
9\%, coverage between 0.94 and 0.99, RMSE of $\hat f_2$ at 0.246,
and RMSE of $\hat\rho$ at oracle level. The components not involved
in the concurvity are insensitive throughout: the RMSE of
$\hat f_1$ and $\hat f_3$ varies by less than $0.005$ across all
arms and cells.

\paragraph{Boosting: selection is robust, and identification
decouples from attribution}
The component-wise boosting estimator at its default stopping value
is nearly immune to the trap on the structural side:
\texttt{BSPA\_SAR\_CFE} with a free $s(u,v)$ base learner keeps
$|$bias$|\leq0.013$ in every cell, because early stopping is itself
a complexity control. Selection is equally robust except exactly
where geography is ambiguous: the three true covariates are selected
in every replication (TPR $=1$) with at least 94\% of boosting
iterations spent on true learners in all cells but S2$\times$C2,
where the TPR drops to 0.53 and the distance candidates capture
about 18\% of the iterations --- the algorithm cannot tell
\emph{which} distance variable is the true one, since
$\tilde x_2$ and $d_1$--$d_3$ are all distances to fixed points.
Remarkably, this covariate-level ambiguity is independent of
$\rho_0$ and leaves $\hat\rho$ unharmed: identification of the
spatial parameter and attribution among geographic covariates are
distinct problems, and only the latter is affected.

\begin{figure}[htbp]
\centering
\includegraphics[width=0.95\linewidth]{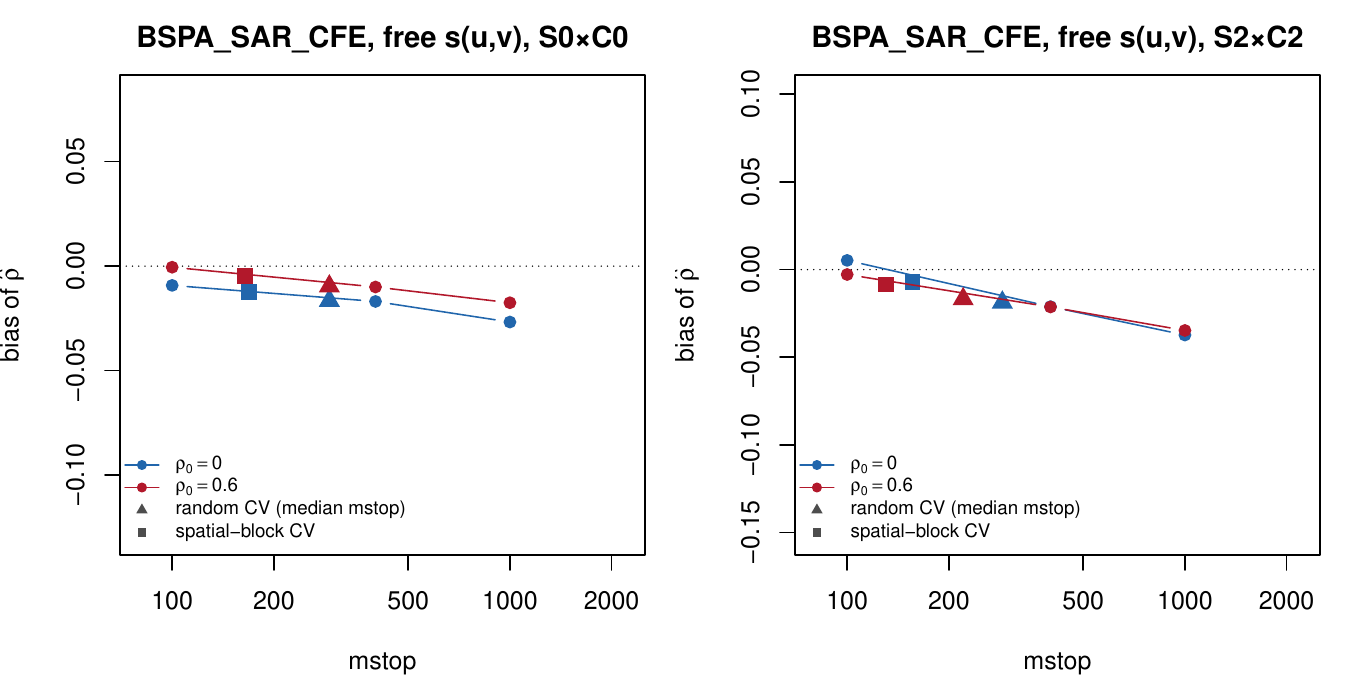}
\caption{\texttt{mstop} sub-experiment (boosted estimator
\texttt{BSPA\_SAR\_CFE} with free $s(u,v)$ learner, 250 replications): bias of $\hat\rho$ against fixed
\texttt{mstop} values (lines), with the median \texttt{mstop}
selected by cross-validation under random folds (triangles) and
spatial-block folds (squares).}
\label{fig:E3p_mstop}
\end{figure}

\paragraph{The \texttt{mstop} sub-experiment: prediction-based
tuning walks into the trap; spatial folds protect}
The apparent immunity of boosting is a property of the dial
position, not of the method. Figure~\ref{fig:E3p_mstop} shows the
dose-response: as \texttt{mstop} grows from 100 to 1000, the bias
of $\hat\rho$ deepens and the empirical size at $\rho_0=0$ rises
from 4.4\% to 15.6\% \emph{even in the control cell}. Tuning
\texttt{mstop} by standard cross-validation with random folds lands
on the harmful part of this curve (median selected
\texttt{mstop}\,$\approx$\,290, size roughly doubled): held-out
neighbours let the spatial learners ``cheat'' by continuity, and
the prediction criterion, indifferent to the allocation of spatial
signal, keeps the dial high. Cross-validation with spatial-block
folds selects a complexity half as large (median
\texttt{mstop}\,$\approx$\,130--168) and keeps bias below $0.012$
and size at 5.6--8\%. The pattern mirrors the REML failure for
$\mu_s$: \emph{any} prediction-targeted tuning of a complexity dial
reabsorbs the autoregressive signal, while identification-oriented
choices (spatial folds, the $\mu_s$ plateau, or the cross-fitted
score) do not. Spatial cross-validation is accordingly used as the
\texttt{mstop} protocol for all boosting arms of the E1 and E2
tables: leave-one-cluster-out blocks for the boosted
estimators, scattered k-means cells for plain \texttt{gamboost}.

\paragraph{Growing $n$ does not repair the test}
Replicating the control and adversarial cells at $n=5000$ confirms
the fixed-basis prediction that the trap is a finite-sample
phenomenon of order $K/n$: the F2 bias shrinks from $-0.064$ to
$-0.011$ at S0$\times$C0 ($\rho_0=0$) and from $-0.120$ to $-0.019$
at S2$\times$C2 ($\rho_0=0.6$). The size distortion, however,
persists --- 8.8\% and 22.8\% respectively at $n=5000$ --- because
the bias decreases more slowly than the standard error of
$\hat\rho$: larger samples shrink the estimation error faster than
the confounding, so the invalidity of the test of $\rho=0$ is not
cured by sample size. The combined remedy stays at oracle level at
both sample sizes.

In empirical applications these results translate into a simple
protocol, applied to the house-sales data of the next section:
estimate with the free smooth and with the spectrally penalized,
cross-fitted specification; read $\bar\pi$ and the gap between the
two $\hat\rho$ as the diagnostic; and never tune the geographic
smoother --- or \texttt{mstop} --- by a purely predictive criterion.
One caveat carries over from the fixed-$W$ setting of the
simulations: under substantial misspecification of $W$, the
geographic smooth can legitimately compensate for spatial
heterogeneity that $\rho Wy$ cannot represent, so the remedies
should be read as protecting the interpretation of $\hat\rho$ under
a reasonably well-specified $W$ rather than as a licence to drop
the smoother altogether.

\section{Empirical application: the Lucas County housing market}
\label{sec:empirical}

The Monte Carlo experiments of Section~\ref{sec:simulations} document the
finite-sample properties of the determinant-free estimators under controlled
data-generating processes. We now exercise the complete workflow on
the Lucas County (Ohio) house-sales data ---
$n=25{,}357$ single-family transactions recorded between 1993 and 1998 in
the Toledo metropolitan area --- introduced to the spatial-econometrics
literature by \citet{lesage2004models} and distributed with the
\textit{spData} R package. These transaction-level data combine the
ingredients the framework is built for: strongly nonlinear hedonic
effects --- most prominently the \emph{vintage} profile of the
construction year --- a heterogeneous mix of continuous and categorical
candidates for the selection step, and a strong, smooth spatial price
gradient that the geoadditive surface $s(u,v)$, spatially structured
regressors, and the autoregressive term $\rho Wy$ all compete to
absorb.

The object of the application is the estimation of the spatial parameter
and its decomposition diagnostics. We follow
the safeguarded workflow of Sections~\ref{sec:remedies}
and~\ref{sec:mstop}: a candidate set that
includes the geoadditive surface and distance-type regressors, a small
menu of spatial weight matrices fixed \emph{a priori}, the spectral
diagnostic $\pi_k$, the spectral penalty evaluated along the $\mu_s$ path
with the plateau rule, spatial cross-fitting, the stopping iteration
selected by spatial-block cross-validation, and closed-form, HC0 and SHAC
variances for $\hat\rho$.

\subsection{Data and candidate specification}
\label{sec:emp_data}

The dependent variable is the logarithm of the transaction price
(median \$65{,}500; range \$2{,}000--\$875{,}000). The structural
covariates describe the dwelling: year of construction (\texttt{yrbuilt},
1835--1998), total living area (\texttt{TLA}, median $1{,}318$ sq.\,ft.),
lot size (median $6{,}800$ sq.\,ft.), lot frontage and depth, numbers of
rooms, bedrooms, full and half bathrooms, garage area, and three
categorical attributes --- exterior wall material, garage type, and
number of stories, with levels representing fewer than 300 sales merged
into their nearest category (five, four and five levels retained,
respectively). Sale-year
indicators (1993--1998; annual sale counts rising from $3{,}260$ to about
$5{,}000$) absorb market-wide price drift. Coordinates are projected
(Lambert conformal conic, metres). To this hedonic core we add the
distance to the Toledo central business district computed from the
projected coordinates --- a regressor of the ``looks like geography''
type that the factorial design of Section~\ref{sec:E3p_results}
identified as the active ingredient of the geoadditive trap.

The candidate set given to the estimators mirrors the practice the paper
targets. Continuous covariates enter as penalized-spline base learners
(with a linear base learner alongside, so that the selection step can
choose the linear representation when it suffices); categorical
attributes and sale-year indicators enter as linear base learners; and
the set includes the bivariate geoadditive surface $s(u,v)$, both
\emph{free} --- the configuration exposed to the trap --- and under the
spectral penalty of Section~\ref{sec:remedies}. The vintage profile of
\texttt{yrbuilt}, a strongly nonlinear and non-monotonic effect that
linear hedonic specifications truncate, serves as the showcase nonlinear
component of the application.

The menu of spatial weight matrices is fixed \emph{a priori} and kept
deliberately small: a $4$-nearest-neighbour matrix and a
$10$-nearest-neighbour matrix built on the projected coordinates, and a
distance-band matrix whose
threshold is the smallest radius leaving no observation without a
neighbour; all three are row-normalized. The menu serves to verify that
$\hat\rho$, its sandwich standard errors, and the spectral diagnostics
are stable across reasonable neighbourhood definitions.

\subsection{The spatial parameter under the safeguarded workflow}
\label{sec:emp_rho}

Table~\ref{tab:lucas_rho} collects, for each matrix of the menu, the
estimates of the spatial parameter, their closed-form and
\texttt{SHAC} standard errors, and the spectral diagnostic $\bar\pi$,
for the sequence of specifications that Section~\ref{sec:remedies}
motivates --- from the free geoadditive surface to the retained
combination of the plateau value $\mu_s^\star$ with spatial
cross-fitting. The boosted estimator (\texttt{BSPA\_SAR\_CFE}) runs with a free
$s(u,v)$ base learner and spatial-block cross-validation of its
stopping iteration.

\begin{table}[htbp]
\centering
\caption{Lucas County: estimates of the spatial parameter across
specifications and spatial weight matrices ($n=25{,}357$). Standard
errors in parentheses: closed-form, then \texttt{SHAC}. $\bar\pi$:
mean spectral projection of the fitted surface on the 25 dominant
eigenvectors of $W$. Configuration codes as in axis F of the Monte
Carlo design. The retained specification (F2pc) combines the spectral
penalty at the plateau value $\mu_s^\star$ ($10^4$ for the kNN
matrices, $10^5$ for the distance band) with spatial cross-fitting. \texttt{BSPA\_SAR\_CFE}: free $s(u,v)$ surface,
stopping iteration by spatial-block cross-validation (2624, 3794, 434).
The distance-band matrix uses the smallest radius leaving no
observation isolated (median of 488 neighbours per observation).
}
\label{tab:lucas_rho}
\resizebox{\linewidth}{!}{%
\begin{tabular}{llccccccccc}
\toprule
 & & \multicolumn{3}{c}{\textbf{$W$: 4-NN}} & \multicolumn{3}{c}{$W$: 10-NN} & \multicolumn{3}{c}{$W$: dist. band}\\
\cmidrule(lr){3-5}\cmidrule(lr){6-8}\cmidrule(lr){9-11}
Estimator & Configuration & $\hat\rho$ & (se, se$_{\mathrm{SHAC}}$) & $\bar\pi$ & $\hat\rho$ & (se, se$_{\mathrm{SHAC}}$) & $\bar\pi$ & $\hat\rho$ & (se, se$_{\mathrm{SHAC}}$) & $\bar\pi$ \\
\midrule
\multirow[t]{5}{*}{\texttt{GAM\_SAR\_CFE}} & F2 (free $s(u,v)$) & 0.407 & (0.0060, 0.0108) & 0.060 & 0.511 & (0.0077, 0.0134) & 0.264 & 0.546 & (0.0150, 0.0602) & 0.763\\
 & F2c (F2 + cross-fit) & 0.478 & (0.0050, 0.0085) & -- & 0.542 & (0.0060, 0.0152) & -- & 0.437 & (0.0063, 0.0431) & --\\
 & F2p, REML $\mu_s$ & 0.382 & (0.0061, 0.0122) & 0.062 & 0.488 & (0.0080, 0.0157) & 0.241 & 0.483 & (0.0180, 0.0969) & 0.718\\
 & F2p, $\mu_s=\mu_s^\star$ & 0.472 & (0.0057, 0.0089) & -- & 0.574 & (0.0071, 0.0108) & -- & 0.594 & (0.0115, 0.0633) & --\\
 & \textbf{F2pc (retained)} & \textbf{0.496} & \textbf{(0.0055, 0.0089)} & \textbf{0.013} & 0.592 & (0.0068, 0.0115) & 0.058 & 0.622 & (0.0107, 0.0599) & 0.232\\
\addlinespace[0.6ex]
\texttt{BSPA\_SAR\_CFE} & F2, spatial-CV \texttt{mstop} & 0.427 & (0.0059, --) & -- & 0.525 & (0.0076, --) & -- & 0.592 & (0.0126, --) & --\\
\addlinespace[0.6ex]
SAR (ML) & linear covariates & 0.600 & (0.0039, --) & -- & -- & -- & -- & -- & -- & --\\
\bottomrule
\end{tabular}}
\end{table}
 
The two channels of the geoadditive trap, established on controlled
designs in Section~\ref{sec:E3p_results}, are both visible --- and
separable --- on these transaction data. Start with the 4-NN matrix.
With a free surface, $\hat\rho=0.407$. Cross-fitting alone, which
corrects the finite-sample own-observation channel and touches nothing
else, lifts the estimate to $0.478$. The spectral penalty, which
corrects the population channel, moves $\hat\rho$ along the $\mu_s$
path from $0.382$ to $0.473$, with the increase concentrated between
$\mu_s=10^{2}$ and $10^{4}$ and a genuine interior plateau beyond:
between $\mu_s=10^4$ and $10^6$ the estimate moves by less than
$0.001$. REML-selected $\mu_s$ leaves the estimate at the bottom of
the path ($0.382$): the structural indifference of prediction-targeted
criteria, derived in Section~\ref{sec:remedies} and documented in the
factorial design, is reproduced on real data. The retained
specification --- plateau value $\mu_s^\star=10^4$ plus cross-fitting
--- yields $\hat\rho=0.496$. The gap between the naive and the
safeguarded estimate, $0.09$, is roughly sixteen closed-form standard
errors: on these data the geoadditive trap is not a small-sample
curiosity but a first-order feature of the estimate.

Read predictively, the same data would not sound any alarm. The
free-surface fit is in fact the \emph{better} one in sample (RMSE
$0.278$ against $0.286$ for the retained specification): any
criterion that targets fit --- in sample or through random-fold
cross-validation --- selects the trapped specification, and REML
does. Spatial-block cross-validation pushes in the opposite
direction, but as a change of prediction regime whose verdict
depends on the block geometry, not as an identification criterion:
it penalizes the legitimate share of the surface along with the
redundant one. This is why the workflow entrusts it with
\texttt{mstop} only, and decides $\mu_s$ on the stability of
$\hat\rho$.

The 10-NN matrix repeats the pattern with larger amplitude, from
$0.511$ (free) to $0.592$ (retained). The level of $\hat\rho$ itself
is not comparable across weight matrices --- a coarser neighbourhood
averages over more units and mechanically shifts the autoregressive
parameter --- but the \emph{geometry} of the safeguards is the same:
cross-fitting and the spectral penalty push in the same direction, the
path plateaus at the same $\mu_s^\star$, and the relative correction
is of the same order. The boosted estimates sit between the free
and the retained GAM figures ($0.427$ and $0.525$), with stopping
iterations selected well inside the admissible range (2624 and 3794).
This is exactly where the evidence of
Section~\ref{sec:E3p_results} places a boosting estimator whose
surface base learner is free: spatial-block cross-validation of
\texttt{mstop} protects partially, and the full correction belongs to
the penalized, cross-fitted variant. The table reports a single
boosted specification --- free surface, no spectral penalty, no
cross-fitting --- and this is by design: as
Section~\ref{sec:remedies} explains, the spectral penalty lives in
the penalized-spline representation, the only identification-aware
dial on the boosting side is the stopping iteration \texttt{mstop},
and cross-fitting the three boosted first stages would entangle
fold-specific selection paths.
The division of labour in the workflow follows: the boosted estimator
performs the specification search --- which covariates, which
functional forms --- under the protected stopping rule, and the
structural estimate with full safeguards is then obtained by
re-estimating the retained specification with the penalized,
cross-fitted \texttt{GAM} variant, mirroring the two-step workflow
recommended for large samples in Section~\ref{sec:simulations}.

The distance-band matrix stresses the workflow rather than merely
replicating it: its minimal-connectivity radius ($1{,}324$~m) yields
a median of 488 neighbours per sale, so that $Wy$ averages over
entire neighbourhoods. The free surface then absorbs the dominant
spectrum almost completely ($\bar\pi=0.76$), the $\mu_s$ path only
settles at $10^{5}$, and cross-fitting corrects \emph{downwards}
($0.546$ to $0.437$) where it corrected upwards under the sparse
matrices. The safeguarded estimate ($0.622$) lines up with the kNN
pattern, but its \texttt{SHAC} standard error is nearly six times the
closed-form one --- with two-hop neighbourhoods covering a large
share of the sample, the independence-based variance is no longer a
reasonable first approximation. $\bar\pi$ and the
\texttt{SHAC}-to-closed-form ratio thus flag, on their own, a
qualitatively different smoothing device.

Across the kNN specifications the \texttt{SHAC} standard error exceeds
the closed-form one by 55--65\,\% (and by a factor of six under the
dense distance band), a reminder that the independence-based
variance is optimistic about the score correlation induced by the
neighbourhood graph on real data; the \texttt{HC0} correction, by
contrast, is negligible here. Inference on $\hat\rho$ should
accordingly be read off the \texttt{SHAC} column.

The same diagnostics motivate the emphasis on the 4-NN column in
Table~\ref{tab:lucas_rho}. It is the matrix under which the surface
and the autoregressive term compete least for the dominant spectrum
($\bar\pi=0.06$ under the free surface, against $0.26$ and $0.76$),
the one whose \texttt{SHAC}-to-closed-form ratio stays closest to
one, and the one that matches the mechanism plausibly at work in a
transaction-level housing market, where prices form on a handful of
nearest comparable sales rather than on neighbourhood averages.
$\bar\pi$ alone would be a circular selection criterion for $W$ ---
Section~\ref{sec:E3p_results} shows it can be driven to oracle-like
values by a badly biased arm; it is its convergence with the
inferential and economic arguments that singles out the sparse
matrix.

\begin{table}[htbp]
\centering
\caption{Lucas County: average impact multipliers implied by
$\hat\rho$ under the 4-NN matrix. Direct: $n^{-1}\mathrm{tr}\,(I-\hat\rho W)^{-1}$
(Hutchinson estimator, 20 probes); total: $1/(1-\hat\rho)$;
indirect: total minus direct. Configuration codes as in axis F.
Any covariate effect scales by these
factors: the average marginal effect of living area in the retained
specification (F2pc), 0.0375 log point per 100 sq.\,ft., splits into
0.0402 direct and 0.0342 indirect.}
\label{tab:lucas_impacts}
\begin{tabular}{lccccc}
\toprule
Configuration & $\hat\rho$ & Direct & Indirect & Total & Indirect share\\
\midrule
F2 (free $s(u,v)$) & 0.407 & 1.044 & 0.644 & 1.687 & 38\%\\
F2pc (retained) & 0.496 & 1.071 & 0.911 & 1.983 & 46\%\\
linear SAR (ML) & 0.600 & 1.120 & 1.377 & 2.497 & 55\%\\
\bottomrule
\end{tabular}
\end{table}
 
The difference matters economically, not only statistically.
Table~\ref{tab:lucas_impacts} translates the estimates into average
impact multipliers \`a la LeSage--Pace. Under the free surface the
indirect (spillover) share of any covariate effect is 38\,\%; under
the retained specification (F2pc) it is 46\,\%; and
the linear SAR benchmark
--- whose $\hat\rho=0.600$ absorbs the nonlinearities its
specification truncates --- pushes it to 55\,\%, with a total
multiplier of 2.50 against 1.98. For the living-area effect, the
average marginal impact of 100 additional square feet
($0.0375$ log point) splits into $0.040$ direct and $0.034$
transmitted to neighbouring sales. The two misspecifications thus
push the spatial parameter in opposite directions --- free
flexibility understates spillovers, truncated flexibility
overstates them --- the empirical
counterpart of the two failure directions documented in the
factorial design (free surface versus the over-penalized
RSR-type arm).

\subsection{Spectral diagnostics and the $\mu_s$ path}
\label{sec:emp_pik}

\begin{figure}[htbp]
\centering
\includegraphics[width=0.9\linewidth]{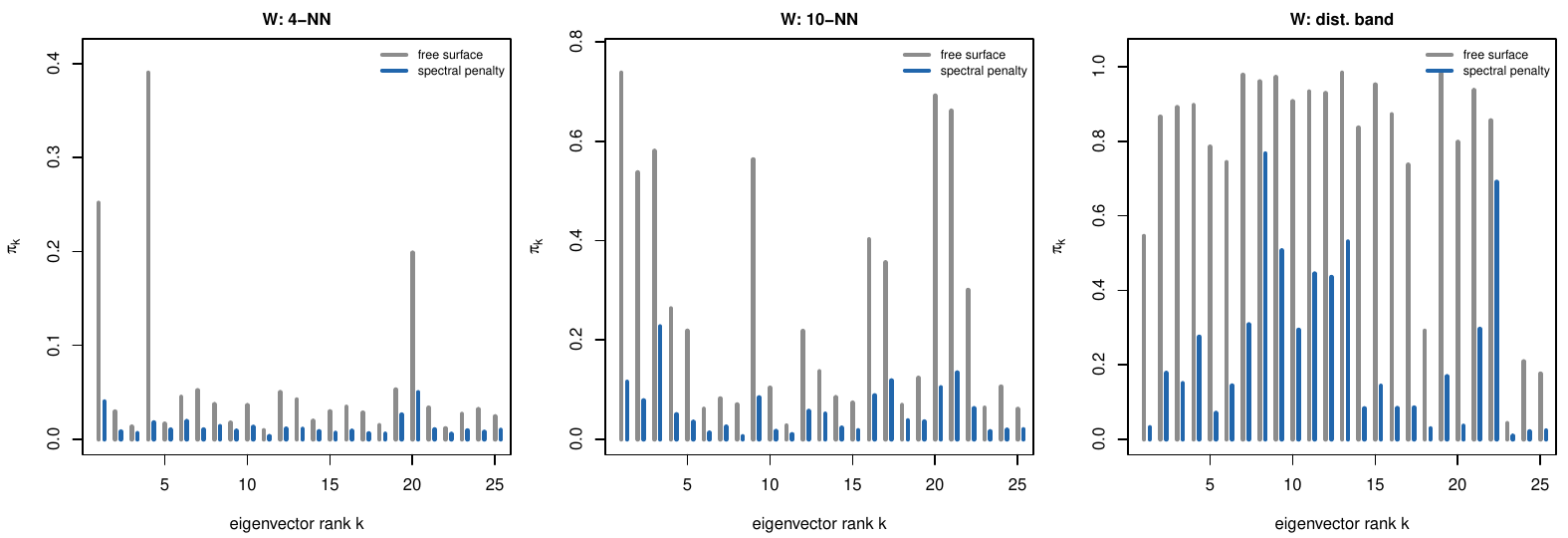}
\caption{Spectral projection diagnostic $\pi_k$ on the 25 dominant
eigenvectors of $W$ for the free surface (grey) and the spectrally
penalized surface at $\mu_s^\star$ (blue), Lucas County data.}
\label{fig:lucas_pik}
\end{figure}

Figure~\ref{fig:lucas_pik} explains \emph{why} the two matrices differ
in trap exposure. Under the 4-NN matrix the free surface projects
moderately on the dominant spectrum ($\bar\pi=0.06$, with isolated
spikes at $0.25$ and $0.39$ on the first and fourth eigenvectors).
Under the 10-NN matrix the overlap is massive: the first three
eigenvectors are absorbed at $0.74$, $0.54$ and $0.58$, and the mean
projection reaches $0.26$. Under the dense distance-band matrix the
absorption is nearly total ($\bar\pi=0.76$). The gradient across the
menu has a simple reading: the dominant eigenvectors of a coarser
neighbourhood matrix are smoother functions of space, and a bivariate
smooth with 60 degrees of freedom reproduces them almost exactly;
trap exposure is therefore a property of the pair (candidate set,
$W$), observable before any structural estimate is computed. The
spectral penalty empties the overlap under the kNN matrices
($\bar\pi=0.013$ and $0.058$) and reduces it by more than two thirds
under the distance band ($\bar\pi=0.23$), while leaving the surface
in the model.

\begin{figure}[htbp]
\centering
\includegraphics[width=0.85\linewidth]{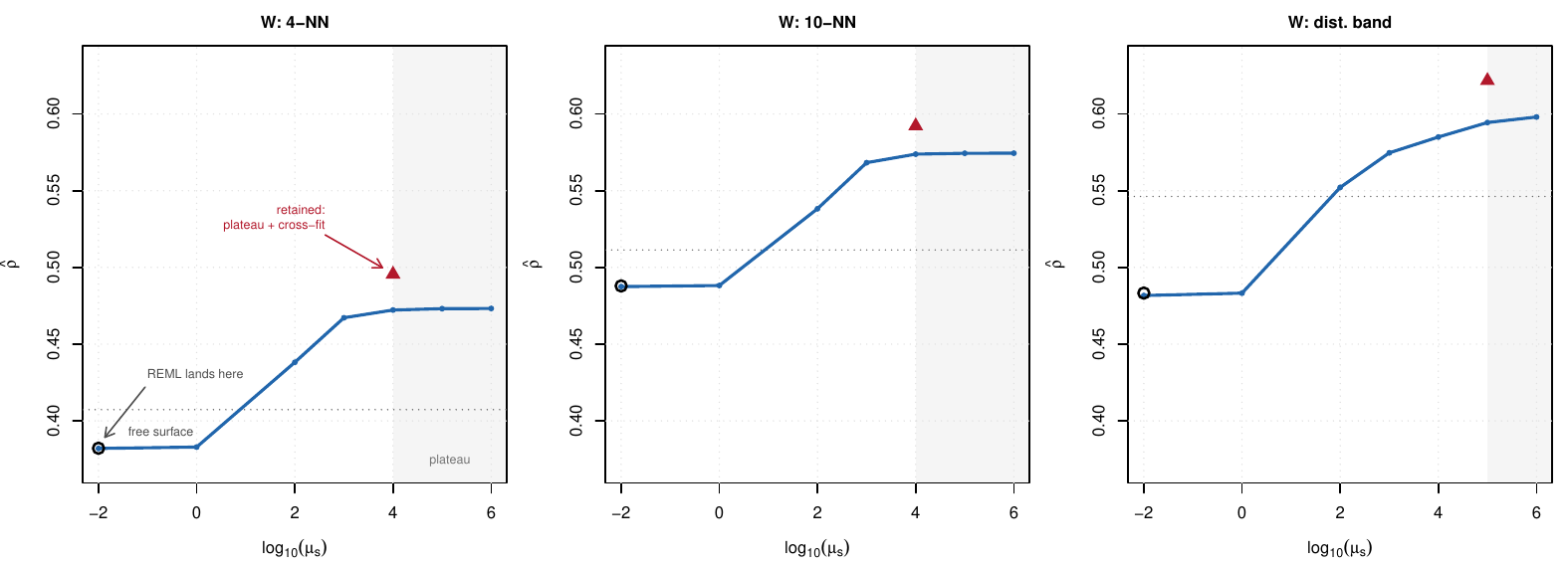}
\caption{$\hat\rho$ along the spectral-penalty path
(\texttt{GAM\_SAR\_CFE}, Lucas County data), one panel per weight
matrix. Blue curve: F2p along the $\mu_s$ grid (log scale). Dotted
line: free-surface estimate (F2). Open circle: REML-selected
$\mu_s$. Red triangle: the retained specification F2pc (plateau
value $\mu_s^\star$ plus cross-fitting). Shaded area: the plateau
targeted by the selection rule.}
\label{fig:lucas_mu_path}
\end{figure}

Figure~\ref{fig:lucas_mu_path} makes the selection rule readable
panel by panel: the path leaves the free-surface level over
$\mu_s=10^{2}$--$10^{4}$ --- the penalty handing the low-frequency
variation back to the autoregressive term --- and settles on the
shaded plateau. The REML circles sit on the flat initial segment,
and the cross-fitting increment (triangles above the plateau) is
visibly additive to the spectral correction, consistent with the two
channels being distinct.

\begin{figure}[htbp]
\centering
\includegraphics[width=\linewidth]{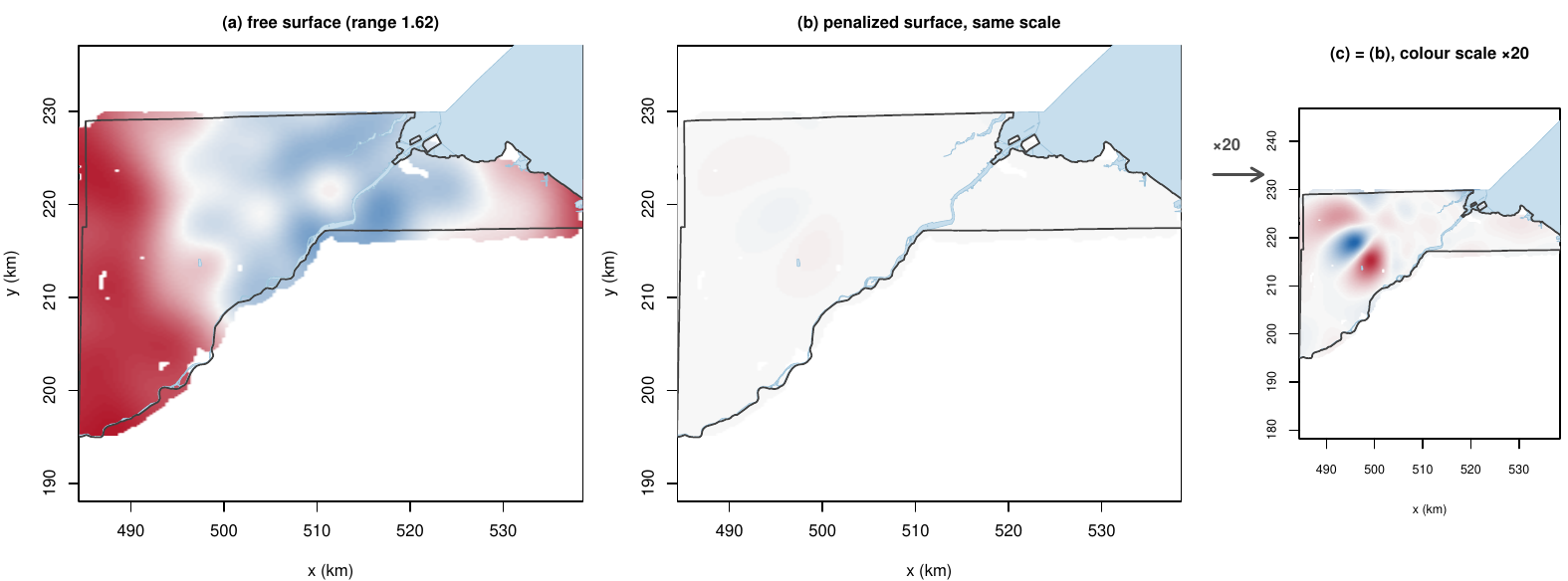}
\caption{Fitted geoadditive surface $\hat s(u,v)$ under the 4-NN
matrix: (a) free specification; (b) spectrally penalized
specification at $\mu_s^\star$, on the same colour scale as (a);
(c) the same surface as (b) with the colour scale magnified twenty
times (inset). The grid is masked beyond 1.5~km of any sale. County
boundary and main water bodies (Maumee River and Bay, Lake Erie)
from the US Census TIGER files.}
\label{fig:lucas_surfaces}
\end{figure}

Figure~\ref{fig:lucas_surfaces} makes the same diagnostic visible in
physical space, and it carries a message that goes beyond amplitude.
The nominal role of $s(u,v)$ is to approximate the \emph{unobserved}
spatially structured heterogeneity --- omitted neighbourhood quality,
amenities, anything the covariates miss. The two fits give completely
different answers to what that unobserved heterogeneity is. The free
surface carries a west--east gradient of amplitude $1.6$ log points
--- essentially the low-frequency price geography of the entire
metropolitan area, which is exactly the variation the autoregressive
term claims: read at face value, it would tell the analyst that
``unobserved heterogeneity'' spans a five-fold price factor and is
organised at the scale of the county. Under the spectral penalty the
same surface, in the same model on the same data, is flat at that
scale (amplitude $0.08$); what remains, visible at its own scale, is
a localized dipole around the central districts --- pockets of
genuinely local heterogeneity that neither $\rho Wy$ nor the
covariates account for, which is what an omitted-heterogeneity proxy
should look like once the spillover field is accounted for. An
analyst who maps the free $\hat s(u,v)$ and interprets it as
unobserved neighbourhood quality is, for the most part, mapping the
confounded spillover field. The penalty does not delete the surface:
it hands the $W$-dominant variation back to the autoregressive term
and leaves $s(u,v)$ with the part of the job it can identifiably do.

\subsection{Two nonlinear showcases: vintage and the CBD gradient}
\label{sec:emp_vintage}

\begin{figure}[htbp]
\centering
\includegraphics[width=\linewidth]{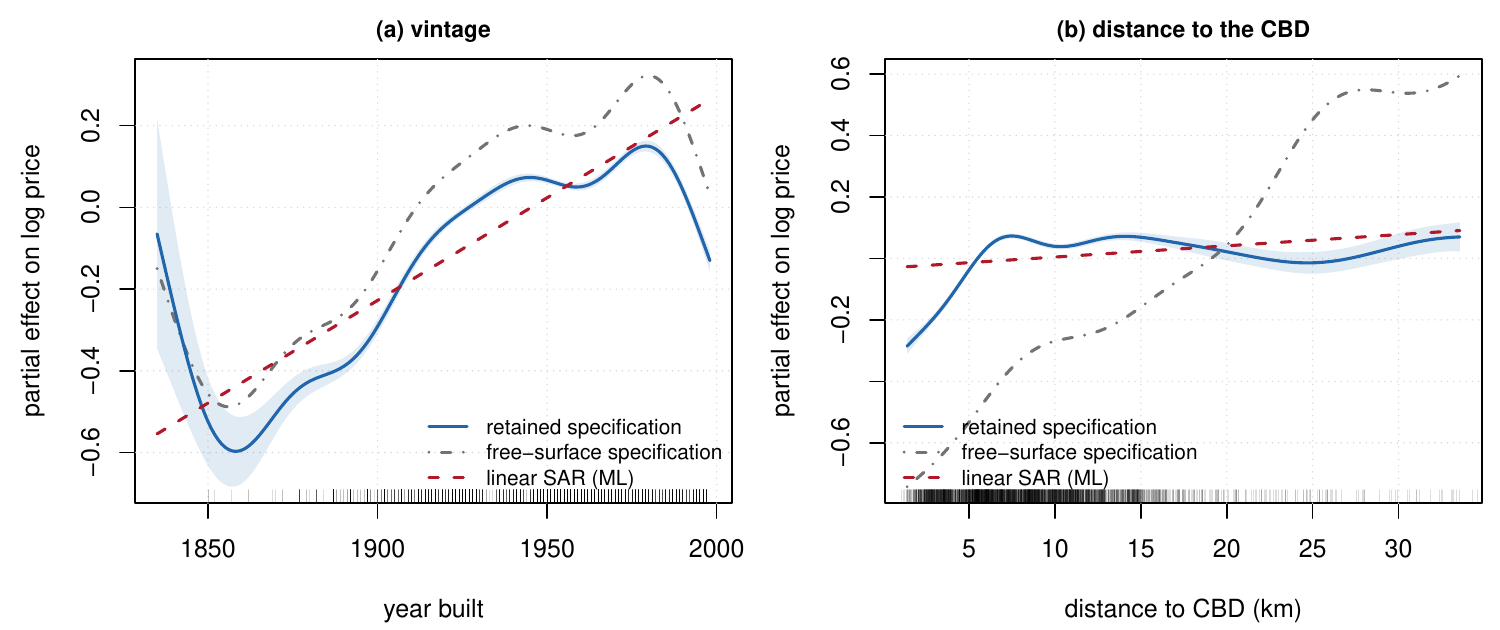}
\caption{Partial effects on log price in the retained specification
(4-NN, spectral penalty at $\mu_s^\star$), with 95\,\% pointwise
bands and rugs of the data. Panel (a): year of construction (vintage
curve). Panel (b): distance to the Toledo central business district.
Dash-dotted grey curves: the same effects estimated under the
\emph{free} surface; dashed lines: the corresponding linear SAR (ML)
effects.}
\label{fig:lucas_effects}
\end{figure}

Figure~\ref{fig:lucas_effects} shows the two nonlinear components
with the largest nonlinear content in the fit, both measured net of
their best linear approximation weighted by the sales density. Panel
(a) is the \emph{vintage} curve, the nonlinearity most discussed in
the hedonic literature on construction year, where time, age and
vintage effects are classically disentangled only under
functional-form restrictions \citep{coulson2008time}. The profile
spans about $0.75$ log points: a steep discount for the
mid-nineteenth-century stock (trough around 1860), an almost monotone
rise through the first half of the twentieth century, a pause for the
1950--1965 cohorts, a peak around 1980 at $+0.15$, and a decline for
the newest vintages sold over 1993--1998. The superimposed linear age
effect of the SAR benchmark --- age enters linearly or through
low-order polynomials in standard treatments of these data, from
\citet{lesage2004models} onwards --- cuts straight through all four
features. Panel (b) is the distance to the central business district,
the regressor that ``looks like geography'', whose flexible treatment
has a long tradition in urban economics
\citep{mcmillen1996one}: a steep urban-core discount of about $0.35$
log point over the first five kilometres, followed by undulations
that track the ring structure of the metropolitan area. The linear
benchmark averages discount and recovery into a slope close to zero
($0.004$ per km): the gradient is not merely truncated but erased.
The grey curves re-estimate both effects under the \emph{free}
surface, and their contrast replays the factorial design at the
covariate level. The vintage profile is essentially unchanged
(amplitude $0.81$ against $0.75$, same shape): construction year is
not a function of the coordinates, so the surface has nothing to
take from it. The distance gradient, by contrast, is no longer
interpretable under the free surface: its amplitude reaches $1.34$
log points (against $0.36$ in the retained specification) because
$s(u,v)$, the distance smooth and $\rho Wy$ split the same geography
arbitrarily --- only the urban-core discount, identified by the
dense ring of nearby sales, survives in both fits ($0.21$ versus
$0.24$ over the first five kilometres). The spectral penalty is what
turns the distance smooth into a stable, interpretable gradient.
Both profiles are estimated with
full flexibility in the most heavily penalized specification: by
construction the spectral penalty constrains only the geoadditive
surface, so protecting the structural parameter costs nothing in
terms of the nonlinearities --- including a distance effect that the
free surface would otherwise absorb.

\subsection{Practical summary}
\label{sec:emp_guidance}

The application condenses into a short recipe: fix a small menu of
weight matrices \emph{a priori} and keep the geoadditive surface in
the candidate set; read $\pi_k$ before any structural estimate;
select $\mu_s$ by the stability of $\hat\rho$ along the penalty path
--- never by a prediction-targeted criterion --- and add spatial
cross-fitting; report \texttt{SHAC} standard errors. Determinant-free
estimation makes the workflow inexpensive: at $n=25{,}357$ each
specification costs a few seconds on a laptop.

\section{Conclusion}
\label{sec:conclusion}

This paper develops a determinant-free estimation framework for
nonlinear spatial autoregressive models. The central theoretical
result is that the Closed-Form Estimator (CFE) remains valid when
the regression function is estimated nonparametrically through
additive models. For penalized spline estimators admitting a
smoothing matrix representation, we establish formally that the
residual projection structure underlying the CFE estimating
equation is preserved asymptotically: under standard regularity
conditions, the estimation error of the nonlinear component does
not affect the asymptotic behaviour of the spatial autoregressive
parameter, and the variance formula of the CFE estimator retains
asymptotic validity under the same conditions as standard GAM
estimation in a nonspatial setting. For component-wise gradient
boosting estimators, formal extension of this result requires
new theoretical arguments; the Monte Carlo experiments
consistently show, however, that boosting-based and spline-based
CFE estimators yield virtually identical finite-sample
performance, providing empirical support for the same asymptotic
behaviour.

An important implication of the proposed framework is computational
efficiency: because the CFE estimator avoids repeated evaluation of
the spatial log-determinant, the estimators remain scalable to large
datasets. Monte Carlo experiments confirm that the CFE variants match
maximum-likelihood accuracy at a fraction of its cost, and that the
boosted estimators provide reliable automatic variable selection with
only a minor efficiency loss relative to their penalized
counterparts.

The simulation experiments also yield concrete guidance on the
number of boosting iterations. Cross-validation with random folds
selects too many iterations as soon as covariates are spatially
structured, whereas spatial-block cross-validation stops the
algorithm markedly earlier and protects the spatial parameter; it is the
protocol adopted for all boosting estimators in this paper.
For SEM specifications, full ML remains the recommended default near
the boundary, with an automatic switching rule limiting the upward
bias in the spatial parameter $\hat\lambda$ at high spatial dependence.

The simulation experiments further highlight the risks associated
with formula specification in the presence of spatially structured
covariates. When geographic proximity variables are included as
controls --- a common practice in regional and urban economics ---
false positive selection rates remain modest at low numbers of
boosting iterations and do not materially affect the estimation
of the spatial parameter $\hat\rho$. However, including a bivariate geographic smoother
in the candidate specification poses a serious risk of
identification failure --- the geoadditive form of spatial
confounding studied in this paper --- and the factorial design
locates that risk precisely: the estimators are robust to spatial
\emph{autocorrelation} of the covariates per se, and the collapse is
driven by covariates that are direct functions of the coordinates
--- distances to fixed points, geographic trends --- whose
representation overlaps the dominant spectrum of $W$. The spectral diagnostic
$\pi_k$ of Section~\ref{sec:remedies} provides the operational
\emph{ex-ante} test, and the spectral penalty --- whose weight
$\mu_s$ controls how strongly the geoadditive surface is kept off
the dominant eigenvectors of $W$ --- evaluated along the $\mu_s$
path and combined with spatial cross-fitting, provides the remedy
that keeps the surface in the model.

The empirical application to the Lucas County transaction data shows
that these warnings and their remedies are not confined to controlled
designs. On $25{,}357$ house sales, the two channels of the trap are
visible and separable: spatial cross-fitting alone lifts $\hat\rho$ by
about $0.07$, the spectral penalty moves it along the $\mu_s$ path
to a genuine interior plateau, and the combined correction shifts the
estimate by roughly sixteen closed-form standard errors relative to
the free-surface fit. REML-selected $\mu_s$ reproduces the free
surface, confirming on real data that prediction-targeted criteria
cannot arbitrate the allocation of low-frequency variation. The
\texttt{SHAC} standard error exceeds the closed-form one by over
one half, and the vintage profile of the construction year ---
strongly nonlinear and non-monotonic --- is estimated with full
flexibility in the most heavily penalized specification: structural
protection costs nothing in terms of the nonlinearities the model is
built to capture.

In practice, the framework therefore operates as a two-step
workflow. The boosted estimator performs the specification search ---
which covariates, which functional forms --- under the protected
stopping rule of spatial-block cross-validation, the only
identification-aware dial available on the boosting side. The
structural estimate is then obtained by re-estimating the retained
specification with the penalized, cross-fitted variant, in which the
complexity of the geoadditive surface is governed by an
identification criterion --- the stability of $\hat\rho$ along the
path of the spectral-penalty weight $\mu_s$ ---
rather than by any prediction-targeted rule.

Together, these results show that determinant-free spatial
estimation extends naturally beyond linear models and provides
a practical way to combine flexible nonlinear regression with
structural spatial econometric interpretation.

A first direction for future research concerns the spatial weight
matrix: $W$ is fixed by the analyst in the current framework, and
a natural extension would estimate the spatial parameter jointly
with the weights of a (possibly signed) linear combination of
candidate matrices, in the spirit of \citet{debarsy2018flexible},
within the same determinant-free framework. The empirical
application shows why this is more than a computational extension:
once a geoadditive surface is in the model, the combination weights
and the surface compete for the same low-frequency variation, so
that estimating $W$ raises the identification question studied here
in a new form.

The selection rule for the spectral penalty --- take the
smallest $\mu_s$ on the plateau of the $\hat\rho$ path --- is
supported here by the factorial design and the application, not by
theory. Formal guarantees --- conditions under which the plateau
exists, consistency of $\hat\rho$ under plateau-selected $\mu_s$,
and valid inference after that selection --- are the direct
continuation of the impossibility result: once prediction-targeted
criteria are ruled out, the properties of the criteria that can
arbitrate the allocation remain to be established.

Finally, the asymptotic theory of
Appendix~\ref{appendix_asymptotic_cfe} covers component-wise
gradient boosting conditionally on the selection path
(Corollary~\ref{cor:boosting}); removing that conditioning --- a
path-stability argument for the greedy updates, and a formal
treatment of the cross-validated stopping iteration --- remains an open
theoretical question. Tree-based learners raise a more fundamental
difficulty: they naturally capture covariate interactions but do
not operate on a fixed basis, so that neither the collapse
argument of Lemma~\ref{lem:collapse} nor a smoothing-matrix
representation is available, and extending the determinant-free
asymptotics to that setting would require new theoretical
arguments.

\ifpreprint
\section*{Acknowledgments}
An earlier version of this work was presented at the Spatial
Econometrics Workshop (SEW 2026); the author thanks the participants,
and in particular J.~Paul Elhorst, for their helpful comments.
\fi

\section*{Data availability statement}
The house-sales data analysed in Section~\ref{sec:empirical} are
publicly available in the R package \textit{spData} (dataset
\texttt{house}); the county boundary and water features come from the
public US Census TIGER files.
\ifpreprint
All estimators are implemented in the \textit{spboost} R package
(version 0.7.0). The source package and the replication scripts for
the Monte Carlo designs and for the empirical workflow are available at
\url{https://github.com/ggeniaux/replication_spboost_geniaux_2026}.
\else
All estimators are implemented in the
\textit{spboost} R package, and the replication scripts for the Monte
Carlo designs and for the empirical workflow are distributed with the
package; they will additionally be deposited in a public repository
upon acceptance.
\fi

\pagebreak
\appendix

\section{Relationship between the CFE estimator and simplified
ratio and IV estimators in linear spatial models}
\label{appendix_cfe_linear}

This appendix clarifies the relationship between the Closed-Form
Estimator (CFE) proposed by \citet{smirnov2020} and the hierarchy of 
simplified estimators used in the \textit{spboost} R package. While 
these derivations follow the linear framework of \citet{smirnov2020}, 
they provide the geometric foundations and the orthogonality 
conditions underlying the quasi-linear decomposition of
Section~\ref{sec:asymptotic_nonlinear} and the asymptotic analysis of
Appendix~\ref{appendix_asymptotic_cfe}. We treat both the SEM and SAR
cases, providing a geometric interpretation of the approximations used
when exact quadratic solutions are numerically unstable.

\subsection{The linear SEM model}

Consider the spatial error model $y = X\beta + u$, with
$u = \lambda W u + \varepsilon$, where $E(\varepsilon)=0$ and
$\mathrm{Var}(\varepsilon)=\sigma^2 I$.

Following \citet{smirnov2020}, define the auxiliary residuals
$e_0 = M_Z y$ and $e_1 = M_Z Wy$, where $Z=[X,WX]$ and
$M_Z = I - Z(Z'Z)^{-1}Z'$.

The CFE estimator of $\lambda$ is obtained from the quadratic equation
$a\lambda^2 - b\lambda + c = 0$, where $a = e_1'We_1$,
$b = e_0'We_1 + e_1'We_0$, and $c = e_0'We_0$; since a quadratic form
only involves the symmetric part of $W$, $b = e_0'(W+W')e_1$. The admissible root is
\[
\hat{\lambda}_{CFE} = \frac{b-\sqrt{b^2-4ac}}{2a}.
\]

\begin{proposition}
\label{prop:sem_ratio}
Define the ratio estimator $\tilde{\lambda} = \frac{e_1'We_0}{e_1'We_1}$.
Then:
\begin{enumerate}
\item $\tilde{\lambda}$ minimizes the quadratic form
  $q(\lambda)=(e_0-\lambda e_1)'W(e_0-\lambda e_1)$.
\item If the discriminant $D=b^2-4ac \ge 0$, then
  $\tilde{\lambda}=\frac{\lambda_1+\lambda_2}{2}$, where $\lambda_1$
  and $\lambda_2$ denote the two roots of the CFE quadratic equation.
\item $\tilde{\lambda}-\hat{\lambda}_{CFE} = \frac{\sqrt{D}}{2a}$.
\end{enumerate}
\end{proposition}

\noindent\textit{Proof.}
Expanding the quadratic form yields
$q(\lambda) = a\lambda^2-b\lambda+c$.
Differentiating with respect to $\lambda$ gives $q'(\lambda)=2a\lambda-b$.
Setting the derivative equal to zero implies $\lambda=b/(2a)$.
With $\widetilde W=(W+W')/2$, $b/(2a)=e_1'\widetilde We_0/e_1'\widetilde We_1$,
which coincides with $\tilde{\lambda}=e_1'We_0/e_1'We_1$ whenever the
two bilinear forms agree (exactly so for symmetric $W$), proving that
$\tilde{\lambda}$ minimizes $q(\lambda)$ in that case.
If $D\ge0$, the roots are $\lambda_{1,2}=(b\pm\sqrt{D})/(2a)$, whose
midpoint is $b/(2a)=\tilde{\lambda}$.
Finally, $\tilde{\lambda}-\hat{\lambda}_{CFE} = b/(2a)-(b-\sqrt{D})/(2a)
= \sqrt{D}/(2a)$.
\hfill$\square$

\begin{corollary}
\label{cor:sem_asymptotic}
Let $e_u = e_0-\hat{\lambda}_{CFE}e_1$. If
$\frac{1}{n}e_1'We_1 \rightarrow \kappa>0$ and
$\frac{1}{n}e_1'We_u \rightarrow 0$, then
$\tilde{\lambda}-\hat{\lambda}_{CFE}\rightarrow 0$.
Thus, the ratio estimator is asymptotically equivalent to the CFE
estimator when the spatial cross-moment between $e_1$ and the
structural residual vanishes.
\end{corollary}

To ensure numerical stability in finite samples, especially when both 
real roots are admissible, we implement a robust root-selection rule. 
Among the candidates $\lambda_{-}$ and $\lambda_{+}$, we retain the 
one minimizing the filtered residual variance proxy:
\[
\hat\lambda = \arg\min_{\lambda\in\{\lambda_{-},\lambda_{+}\}\cap\mathcal{R}} \|e_0-\lambda e_1\|_2^2,
\]
where $\mathcal{R}$ denotes the admissible parameter interval. This 
modification preserves the closed-form nature of the CFE while 
improving robustness to nonlinear model structures and potential 
numerical instability near $D \approx 0$.

\subsection{Robust root selection for CFE in the SEM case}

When the CFE discriminant satisfies $D \ge 0$, two real closed-form candidates are available:
\[
\lambda_{-}=\frac{b-\sqrt{D}}{2a},\qquad
\lambda_{+}=\frac{b+\sqrt{D}}{2a}.
\]

In baseline implementations, selecting a fixed branch (typically $\lambda_{-}$) may induce instability in finite samples, especially when both roots are admissible.  
We therefore use a robust root-selection rule: among admissible roots (after enforcing parameter bounds), we retain the one minimizing
\[
S(\lambda)=\|e_0-\lambda e_1\|_2^2
      =\sum_{i=1}^n (e_{0i}-\lambda e_{1i})^2.
\]
Equivalently, this selects the root with the smallest proxy of filtered residual variance,
\[
\hat{\sigma}^2(\lambda)=\frac{1}{n}S(\lambda).
\]

Hence, for $D\ge 0$, the estimator is
\[
\hat\lambda
=
\arg\min_{\lambda\in\{\lambda_{-},\lambda_{+}\}\cap\mathcal{R}}
\|e_0-\lambda e_1\|_2^2,
\]
where $\mathcal{R}$ denotes the admissible parameter interval (e.g., $[-0.99,0.99]$).  
If no admissible exact root is available, the method falls back to the approximation rule described in the main text.

This robust two-root check is motivated by practical departures from
the ideal linear-asymptotic setting: finite-sample effects, the use
of estimated (boosting-regularized) residual components in place of
population moments, nonlinear or approximate model structure,
numerical instability when $D\approx 0$ or $a\approx 0$, and
possible heteroskedastic or non-Gaussian disturbances. This
modification preserves the closed-form nature of the CFE step while
improving numerical robustness and reducing root-branch sensitivity
in moderate samples.

\subsection{The linear SAR model}

Consider the spatial autoregressive model
$y = \rho Wy + X\beta + \varepsilon$.
Define the auxiliary residuals $e_0=M_Z y$, $e_1=M_Z Wy$, and
$e_2=M_Z W^2y$, with the same $Z$ and $M_Z$ as above.
The filtered residual at $\rho$ is $u(\rho) = e_0 - \rho e_1$.

Following \citet{smirnov2020}, $\rho$ satisfies the orthogonality
condition $u(\rho)'(e_1-\rho e_2)=0$, which expands to
$a\rho^2-b\rho+c=0$ with $a=e_1'e_2$, $b=e_0'e_2+e_1'e_1$, and
$c=e_0'e_1$. The CFE estimator is
\[
\hat{\rho}_{CFE} = \frac{b-\sqrt{b^2-4ac}}{2a}.
\]

\begin{proposition}
\label{prop:sar_ratio}
Define $\tilde{\rho}_R = \frac{e_0'e_1}{e_0'e_2+e_1'e_1}$.
If $\delta = \frac{4ac}{b^2}$ is small, then
$\hat{\rho}_{CFE} = \tilde{\rho}_R + O(\delta)$.
\end{proposition}

\noindent\textit{Proof.}
Using $\sqrt{b^2-4ac}=b\sqrt{1-\delta}$ and the Taylor expansion
$\sqrt{1-\delta} = 1-\delta/2+O(\delta^2)$:
\[
\hat{\rho}_{CFE}
  = \frac{b}{2a}\!\left(\frac{\delta}{2}+O(\delta^2)\right)
  = \frac{c}{b}+O\!\left(\frac{ac^2}{b^3}\right).
\]
Since $c/b=\tilde{\rho}_R$, the result follows.
\hfill$\square$

\begin{corollary}
When the quadratic coefficient $a$ is small relative to $b^2$, the CFE
estimator is well approximated by the ratio estimator $\tilde\rho_R$.
\end{corollary}

\subsection{Geometric interpretation and IV connection}

The Wald-type estimator $\tilde\rho_{IV} = e_1'e_0/e_1'e_1$ arises
as a further simplification obtained by replacing the instrument
$e_1-\rho e_2$ by $e_1$ in the CFE moment condition. It admits a
natural least-squares interpretation: $\tilde\rho_{IV}$ minimizes
$Q(\rho) = \|e_0 - \rho e_1\|^2$ and is therefore the scalar
projection of $e_0$ onto $e_1$ in the filtered residual space.

This interpretation connects the hierarchy to the instrumental-variables
literature in spatial econometrics. The spatial two-stage least squares
estimators of \citet{kelejian1998,kelejian1999} use spatially lagged
regressors $WX$ and $W^2X$ as instruments for $Wy$; the GMM estimator
of \citet{lee2003} and the framework of \citet{kelejian2010} rely on
moment conditions involving spatially filtered variables conceptually
close to those used here. The estimator $\tilde\rho_{IV}$ can therefore
be interpreted simultaneously as a first-order approximation to the CFE
moment condition and as a Wald-type just-identified IV estimator in the
filtered SAR representation.

\section{Formal bound on the quasi-linear approximation error under
GCV smoothing parameter estimation}
\label{appendix_formal_bound}

\paragraph{Scope of this appendix}
The results established here --- the Lipschitz bound on the
smoothing operator (Proposition~\ref{prop:lipschitz}), the
differentiability of the GCV-optimal smoothing parameter via the
implicit function theorem (Section~\ref{sec:ift}), and the main
approximation error bound (Proposition~\ref{prop:bound}) ---
apply to additive estimators based on penalized splines with GCV
or REML smoothing parameter selection. They cover the
\texttt{GAM\_SAR\_CFE} and \texttt{GAM\_SEM\_CFE} estimators and
complement the asymptotic analysis of
Appendix~\ref{appendix_asymptotic_cfe}, whose results are
established directly for the implemented fixed-smoother estimator
and do not rely on the bounds below. Component-wise gradient
boosting (the boosted \texttt{BSPA} variants) does not produce a fixed
smoothing matrix and the implicit function theorem argument does
not apply directly; the validity of the CFE approach for
boosting-based estimators is addressed in
Appendix~\ref{appendix_asymptotic_cfe}
(Corollary~\ref{cor:boosting}), conditionally on the selection
path, and supported by the Monte Carlo evidence of
Section~\ref{sec:simulations}.

This appendix establishes the formal $O_p$ bound on the approximation
error induced by the endogeneity of the smoothing parameter $\hat\mu(\rho)$.
In nonlinear settings, the smoothing operator itself depends on the 
spatial parameter $\rho$, potentially distorting the CFE estimating 
equations. By proving that this perturbation is asymptotically 
negligible (Proposition~\ref{prop:bound}), we provide the theoretical 
justification for the consistency and asymptotic normality of the 
determinant-free estimator in nonparametric boosting contexts.

\subsection{Setup and approximation error}

For a single-component penalized spline model (the additive extension
follows component-wise by the same argument), the filtered response is
$\mathbf{y}_\rho = (\mathbf{I}-\rho\mathbf{W})\mathbf{y}$
and the residual operator at the GCV-optimal smoothing parameter is
$\mathbf{M}_{\hat\mu(\rho)} = \mathbf{I} - \mathbf{S}_{\hat\mu(\rho)}$,
where
\[
\mathbf{S}_\mu
  = \mathbf{B}(\mathbf{B}^\top\mathbf{B}+\mu\mathbf{D})^{-1}\mathbf{B}^\top
\]
denotes the smoothing matrix, $\mathbf{B}\in\mathbb{R}^{n\times K}$
is the fixed spline design matrix, and $\mathbf{D}\in\mathbb{R}^{K\times K}$
is the fixed penalty matrix.

The quasi-linear decomposition $e(\rho)\approx e_0-\rho e_1$, with
$e_0=\mathbf{M}_{\hat\mu(0)}\mathbf{y}$ and
$e_1=\mathbf{M}_{\hat\mu(0)}\mathbf{W}\mathbf{y}$, replaces
$\hat\mu(\rho)$ by its value at $\rho=0$. The resulting
approximation error is
\[
\Delta(\rho)
  = -\bigl[\mathbf{S}_{\hat\mu(\rho)}-\mathbf{S}_{\hat\mu(0)}\bigr]
    \mathbf{y}_\rho.
\]
We bound $\|\Delta(\rho)\|$ under the following assumptions.

\begin{description}
\item[(H1)] \emph{Regularity of the spline basis and penalty.}
  $\mathbf{B}$ has full column rank $K$ and $\mathbf{D}$ is positive
  semi-definite with null-space of dimension $K_0<K$.
\item[(H2)] \emph{Non-degeneracy of the response.}
  In the Demmler--Reinsch basis of $(\mathbf{B},\mathbf{D})$
  introduced below,
  $\alpha_k(\rho)=\mathbf{u}_k^\top\mathbf{y}_\rho\neq 0$ for at
  least one $k$ with $d_k>0$, almost surely, for every
  $\rho\in[-1+\varepsilon,1-\varepsilon]$.
\item[(H3)] \emph{Interior optimum.}
  $\hat\mu(\rho)\in[\mu_{\min},\mu_{\max}]$ almost surely
  for constants $0<\mu_{\min}<\mu_{\max}<\infty$, uniformly
  over $\rho\in[-1+\varepsilon,1-\varepsilon]$.
\item[(H4)] \emph{Moment conditions.}
  $\|\mathbf{y}\|^2=O_p(n)$ and $\|\mathbf{W}\mathbf{y}\|^2=O_p(n)$.
  This condition is implied by (A1) and (A4) of
  Appendix~\ref{appendix_asymptotic_cfe}: the spectral bound on
  $(I-\rho W)^{-1}$ from (A1) combined with the moment condition
  on $\varepsilon$ from (A4) yields $n^{-1}\|\mathbf{y}\|^2=O_p(1)$,
  and row-standardisation of $\mathbf{W}$ then gives
  $n^{-1}\|\mathbf{W}\mathbf{y}\|^2=O_p(1)$.
\item[(H5)] \emph{Unique, non-degenerate GCV optimum.}
  For every $\rho\in[-1+\varepsilon,1-\varepsilon]$, the global
  minimiser $\hat\mu(\rho)$ of the GCV criterion is almost surely
  unique, and there exists $c>0$ such that
  $\inf_{\rho}\, n^{-1}\,\partial F/\partial\mu\,(\hat\mu(\rho),\rho)
  \geq c$ with probability tending to one, where
  $F(\mu,\rho)$ denotes the first-order condition defined below.
\end{description}

Assumption (H1) is satisfied by construction for standard B-spline
bases. (H2) holds almost surely whenever the error term has an
absolutely continuous distribution. (H3) is a standard identifiability
condition excluding degenerate signal-to-noise ratios. (H4) follows
from row-standardisation of $\mathbf{W}$ and second-moment conditions
on the disturbances. (H5) rules out ties and degenerate flat regions
of the GCV profile; it is required because the implicit function
theorem only tracks a locally unique, non-degenerate solution branch,
while the GCV criterion may in general possess multiple local minima.

\subsection{Lipschitz continuity of the smoothing matrix}

\begin{proposition}[Lipschitz continuity of $\mathbf{S}_\mu$]
\label{prop:lipschitz}
Under \emph{(H1)} and \emph{(H3)}, there exists a constant $L<\infty$
such that for all $\mu,\mu'\in[\mu_{\min},\mu_{\max}]$,
\[
\|\mathbf{S}_\mu-\mathbf{S}_{\mu'}\|_2 \leq L\,|\mu-\mu'|.
\]
\end{proposition}

\noindent\textit{Proof.}
Let $\{(\sigma_k^2,d_k,\mathbf{u}_k)\}_{k=1}^K$ denote the
Demmler--Reinsch decomposition of $(\mathbf{B},\mathbf{D})$ (the
smoother $\mathbf{S}_\mu$ is symmetric since $\mathbf{D}$ is), so that
$\mathbf{S}_\mu = \sum_k s_k(\mu)\,\mathbf{u}_k\mathbf{u}_k^\top$
with $s_k(\mu) = \sigma_k^2/(\sigma_k^2+\mu d_k)$ for $d_k>0$.
Since the $\mathbf{u}_k$ form an orthonormal system,
$\|\mathbf{S}_\mu-\mathbf{S}_{\mu'}\|_2
= \max_{k:\,d_k>0}|s_k(\mu)-s_k(\mu')|$.
By the mean value theorem,
$|s_k(\mu)-s_k(\mu')|\leq|s_k'(\xi_k)|\,|\mu-\mu'|$
where
$|s_k'(\mu)|
= \sigma_k^2 d_k/(\sigma_k^2+\mu d_k)^2
\leq \sigma_k^2 d_k/(\sigma_k^2+\mu_{\min}d_k)^2 \equiv L_k$.
Setting $L=\max_k L_k<\infty$ completes the proof. Writing
$\gamma_k=d_k/\sigma_k^2$, note moreover that
$|s_k'(\mu)|=\gamma_k/(1+\mu\gamma_k)^2\leq 1/(4\mu_{\min})$,
so that $L\leq 1/(4\mu_{\min})$ uniformly in the basis dimension
$K$ and in $n$.
\hfill$\square$

\subsection{Differentiability of $\hat\mu(\rho)$ via the implicit
function theorem}

The GCV criterion evaluated at $\mathbf{y}_\rho$ is
\[
\operatorname{GCV}(\mu,\rho)
  = \frac{\|\mathbf{M}_\mu\mathbf{y}_\rho\|^2/n}
         {\bigl[1-\tau(\mu)/n\bigr]^2},
\]
where $\tau(\mu)=\operatorname{tr}(\mathbf{S}_\mu)$ depends only
on $\mu$. The first-order condition is
$F(\mu,\rho) \equiv r'(\mu,\rho)\,d(\mu)
- 2\,r(\mu,\rho)\,d'(\mu) = 0$,
where $r(\mu,\rho)=\|\mathbf{M}_\mu\mathbf{y}_\rho\|^2$ and
$d(\mu)=1-\tau(\mu)/n$.

We apply the implicit function theorem to $F(\mu,\rho)=0$ at
$(\hat\mu(0),0)$ by verifying two conditions.

\paragraph{$C^1$ regularity in $\rho$}
Since $\mathbf{y}_\rho=\mathbf{y}-\rho\mathbf{W}\mathbf{y}$ is linear
in $\rho$, the function $r(\mu,\rho)$ is a degree-two polynomial
in $\rho$ at each fixed $\mu$, so $F$ is $C^\infty$ in $\rho$.

\paragraph{Non-degeneracy of the GCV minimum}
In the spectral representation,
$r'(\mu,\rho) = \sum_{k:\,d_k>0}
2(1-s_k(\mu))(-s_k'(\mu))\alpha_k^2(\rho) > 0$
almost surely under (H2), where $\alpha_k(\rho)=\mathbf{u}_k^\top\mathbf{y}_\rho$.
Non-degeneracy of the minimum is guaranteed by (H5), which requires
$n^{-1}\partial F/\partial\mu$ to be bounded away from zero at the
optimum, uniformly in $\rho$; (H5) also ensures that the global
minimiser coincides with the locally tracked solution branch.

\paragraph{Application of the implicit function theorem}
\label{sec:ift}
Both conditions are satisfied, so $\hat\mu(\rho)$ is $C^1$ in $\rho$
and
$\partial\hat\mu/\partial\rho
= -(\partial F/\partial\rho)/(\partial F/\partial\mu)$.
Since $\partial F/\partial\rho = O_p(n)$ under (H4) and
$n^{-1}\partial F/\partial\mu \geq c > 0$ with probability tending
to one under (H5), uniformly in $\rho$, we obtain
\[
\sup_{\rho\in[-1+\varepsilon,1-\varepsilon]}
\left|\frac{\partial\hat\mu}{\partial\rho}\right| = O_p(1).
\]

\subsection{Main result}

\begin{proposition}[Quasi-linear approximation error]
\label{prop:bound}
Under assumptions \emph{(H1)--(H5)}, for all
$\rho\in[-1+\varepsilon,1-\varepsilon]$,
\[
\frac{1}{\sqrt{n}}\|\Delta(\rho)\|
  \;\leq\; \frac{1}{4\mu_{\min}}\,
  \bigl|\hat\mu(\rho)-\hat\mu(0)\bigr|\cdot
  \frac{\|\mathbf{y}_\rho\|}{\sqrt{n}}
  \;=\; O_p\!\bigl(|\hat\mu(\rho)-\hat\mu(0)|\bigr),
\]
and the induced perturbation of the CFE quadratic coefficients
satisfies
$|n^{-1}e_i^\top\Delta_j|
 = O_p\!\bigl(|\hat\mu(\rho)-\hat\mu(0)|\bigr)$.
The quantity $|\hat\mu(\rho)-\hat\mu(0)|$ is observable and can be
reported as an empirical diagnostic. The asymptotic results of
Appendix~\ref{appendix_asymptotic_cfe} are established directly for
the estimator with smoothing parameters held fixed, as implemented,
and do not rely on this bound.
\end{proposition}

\noindent\textit{Proof.}
By Proposition~\ref{prop:lipschitz} with the uniform constant
$L\leq 1/(4\mu_{\min})$,
$\|\mathbf{S}_{\hat\mu(\rho)}-\mathbf{S}_{\hat\mu(0)}\|_2
\leq (4\mu_{\min})^{-1}\,|\hat\mu(\rho)-\hat\mu(0)|$.
Therefore
$\|\Delta(\rho)\|
\leq (4\mu_{\min})^{-1}|\hat\mu(\rho)-\hat\mu(0)|\cdot
\|\mathbf{y}_\rho\|$,
with $\|\mathbf{y}_\rho\|=O_p(\sqrt{n})$ under (H4); dividing by
$\sqrt{n}$ gives the first display. By the implicit function
theorem argument above, under (H5),
$|\hat\mu(\rho)-\hat\mu(0)|
\leq \sup_\xi|\partial\hat\mu/\partial\rho|_\xi\,|\rho|
= O_p(|\rho|)$; whether this quantity vanishes asymptotically
depends on the behaviour of $\partial\hat\mu/\partial\rho$ as
$n$ grows, which is not further quantified here --- hence the
bound is stated in terms of the observable
$|\hat\mu(\rho)-\hat\mu(0)|$.
For the CFE coefficients,
$|n^{-1}e_i^\top\Delta_j|
\leq n^{-1}\|e_i\|\,\|\Delta_j\|
= O_p\!\bigl(|\hat\mu(\rho)-\hat\mu(0)|\bigr)$.
\hfill$\square$

\subsection{Extension to REML and to additive models}

For REML, the same conclusion holds because the REML log-likelihood is
smooth and strictly concave in $\mu$ under (H2) and (H3)
\citep{wood2011fast}, so the implicit function theorem applies
identically and yields $\partial\hat\mu_{\mathrm{REML}}/\partial\rho
= O_p(1)$.

For a $p$-component additive model, each $\hat\mu_j(\rho)$ satisfies
an implicit equation applied to the partial residuals of component $j$.
Since partial residuals are linear functions of $\mathbf{y}_\rho$, the
same arguments apply component-wise, yielding
$\partial\hat\mu_j/\partial\rho=O_p(1)$ for each $j$ and the same
overall $O_p(|\rho|/\sqrt{n})$ bound on the total approximation error.

\subsection{Spectral interpretation and empirical diagnostic}

The bound in Proposition~\ref{prop:bound} is derived through an analytic
argument and does not exploit the spatial structure of the problem. We
now show that the geometry of spatial autoregressive models provides an
independent and complementary perspective that illuminates the mechanism
underlying the approximation and suggests an empirically verifiable
diagnostic.

\paragraph{Spectral decomposition of the approximation error}

Let $\{\boldsymbol{\phi}_k, \omega_k\}_{k=1}^n$ denote the eigenvectors
and eigenvalues of the symmetrised spatial weights matrix $\mathbf{W}$,
ordered so that $\omega_1\geq\omega_2\geq\cdots\geq\omega_n$.
The filtered response decomposes as
$\mathbf{y}_\rho = \sum_k (1-\rho\omega_k)\,y_k\,\boldsymbol{\phi}_k$,
so that spatial filtering attenuates the low-frequency modes (large
$\omega_k$) most strongly for $\rho>0$.

Define the \emph{spectral projection coefficient}
\[
\pi_k(\mu)
  = \boldsymbol{\phi}_k^\top\mathbf{S}_\mu\boldsymbol{\phi}_k
  = \sum_{j:\,d_j>0} s_j(\mu)
    \bigl(\mathbf{u}_j^\top\boldsymbol{\phi}_k\bigr)^2
  \in [0,1],
\]
which measures the fraction of the $k$-th spatial eigenvector lying in
the column space of the smoothing operator. The total derivative of
$\mathbf{S}_{\hat\mu(\rho)}\mathbf{y}_\rho$ with respect to $\rho$
decomposes as
\begin{equation}
\frac{d}{d\rho}\mathbf{S}_{\hat\mu(\rho)}\mathbf{y}_\rho
  = \underbrace{-\mathbf{S}_{\hat\mu(\rho)}\mathbf{W}\mathbf{y}}_{\Delta_{\mathrm{geom}}(\rho)}
  + \underbrace{\mathbf{S}'_{\hat\mu(\rho)}
    \frac{\partial\hat\mu}{\partial\rho}\,\mathbf{y}_\rho}_{\Delta_{\mathrm{smooth}}(\rho)}.
\label{eq:decomp}
\end{equation}
The first term $\Delta_{\mathrm{geom}}$ is the variation of the fit
at a fixed smoother; it is not part of the approximation error
$\Delta(\rho)$, being already accounted for by the term $-\rho e_1$
of the quasi-linear decomposition. Only the second term contributes
to $\Delta(\rho)$, and it is the one bounded in
Proposition~\ref{prop:bound}.

\paragraph{Spectral separation and its consequences}

Define the index set of low-frequency modes as
$\mathcal{L}=\{k:\omega_k\geq\bar\omega\}$ for a threshold
$\bar\omega\in(0,1)$.

\begin{description}
\item[(HSS)] \emph{Spectral separation.}
  There exists $\delta\in(0,1)$ such that
  $\pi_k(\hat\mu)\leq\delta$ for all $k\in\mathcal{L}$.
\end{description}

Assumption (HSS) formalises the idea that the spline base learners do
not project substantially onto the dominant spatial eigenvectors of
$\mathbf{W}$. Under (HSS),
\[
\|\Delta_{\mathrm{geom}}(\rho)\|
  \leq |\rho|\left(
    \sqrt{\delta}\sum_{k\in\mathcal{L}}|\omega_k|\,|y_k|
    + \sum_{k\in\mathcal{H}}|\omega_k|\,|y_k|
  \right),
\]
so the contribution of dominant spatial modes to the fitted values
is attenuated by $\sqrt{\delta}$. The quantity $\pi_k(\hat\mu)$ is
observable and provides an empirical diagnostic of the overlap
between the smoother and the dominant spatial modes; it is not used
in the proofs of Propositions~\ref{prop:consistency_main}
and~\ref{prop:variance_main}. Large values of $\pi_k(\hat\mu)$ on
low-frequency modes signal spatial confounding and flag
configurations in which the identification condition (A3) deserves
empirical scrutiny.

\paragraph{Connection with the spatial confounding literature}

Assumption (HSS) is closely related to the concept of spatial
confounding studied in geostatistics and spatial regression
\citep{hanks2015restricted,reich2006effects}. The quantity $\pi_k$
can be read as the measured counterpart of the Moran-basis
construction of \citet{hughes2013}, which enforces zero overlap
between the spatial basis and the covariates by design; it also
gives a matrix form to the scale-based analyses of spatial
confounding of \citet{paciorek2010} and \citet{keller2020}, with the
distinctive feature of being computed on the column space of the
fitted covariate smoother $\mathbf{S}_{\hat\mu}$ rather than on the
covariates themselves. Spatial confounding
arises precisely when $\pi_k$ is large for dominant spatial modes,
i.e.\ when covariates are strongly spatially structured and their
spline representations project substantially onto the low-frequency
eigenvectors of $\mathbf{W}$.

\begin{figure}[htbp]
\centering
\includegraphics[width=0.85\textwidth]{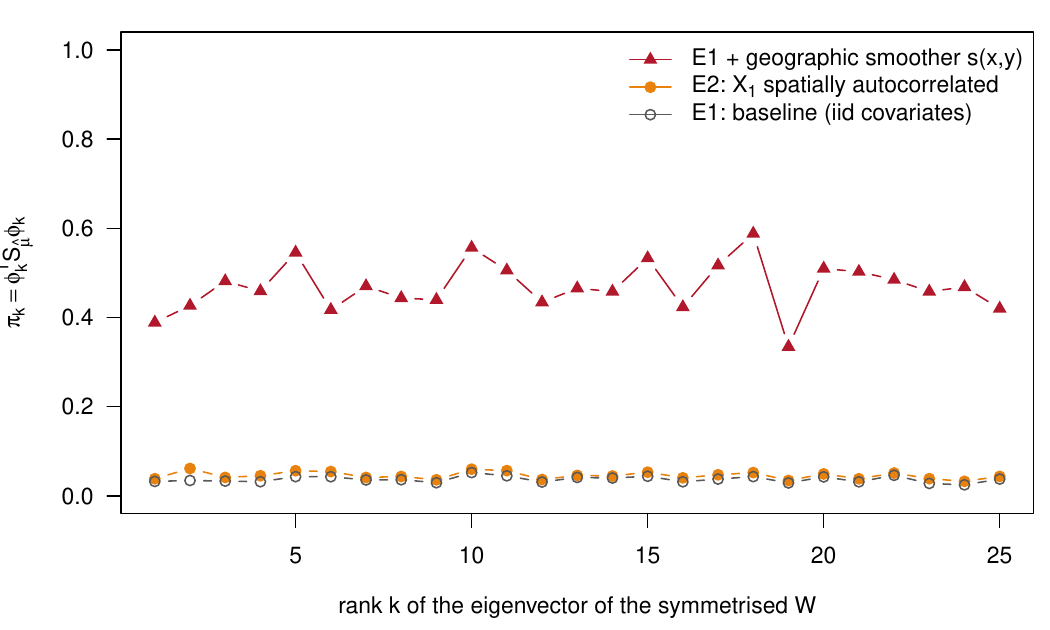}
\caption{Spectral projection diagnostic
$\pi_k=\boldsymbol{\phi}_k^\top\mathbf{S}_{\hat\mu}\boldsymbol{\phi}_k$
on the 25 dominant eigenvectors of the symmetrised $\mathbf{W}$
($n=1000$, 4-nearest-neighbour weights, $\rho_0=0.6$,
$\mathrm{SNR}=0.7$; averages over 10 replications). The covariate
smoother $s(x_1)+s(x_2)+s(x_3)$ leaves the dominant spatial modes
essentially untouched in the baseline design E1 and in design E2 with
a spatially autocorrelated $x_1$, supporting the spectral separation
condition (HSS). Adding a bivariate geographic smoother $s(u,v)$
raises $\pi_k$ sharply: the geoadditive specification spans the
low-frequency spatial modes --- the spectral signature of the
geoadditive trap.}
\label{fig:pi_k}
\end{figure}

\paragraph{Two complementary diagnostics}
The \textit{spboost} package implements two diagnostics that probe
spectral separation at different stages of the analysis. The first,
available through the \texttt{spectral\_overlap} control option, is a
\emph{data-level screening} tool: it compares the spectral energy
profiles of each covariate and of the spatial lag component on the
dominant eigenvectors of the symmetrised $\mathbf{W}$, and flags,
covariate by covariate, those living on the same spatial wavelengths
as the autoregressive signal. Because it only requires $\mathbf{W}$
and the data, it can be computed before any model is fitted and
applies identically to GAM, boosting, and tree-based learners. The
second, the \texttt{pi\_k\_diagnostic()} function, is the
\emph{model-level check} corresponding exactly to condition (HSS): it
measures the fraction $\pi_k$ of each dominant spatial mode
reproducible by the \emph{fitted} smoother $\mathbf{S}_{\hat\mu}$,
thereby accounting for the joint effect of all covariates and for the
amount of penalization actually selected by GCV or REML; it requires
a smoothing-matrix representation and is therefore available for the
penalized-spline (GAM) estimators. The two are complementary: a large
covariate-level overlap signals a \emph{potential} risk that the
fitted model may or may not realise --- a heavily penalized smooth
absorbs little spatial structure even when its covariate is spatially
organised --- while a large $\pi_k$ establishes that the fitted model
does span the dominant spatial modes. In practice we recommend
screening with \texttt{spectral\_overlap} for any learner, and
confirming with $\pi_k$ on the penalized-spline fit. The Monte Carlo experiment E2, which
introduces spatial autocorrelation in covariate $x_1$, is precisely the
setting where (HSS) is most at risk. Figure~\ref{fig:pi_k} makes the
diagnostic operational: on simulated E1 and E2 designs ($n=1000$,
$\rho_0=0.6$, $\mathrm{SNR}=0.7$), the additive smoother of the
covariates captures almost none of the dominant spatial modes
($\pi_k\leq 0.07$ on the 25 leading eigenvectors, with only a slight
increase from E1 to E2), so that (HSS) holds comfortably in both
designs --- direct evidence consistent with the stability of the
estimators observed in E2. By contrast, augmenting the model with a
bivariate geographic smoother $s(u,v)$ raises $\pi_k$ to between
roughly $0.3$ and $0.6$ on the dominant modes: the geoadditive
specification directly absorbs low-frequency spatial structure, which
is the spectral signature of the geoadditive identification trap
analysed in Section~\ref{sec:E3p_results}. The diagnostic is
implemented in the \texttt{pi\_k\_diagnostic()} function of the
\textit{spboost} package --- computed without forming
$\mathbf{S}_{\hat\mu}$, by refitting the additive model on each
eigenvector with smoothing parameters held fixed --- and
Figure~\ref{fig:pi_k} is reproduced by the script
\ifpreprint
\path{replication_scripts/fig_pi_k_E1_E2.R} of the replication
repository \url{https://github.com/ggeniaux/replication_spboost_geniaux_2026}.
\else
\texttt{inst/scripts/fig\_pi\_k\_E1\_E2.R}.
\fi

Table~\ref{tab:spectral_overlap} reports the data-level diagnostic on
the same designs as Figure~\ref{fig:pi_k}. For unstructured variables
the overlap fluctuates around its chance level of roughly $0.4$; the
spatially autoregressive covariate $\tilde x_1$ of design E2 is
clearly flagged ($0.58$, against $0.44$ in E1), while $x_2$, $x_3$
and the coordinates remain at the chance level. Read together with
Figure~\ref{fig:pi_k}, the two diagnostics illustrate their
complementarity in both directions. The covariate-level screen flags
$\tilde x_1$ as a potential risk that the fitted smoother does not,
in fact, realise ($\pi_k\leq 0.07$ in E2); conversely, it leaves the
raw coordinates unflagged, whereas the geoadditive smoother $s(u,v)$
built from them is precisely the specification that spans the
dominant spatial modes ($\pi_k$ up to $0.59$ in
Figure~\ref{fig:pi_k}). Screening on the data alone is therefore
neither sufficient nor necessary for spectral overlap of the fitted
model: the model-level check $\pi_k$ is the operative one, and the
data-level screen is best used to decide which variables deserve a
constrained specification before fitting.

\begin{table}[htbp]
\centering
\caption{Data-level spectral overlap diagnostic
(\texttt{spectral\_overlap} option of \textit{spboost}) between each
variable and the spatial lag component, on the 25 dominant
eigenvectors of the symmetrised $\mathbf{W}$ ($n=1000$, $\rho_0=0.6$,
$\mathrm{SNR}=0.7$; averages over 10 replications, same designs as
Figure~\ref{fig:pi_k}). Values near $0.4$ correspond to the chance
level of the profile-intersection metric; only departures from it
are informative.}
\label{tab:spectral_overlap}
\begin{tabular}{lccccc}
\toprule
 & $x_1$ & $x_2$ & $x_3$ & $u$ & $v$ \\
\midrule
E1 (iid covariates) & 0.44 & 0.39 & 0.41 & 0.39 & 0.40 \\
E2 ($\tilde x_1$ spatially autocorrelated) & 0.58 & 0.36 & 0.44 & 0.39 & 0.42 \\
\bottomrule
\end{tabular}
\end{table}

\section{Asymptotic properties of the CFE estimator in nonlinear
additive models}
\label{appendix_asymptotic_cfe}

This appendix establishes the asymptotic properties of the CFE
estimator \emph{as implemented}: the auxiliary residuals are obtained
from three separate penalized-spline additive regressions of
$\mathbf{y}$, $\mathbf{W}\mathbf{y}$ and $\mathbf{W}^2\mathbf{y}$ on
a common spline basis of fixed dimension, each with its own GCV-
or REML-selected smoothing parameters. The analysis proceeds in two
steps. A collapse lemma (Lemma~\ref{lem:collapse}) first shows that,
because the basis is fixed while the Gram matrix grows with $n$,
the three fitted smoothers converge to the same orthogonal projection
onto the spline span, so that the implemented estimator differs from
a projection-based CFE by an asymptotically negligible perturbation.
The projection-based CFE is then exactly the closed-form estimator of
\citet{smirnov2020} for the linear SAR model with regressor matrix
$\mathbf{Z}=\mathbf{B}$, so that consistency and asymptotic normality
are inherited from the linear case --- together with the documented
conservativeness of the closed-form variance. We focus on the SAR
case; the SEM case is analogous, with the corresponding linear-case
conditions of \citet{smirnov2020}.

\subsection{Setup and assumptions}

Consider the nonlinear SAR model
\[
\mathbf{y} = \rho_0\mathbf{W}\mathbf{y} + f(\mathbf{X}) + \boldsymbol{\varepsilon},
\qquad
f(\mathbf{X}) = \beta_0 + \sum_{j=1}^p f_j(x_j),
\]
where $\boldsymbol{\varepsilon}\sim(0,\sigma^2\mathbf{I})$ is independent
of $\mathbf{X}$, $\rho_0\in(-1,1)$, and $f$ is unknown and additive.
The intercept $\beta_0$ is identified by the centering constraint
$\mathbb{E}[f_j(x_j)]=0$ for each $j$; after centering, $\beta_0$
equals the unconditional mean of the filtered response and does not
affect the asymptotic properties of the spatial parameter estimator.
The filtered response at the true parameter is
$\mathbf{y}_{\rho_0} = f(\mathbf{X})+\boldsymbol{\varepsilon}$.

The CFE estimator $\hat\rho_{CFE}$ is defined as the admissible root of
$a\rho^2-b\rho+c=0$, where
$a = e_1^\top e_2$, $b = e_0^\top e_2+e_1^\top e_1$, $c = e_0^\top e_1$,
and, as implemented,
$e_k = (\mathbf{I}-\mathbf{S}_{\hat\mu_k})\mathbf{W}^k\mathbf{y}$ for
$k=0,1,2$, where
$\mathbf{S}_{\mu}=\mathbf{B}(\mathbf{B}^\top\mathbf{B}
+\mu\mathbf{D})^{-1}\mathbf{B}^\top$ denotes the penalized-spline
smoother on a common fixed design $\mathbf{B}$ and $\hat\mu_k$ the
smoothing parameters selected by GCV or REML in the $k$-th auxiliary
regression (componentwise smoothing parameters are absorbed in the
notation $\mu\mathbf{D}$).

\begin{description}
\item[(A1)] \emph{Spatial regularity.}
  $\mathbf{W}$ has zero diagonal, is row-standardised, and its row
  and column sums are uniformly bounded in absolute value. All
  eigenvalues of $\mathbf{W}$ lie in the closed unit disc, the
  eigenvalue $1$ (associated with the constant eigenvector
  $\boldsymbol{\iota}$) being simple. The true parameter satisfies
  $\rho_0\in\mathcal{P}=[-1+\varepsilon,1-\varepsilon]$, and
  $\sup_{\rho\in\mathcal{P}}
  \|(\mathbf{I}-\rho\mathbf{W})^{-1}\|_{1}\vee
  \|(\mathbf{I}-\rho\mathbf{W})^{-1}\|_{\infty}=O(1)$.

\item[(A2)] \emph{Fixed-basis representation.}
  The spline design matrix $\mathbf{B}\in\mathbb{R}^{n\times K}$ has
  fixed column dimension $K$ and full column rank for $n$ large,
  $n^{-1}\mathbf{B}^\top\mathbf{B}\xrightarrow{p}\mathbf{G}$ with
  $\mathbf{G}$ positive definite, and
  $f(\mathbf{X})=\mathbf{B}\boldsymbol{\beta}_0$ for some fixed
  $\boldsymbol{\beta}_0$. The selected smoothing parameters are
  stochastically bounded, $\hat\mu_k=O_p(1)$ for $k=0,1,2$
  (cf.\ (H3) of Appendix~\ref{appendix_formal_bound}).
  Remark~\ref{rem:approx} quantifies the effect of an approximation
  error when $f$ lies only near the span of $\mathbf{B}$, and
  Remark~\ref{rem:sieve} discusses the connection with the
  nonparametric sieve regime in which the basis grows with $n$. The present appendix
  establishes the formal bound under the penalized spline
  representation; the scope of this result relative to
  boosting-based estimators is discussed in
  Remark~\ref{rem:scope} of the main text.

\item[(A3)] \emph{Identification.}
  (i) $n^{-1}a\xrightarrow{p}\bar a$,
  $n^{-1}b\xrightarrow{p}\bar b$ and
  $n^{-1}c\xrightarrow{p}\bar c$, with finite limits satisfying
  $\bar b-2\rho_0\bar a>0$;
  (ii) $Q_n \equiv n^{-1}(e_1-\rho_0 e_2)^\top(e_1-\rho_0 e_2)
  \xrightarrow{p} Q_0>0$;
  (iii) $\rho_0$ is the root of
  $\bar g(\rho)=\bar a\rho^2-\bar b\rho+\bar c$ selected by the
  root-selection rule of Appendix~\ref{appendix_cfe_linear}, in the
  sense of Lemma~\ref{lem:root} below.

\item[(A4)] \emph{Moment conditions.}
  $\mathbb{E}|\varepsilon_i|^{4+\eta}<\infty$ for some $\eta>0$, and
  $n^{-1}\|\mathbf{W}^k\mathbf{y}\|^2=O_p(1)$ for $k=0,1,2$.

\item[(A5)] \emph{Imported linear-case conditions.}
  The regularity conditions of \citet{smirnov2020} for the
  consistency and asymptotic normality of the linear closed-form
  estimator hold for the linear SAR model
  $\mathbf{y}=\rho_0\mathbf{W}\mathbf{y}
  +\mathbf{B}\boldsymbol{\beta}_0+\boldsymbol{\varepsilon}$ with
  regressor matrix $\mathbf{Z}=\mathbf{B}$.
\end{description}

Assumptions (A1) and (A3) concern exclusively the spatial structure.
(A2) states the fixed-basis regime in which the implemented
estimators operate; its scope relative to boosting-based estimators
is addressed by Corollary~\ref{cor:boosting} and discussed in
Remark~\ref{rem:scope}. (A4) collects standard moment conditions,
and (A5) imports the linear-case regularity conditions of
\citet{smirnov2020} applied to the spline design.

\begin{remark}[Equal-weight matrices]
Assumption~(A1) does not by itself exclude the equal-weight scheme
$w_{ij}=(n-1)^{-1}$ for $i\neq j$
\citep{kelejian2002,baltagi2006,lee2004asymptotic}. In that case,
however, $\mathbf{W}\mathbf{y}=(n\bar y\,\boldsymbol{\iota}
-\mathbf{y})/(n-1)$, so the demeaned spatial lag degenerates and
$Q_n\rightarrow 0$: identification fails through (A3)(ii), in line
with the known inconsistency of quasi-likelihood estimators under
equal weights. More generally, (A3) requires the number of
neighbours per unit to grow more slowly than $n$.
\end{remark}

\subsection{Collapse of the implemented smoothers onto the spline
projection}

\begin{lemma}[Collapse lemma]
\label{lem:collapse}
Let $\mathbf{P}_B=\mathbf{B}(\mathbf{B}^\top\mathbf{B})^{-1}
\mathbf{B}^\top$ denote the orthogonal projection onto the span of
the basis, let
$e_k^{P}=(\mathbf{I}-\mathbf{P}_B)\mathbf{W}^k\mathbf{y}$,
$k=0,1,2$, and let $(a^P,b^P,c^P)$ be the corresponding CFE
coefficients. Under \emph{(A1)}, \emph{(A2)} and \emph{(A4)},
\[
\|\mathbf{S}_{\hat\mu_k}-\mathbf{P}_B\|_2=O_p(n^{-1}),
\qquad
n^{-1}\bigl(|a-a^P|+|b-b^P|+|c-c^P|\bigr)=O_p(n^{-1}),
\]
and, when $\bar b-2\rho_0\bar a>0$ as in \emph{(A3)(i)}, the
selected roots satisfy
$\hat\rho_{CFE}-\hat\rho_{CFE}^{\,P}=O_p(n^{-1})=o_p(n^{-1/2})$.
\end{lemma}

\noindent\textit{Proof.}
Write $\mathbf{B}^\top\mathbf{B}=n\mathbf{G}_n$ with
$\mathbf{G}_n\xrightarrow{p}\mathbf{G}>0$ by (A2). Then
\[
\mathbf{S}_{\mu}-\mathbf{P}_B
=-\mathbf{B}\,(n\mathbf{G}_n+\mu\mathbf{D})^{-1}\mu\mathbf{D}\,
 (n\mathbf{G}_n)^{-1}\mathbf{B}^\top .
\]
Since
$\|\mathbf{B}(n\mathbf{G}_n+\mu\mathbf{D})^{-1}\|_2
 \leq \|\mathbf{B}\|_2\,/\,(n\lambda_{\min}(\mathbf{G}_n))
 = O_p(n^{-1/2})$
and
$\|(n\mathbf{G}_n)^{-1}\mathbf{B}^\top\|_2
 = \lambda_{\min}(\mathbf{B}^\top\mathbf{B})^{-1/2}=O_p(n^{-1/2})$,
we obtain
$\|\mathbf{S}_{\mu}-\mathbf{P}_B\|_2
 \leq \mu\,\|\mathbf{D}\|_2\,O_p(n^{-1})$, which is $O_p(n^{-1})$
at $\mu=\hat\mu_k=O_p(1)$ by (A2). Consequently
$\|e_k-e_k^P\|
\leq\|\mathbf{S}_{\hat\mu_k}-\mathbf{P}_B\|_2\,
\|\mathbf{W}^k\mathbf{y}\|=O_p(n^{-1})\,O_p(n^{1/2})
=O_p(n^{-1/2})$ by (A4). Each CFE coefficient is a bilinear form in
the $e_k$, so for instance
$|a-a^P|\leq\|e_1-e_1^P\|\,\|e_2\|+\|e_1^P\|\,\|e_2-e_2^P\|
=O_p(n^{-1/2})\,O_p(n^{1/2})=O_p(1)$, whence
$n^{-1}|a-a^P|=O_p(n^{-1})$, and similarly for $b$ and $c$. Finally,
on a neighbourhood of $\rho_0$ where $n^{-1}(b-2\rho a)$ is bounded
away from zero in probability (by (A3)(i)), the selected root is a
Lipschitz function of the normalised coefficients $n^{-1}(a,b,c)$,
so a perturbation of order $O_p(n^{-1})$ of the coefficients moves
the root by $O_p(n^{-1})$.
\hfill$\square$

\begin{corollary}[Boosting variant]
\label{cor:boosting}
Let $\mathbf{B}_m$ denote the linear operator associated with $m$
iterations of componentwise $L_2$-boosting with spline base learners
drawn from the same fixed basis, \emph{conditionally on the
selection path}, and assume
\begin{description}
\item[(BG)] \emph{Geometric path convergence.} Along the realised
  selection path, $\|\mathbf{B}_m-\mathbf{P}_B\|_2\leq C\gamma^{m}$
  for some constants $C<\infty$ and $\gamma\in(0,1)$.
\end{description}
If $\texttt{mstop}\geq \log n/\log(1/\gamma)$, then
$\|\mathbf{B}_{m}-\mathbf{P}_B\|_2=O(n^{-1})$ and the conclusions of
Lemma~\ref{lem:collapse}, hence of
Propositions~\ref{prop:consistency_main}
and~\ref{prop:variance_main}, hold for the boosting-based auxiliary
residuals, conditionally on the selection path.
\end{corollary}

\noindent\textit{Proof.} Given (BG) and the choice of
\texttt{mstop}, $\gamma^{m}\leq n^{-1}$; the rest is identical to
the propagation step of Lemma~\ref{lem:collapse}, with
$\|\mathbf{B}_m-\mathbf{P}_B\|_2=O(n^{-1})$ in place of
$\|\mathbf{S}_{\hat\mu_k}-\mathbf{P}_B\|_2=O_p(n^{-1})$. The
premise (BG) itself is \emph{not} proven here: it is the linear
convergence of greedy coordinate descent on a strongly convex
quadratic, which holds for selection sequences that update every
component sufficiently often --- e.g.\ cyclic or Gauss--Southwell
rules on a full-rank basis
\citep{buhlmann2003boosting} --- but has not been established
uniformly over data-driven paths; see
Section~\ref{sec:conclusion}.
\hfill$\square$

\subsection{Consistency of the CFE estimator}

\paragraph{Proof of Proposition~\ref{prop:consistency_main}}
Under assumptions \emph{(A1)--(A4)}, $\hat\rho_{CFE}\xrightarrow{p}\rho_0$.

\begin{lemma}[Root selection]
\label{lem:root}
Let $\bar g(\rho)=\bar a\rho^2-\bar b\rho+\bar c$ have real roots
$\rho_0\neq\rho_1$, and let
$\rho_{IV}=\operatorname{plim}\,e_1^\top e_0/e_1^\top e_1$. If
$|\rho_0-\rho_{IV}|<|\rho_1-\rho_{IV}|$, then the robust
root-selection rule of Appendix~\ref{appendix_cfe_linear}, which
retains the admissible root minimising $\|e_0-\rho e_1\|^2$,
selects a root converging in probability to $\rho_0$.
\end{lemma}

\noindent\textit{Proof.}
$n^{-1}\|e_0-\rho e_1\|^2$ is a quadratic in $\rho$ whose
coefficients converge in probability under (A3)(i) and (A4); its
limit is uniquely minimised at $\rho_{IV}$ and is strictly
increasing in $|\rho-\rho_{IV}|$. The two sample roots converge to
$(\rho_0,\rho_1)$ by continuity of the roots of a quadratic in its
coefficients away from $\bar a=0$ (if $\bar a=0$ the equation
degenerates to a linear one with single root $\rho_0$ and the rule
is trivial). The rule therefore asymptotically retains the root
closest to $\rho_{IV}$, which is $\rho_0$ by hypothesis.
\hfill$\square$

\medskip
\noindent\textit{Proof of the proposition.}
By Lemma~\ref{lem:collapse} it suffices to prove the result for the
projection-based residuals $e_k^P$; we drop the superscript in this
proof. Write
$g_n(\rho)\equiv n^{-1}(e_0-\rho e_1)^\top(e_1-\rho e_2)
= n^{-1}(a\rho^2-b\rho+c)$, an exact identity. Under (A3)(i) the
coefficients converge, so
$\sup_{\rho\in\mathcal{P}}|g_n(\rho)-\bar g(\rho)|
\xrightarrow{p}0$: uniform convergence is automatic for
polynomials with converging coefficients on a compact set.

\emph{Step 1: $\bar g(\rho_0)=0$.}
With $\mathbf{M}^*=\mathbf{I}-\mathbf{P}_B$ (symmetric and
idempotent),
\[
e_0-\rho_0 e_1 = \mathbf{M}^*\mathbf{y}_{\rho_0}
=\mathbf{M}^*\bigl(f(\mathbf{X})+\boldsymbol{\varepsilon}\bigr)
=\mathbf{M}^*\boldsymbol{\varepsilon},
\]
since $f(\mathbf{X})=\mathbf{B}\boldsymbol{\beta}_0$ under (A2) and
$\mathbf{M}^*\mathbf{B}=0$: the smoothing-bias term vanishes
\emph{exactly} in the fixed-basis regime. Hence, with
$e_1-\rho_0 e_2=\mathbf{M}^*\mathbf{W}\mathbf{y}_{\rho_0}$ and
$\mathbf{M}^{*\top}\mathbf{M}^*=\mathbf{M}^*$,
\[
g_n(\rho_0)
= n^{-1}\boldsymbol{\varepsilon}^\top\mathbf{M}^*
  \mathbf{W}f(\mathbf{X})
+ n^{-1}\boldsymbol{\varepsilon}^\top\mathbf{M}^*
  \mathbf{W}\boldsymbol{\varepsilon}.
\]
The first term is centred with variance $O(n^{-1})$ under (A1) and
(A4), hence $O_p(n^{-1/2})$. The second term is a quadratic form
with mean
$\sigma^2 n^{-1}\operatorname{tr}(\mathbf{M}^*\mathbf{W})
=-\sigma^2 n^{-1}\operatorname{tr}(\mathbf{P}_B\mathbf{W})
=O(n^{-1})$,
since $\operatorname{tr}(\mathbf{W})=0$ and
$|\operatorname{tr}(\mathbf{P}_B\mathbf{W})|\leq
K\,\|\mathbf{W}\|_2$ with $K$ fixed, and with variance $O(n^{-1})$
under (A1) and (A4) \citep{kelejian2001}. Hence
$g_n(\rho_0)=O_p(n^{-1/2})=o_p(1)$ and $\bar g(\rho_0)=0$.

\emph{Conclusion.}
By (A3)(iii) and Lemma~\ref{lem:root}, the selected root of the
sample quadratic converges in probability to $\rho_0$, using the
uniform convergence of $g_n$ and the continuity of the roots of a
quadratic in its coefficients \citep[cf.][Theorem~2.1]{newey1994}.
\hfill$\square$

\subsection{Asymptotic validity of the variance formula}

The CFE variance estimator is
\[
\hat\sigma^2(\rho) = \frac{\hat\sigma^2}{b-2\hat\rho_{CFE}\,a},
\]
where $\hat\sigma^2=\hat{e}^\top\hat{e}/n$. This formula is derived
by \citeauthor{smirnov2020}~(\citeyear{smirnov2020}) under a linear
framework (his eqs.~16--19). In the fixed-basis regime the
nonlinear CFE inherits the linear-case asymptotics directly.

\paragraph{Proof of Proposition~\ref{prop:variance_main}}

\noindent\textit{Proof.}
By Lemma~\ref{lem:collapse},
$\sqrt{n}\,(\hat\rho_{CFE}-\hat\rho_{CFE}^{\,P})
=O_p(n^{-1/2})=o_p(1)$, so it suffices to derive the limiting
distribution of the projection-based estimator
$\hat\rho_{CFE}^{\,P}$. Under (A2), the data-generating process is
the linear SAR model
$\mathbf{y}=\rho_0\mathbf{W}\mathbf{y}
+\mathbf{B}\boldsymbol{\beta}_0+\boldsymbol{\varepsilon}$, and the
$e_k^P$ are exactly the OLS auxiliary residuals of the closed-form
estimator of \citet{smirnov2020} with regressor matrix
$\mathbf{Z}=\mathbf{B}$. Under (A5), his consistency and asymptotic
normality results apply verbatim:
$\sqrt{n}\,(\hat\rho_{CFE}^{\,P}-\rho_0)\xrightarrow{d}
\mathcal{N}(0,V_0)$, with $V_0$ the linear-case asymptotic variance,
and
$\hat\sigma^2(\rho)\xrightarrow{p}
\tilde V\equiv\sigma^2\big/\operatorname{plim}n^{-1}(b-2\rho_0 a)$,
the quasi-score information bound of his eq.~18; consistency of
$\hat\sigma^2$ for $\sigma^2$ follows from
$\hat\rho_{CFE}\xrightarrow{p}\rho_0$ and (A2)--(A4).
\hfill$\square$

\begin{remark}[Conservativeness of the closed-form variance]
\label{rem:conservative}
The quasi-score construction trades efficiency for closed-form
computation: the bound $\tilde V$ does not coincide with $V_0$ in
general, and \citet{smirnov2020} documents (his Tables~3 and~6) that
the closed-form variance systematically \emph{over}-estimates the
realized sampling variance of the linear CFE, by 10--20\% for
moderate spatial dependence and by a factor of 2--3 under very
strong dependence, with an efficiency loss relative to maximum
likelihood that grows with $\rho_0$. Our Monte Carlo experiments
(Appendix~\ref{appendix_additional_mc}) reproduce exactly this
pattern in the nonlinear setting --- variance ratios close to one
for $\rho_0\leq 0.2$, around $0.6$ at $\rho_0=0.6$ and $0.25$ at
$\rho_0=0.9$, stable across sample sizes --- indicating that the
wedge is inherited from the linear quasi-score construction rather
than induced by the nonlinear first stage. Confidence intervals
based on $\hat\sigma^2(\rho)$ are therefore conservative, with
coverage at or above the nominal level, increasingly so as $\rho_0$
grows. The sandwich estimators of Remark~\ref{rem:sandwich} and
Section~\ref{sec:remedies} tighten inference under
heteroskedasticity, but do not target this dependence-driven wedge;
when sharp inference on $\rho$ under strong spatial dependence is
required, the likelihood-based variants remain preferable.
\end{remark}

\begin{remark}[Approximation error]
\label{rem:approx}
If $f(\mathbf{X})=\mathbf{B}\boldsymbol{\beta}_0+\mathbf{r}$ with
$n^{-1/2}\|\mathbf{r}\|\leq\epsilon$, the smoothing-bias term
$\mathbf{M}^*f(\mathbf{X})=\mathbf{M}^*\mathbf{r}$ no longer
vanishes: the probability limits of $n^{-1}(a,b,c)$ are shifted by
$O(\epsilon)$ and the probability limit of $\hat\rho_{CFE}$ moves
to some $\rho^*$ with $|\rho^*-\rho_0|=O(\epsilon)$ under (A3). The
results above therefore degrade continuously with the basis
approximation error. For the smooth components of the Monte Carlo
designs and the default \texttt{mgcv} bases, the measured relative
approximation error $\|(\mathbf{I}-\mathbf{P}_B)f_j\|/\|f_j\|$ is
numerically zero for the linear $f_1$, $1.2\times10^{-3}$ for the
sinusoidal $f_3$, and $3.1\times10^{-2}$ for the quasi-V component
$f_2$, whose rounded kink at scale $\delta=0.01$ is the hardest
feature for a fixed basis (values computed by the reproduction
script of Figure~\ref{fig:pi_k}).
\end{remark}

\begin{remark}[Sieve regime]
\label{rem:sieve}
The fixed-basis regime matches the implementation (fixed-size
\texttt{mgcv} bases and \texttt{mboost} base learners). If instead
the basis grows with $n$, $K_n\asymp n^{1/(2s+1)}$, penalized
spline estimators attain the nonparametric rate
$O_p(n^{-2s/(2s+1)})$ for $f_j\in W^{s,2}$
\citep{claeskens2009}. The smoothing bias is then slower than
$n^{-1/2}$ for every finite $s$, and $\sqrt{n}$ inference requires
additional orthogonality arguments of the type developed for
partially linear models \citep{robinson1988}; the regularity
threshold $s>1/2$ reappears through the effective dimension of the
smoother, $\operatorname{tr}(\mathbf{S}_{\hat\mu})\asymp
n^{1/(2s+1)}=o(n^{1/2})$. Extending the present results to the
sieve regime is left for future research.
\end{remark}

\subsection{Neyman orthogonality of the quasi-score and cross-fitting}
\label{app:orthogonality}

The quasi-score of the CFE has a partialling-out structure: the
moment is built from the residuals of three auxiliary regressions, of
$\mathbf{y}$, $\mathbf{W}\mathbf{y}$ and $\mathbf{W}^2\mathbf{y}$ on
the candidate set. This is the structure introduced by
\citet{robinson1988} for the partially linear model, and it implies
that the population moment is \emph{Neyman-orthogonal} with respect
to the three nuisance functions, in the sense of
\citet{chernozhukov2018}: first-order estimation errors in the
nuisance fits do not move the moment. This subsection states and
proves the property, which (i) explains why flexible first-stage
estimators can be plugged into the CFE without redesigning the score,
and (ii) identifies exactly which bias channel survives --- and how
cross-fitting removes it.

Write the moment as a functional of $\rho$ and of a triplet of
nuisance functions $\boldsymbol{m}=(m_0,m_1,m_2)$ acting on the
covariates $x_i$,
\begin{equation}
\label{eq:psi_functional}
\psi(\rho;\boldsymbol{m})
= \mathbb{E}\Bigl[\bigl\{y_i-m_0(x_i)-\rho\,[(\mathbf{W}\mathbf{y})_i-m_1(x_i)]\bigr\}
\bigl\{(\mathbf{W}\mathbf{y})_i-m_1(x_i)-\rho\,[(\mathbf{W}^2\mathbf{y})_i-m_2(x_i)]\bigr\}\Bigr],
\end{equation}
so that $g_n(\rho)$ is the sample analogue of
$\psi(\rho;\boldsymbol{m})$ evaluated at the fitted nuisances. Let
$m_k^*(x_i)$ denote the conditional expectations
$m_0^*(x_i)=\mathbb{E}[y_i\mid x_i]$,
$m_1^*(x_i)=\mathbb{E}[(\mathbf{W}\mathbf{y})_i\mid x_i]$,
$m_2^*(x_i)=\mathbb{E}[(\mathbf{W}^2\mathbf{y})_i\mid x_i]$.

\begin{proposition}[Neyman orthogonality of the CFE quasi-score]
\label{prop:orthogonality}
Assume (A1), $w_{ii}=0$, that $\varepsilon_i$ is independent of
$\mathbf{X}$ with $\mathbb{E}[\varepsilon_i]=0$, and that the
$\varepsilon_i$ are mutually independent (heteroskedasticity is
allowed). Then $\psi(\rho_0;\boldsymbol{m}^*)=0$, and for every
direction $\boldsymbol{h}=(h_0,h_1,h_2)$ with
$\mathbb{E}[h_k(x_i)^2]<\infty$,
\[
\partial_t\,
\psi\bigl(\rho_0;\boldsymbol{m}^*+t\boldsymbol{h}\bigr)\Big|_{t=0}=0 .
\]
\end{proposition}

\paragraph{Proof}
At $\boldsymbol{m}^*$ the first factor of
\eqref{eq:psi_functional} reduces to the innovation: using
$(\mathbf{I}-\rho_0\mathbf{W})\mathbf{y}=f(\mathbf{X})+\boldsymbol{\varepsilon}$
and $m_0^*-\rho_0 m_1^*=\mathbb{E}[y_i-\rho_0(\mathbf{W}\mathbf{y})_i\mid x_i]
= f(x_i)$,
\[
a_i^* \equiv y_i-m_0^*(x_i)-\rho_0[(\mathbf{W}\mathbf{y})_i-m_1^*(x_i)]
=\varepsilon_i .
\]
The second factor becomes the deviation of
$(\mathbf{W}(\mathbf{I}-\rho_0\mathbf{W})\mathbf{y})_i
=(\mathbf{W}f(\mathbf{X}))_i+(\mathbf{W}\boldsymbol{\varepsilon})_i$
from its conditional expectation given $x_i$,
\[
v_i \equiv (\mathbf{W}f(\mathbf{X}))_i
-\mathbb{E}[(\mathbf{W}f(\mathbf{X}))_i\mid x_i]
+(\mathbf{W}\boldsymbol{\varepsilon})_i ,
\qquad \mathbb{E}[v_i\mid x_i]=0 .
\]
\emph{Validity.} $\psi(\rho_0;\boldsymbol{m}^*)
=\mathbb{E}[\varepsilon_i v_i]
=\sum_j w_{ij}\mathbb{E}[\varepsilon_i\varepsilon_j]=0$, by
independence of $\varepsilon_i$ from $\mathbf{X}$, mutual
independence of the $\varepsilon_i$, and $w_{ii}=0$.

\emph{Orthogonality.} Perturbing $m_0$ in the direction $h_0$
changes only the first factor, by $-t\,h_0(x_i)$, so the derivative
is $-\mathbb{E}[h_0(x_i)\,v_i]
=-\mathbb{E}\bigl[h_0(x_i)\,\mathbb{E}[v_i\mid x_i]\bigr]=0$.
Perturbing $m_2$ changes only the second factor, by
$+t\rho_0 h_2(x_i)$, giving
$\rho_0\,\mathbb{E}[h_2(x_i)\,\varepsilon_i]=0$ since
$\mathbb{E}[\varepsilon_i\mid x_i]=0$. Perturbing $m_1$ changes both
factors, by $+t\rho_0 h_1(x_i)$ and $-t\,h_1(x_i)$ respectively, so
the derivative is
$\rho_0\,\mathbb{E}[h_1(x_i)v_i]-\mathbb{E}[h_1(x_i)\varepsilon_i]
=0+0$. Note that the endogeneity of $\mathbf{W}\mathbf{y}$ plays no
role here: every conditioning runs through $x_i$ only.
\hfill$\square$

\begin{remark}[Own-observation bias and cross-fitting]
\label{rem:crossfit}
Orthogonality removes the \emph{first-order functional} sensitivity
of the moment to nuisance errors; it does not remove the
\emph{own-observation} term that arises when the same observations
are used to fit $\widehat m_k$ and to evaluate the residuals:
$\widehat m_1$ depends on $\varepsilon_i$ through
$\mathbf{W}\mathbf{y}$, so
$\mathbb{E}\bigl[\varepsilon_i\,(\widehat m_1-m_1^*)(x_i)\bigr]\neq 0$
in finite samples. This is precisely the finite-sample channel of
the geoadditive identification trap quantified in
Section~\ref{sec:simulations}: with a free bivariate smooth of the
coordinates in the candidate set, the in-sample residual $e_1$ is
depleted of part of the signal of $\mathbf{W}\mathbf{y}$ and
$\hat\rho$ is biased --- including a spurious nonzero $\hat\rho$
when $\rho_0=0$. Computing the residuals of the three auxiliary
regressions \emph{out-of-fold} (cross-fitting, option
\texttt{crossfit} of \texttt{GAM\_SAR\_CFE}, with small spatially
compact blocks scattered across folds) removes the own-observation
term while leaving the interpretable full-sample fit of $f$
unchanged. A formal treatment of cross-fitting under spatial
dependence --- where observations in different folds are not
independent --- is beyond the scope of this paper and is left for
future work; the Monte Carlo evidence of
Section~\ref{sec:simulations} documents its effect.
\end{remark}

\begin{remark}[Relation to the fixed-basis proof]
\label{rem:orth_fixed}
In the fixed-basis regime the fitted nuisances are the
projections $\widehat m_k = \mathbf{P}_B(\cdot)$, and the proof of
Proposition~\ref{prop:consistency_main} exploits the exact
finite-sample counterpart of orthogonality:
$\mathbf{M}^*\mathbf{B}=0$ annihilates the smoothing-bias term
identically, leaving only the $O(n^{-1})$ trace term and the
own-observation channel of Remark~\ref{rem:crossfit}. The
projections differ from the conditional expectations $m_k^*$ by the
non-constant diagonal of $(\mathbf{I}-\rho_0\mathbf{W})^{-1}$; under
(A2) and (A1) this approximation error is second-order and is
absorbed in the remainder terms of the proof.
\end{remark}

\begin{remark}[Heteroskedasticity-robust variance]
\label{rem:sandwich}
Proposition~\ref{prop:orthogonality} allows independent but
heteroskedastic innovations for the \emph{consistency} of
$\hat\rho$; the closed-form variance of
Proposition~\ref{prop:variance_main}, however, targets the
homoskedastic case through the pooled estimate $\hat\sigma^2$. A
robust alternative follows from the delta method on the quasi-score:
with $u_i=(e_0-\hat\rho e_1)_i$, $v_i=(e_1-\hat\rho e_2)_i$,
$m_i=u_iv_i$ and $d=b-2\hat\rho a=-n\,g_n'(\hat\rho)$,
\[
\widehat{\operatorname{var}}_{\mathrm{HC0}}(\hat\rho)
=\frac{\sum_i m_i^2}{d^2},
\qquad
\widehat{\operatorname{var}}_{\mathrm{SHAC}}(\hat\rho)
=\frac{\sum_i m_i^2+\sum_{i\neq j}\kappa_{ij}\,m_im_j}{d^2},
\]
where $\kappa_{ij}=1$ when $i$ and $j$ are within graph distance two
of each other in $W$ and $0$ otherwise --- the moment contributions
$m_i$ and $m_j$ are correlated for overlapping neighbourhoods
because $v$ contains $\mathbf{W}\boldsymbol{\varepsilon}$ --- in the
spirit of the spatial HAC estimator of \citet{kelejian2007}. Both
variants are returned by \texttt{GAM\_SAR\_CFE}
(\texttt{var\_rho\_hc0}, \texttt{var\_rho\_shac}), also under
cross-fitting. In a Monte Carlo check (oracle design, $n=1{,}000$,
$nr=1{,}000$, innovation standard deviations proportional to
$\exp\{1.5(x_1-0.5)\}$, a $4.5{:}1$ ratio), the closed-form variance
remains valid-to-conservative under this heteroskedasticity
(coverage $0.96$--$0.99$), HC0 under-covers ($0.90$--$0.92$) because
it ignores the neighbourhood cross-terms, and SHAC is closest to
nominal ($0.93$--$0.94$); reproduction script
\texttt{mc\_sandwich.R}.
\end{remark}

\subsection{Empirical validation of the variance formula}

Proposition~\ref{prop:variance_main} and
Remark~\ref{rem:conservative} predict a specific empirical pattern:
coverage at or above the nominal level for all $\rho_0$, a variance
ratio $\hat\sigma^2_{MC}/\overline{\hat\sigma^2(\hat\rho)}$ below
one, stable across sample sizes at fixed $\rho_0$, and decreasing in
$\rho_0$ with the same orders of magnitude as in the linear case of
\citet{smirnov2020}. The empirical coverage rates of nominal 95\%
confidence intervals $\hat\rho_{CFE}\pm 1.96\,\hat\sigma(\rho)$
reported in Appendix~\ref{appendix_additional_mc} display exactly
this pattern.

\section{Additional Monte Carlo results}
\label{appendix_additional_mc}

\subsection{Estimators included in the Monte Carlo experiments}

Table~\ref{MCdesign_estimator} lists the estimators evaluated in each
experiment and the R packages used for their implementation.

\begin{table}[htp]
\caption{List of estimators used in Monte Carlo experiments}
\label{MCdesign_estimator}
\begin{center}
\small
\begin{tabular}{|c|c|c|c|}
 \hline
 Estimators / DGP  & SAR NON LINEAR  & SEM NON LINEAR  & R Package\\
 \hline
BSPA\_SAR\_ML  & x& &  spboost\\
BSPA\_SAR\_CFE  & x& &  spboost\\
GAM\_SAR\_ML  & x& &  spboost\\
GAM\_SAR\_CFE  & x& & spboost\\
BSPA\_SEM\_ML  &  & x& spboost\\
BSPA\_SEM\_CFE  &  & x& spboost\\
gam &  x& & mgcv\\
gamboost & x & & mboost\\
lagsarlm & x & & spatialreg\\
errorsarlm & & x & spatialreg\\
 \hline
\end{tabular}
\end{center}
\end{table}

\subsection{Synthetic results for E1 experiments across sample sizes
and signal-to-noise ratios}

Tables~\ref{tab:SAR_E1_synt} and~\ref{tab:SEM_E1_synth}
report the bias and RMSE of the CFE-based estimators for the SAR and
SEM designs respectively, across all combinations of sample size
$n\in\{1000,2000,5000\}$ and signal-to-noise ratio
$\mathrm{SNR}\in\{0.7,0.9\}$. These results complement the
representative configuration reported in Section~\ref{sec:simulations}
and confirm the convergence patterns described there.

\begin{table}[htbp]
\centering
\scriptsize
\caption{Results for experiment E1, nonlinear SAR, varying size and SNR (CFE estimators, 1000 replications; \texttt{mstop} by spatial-block cross-validation).}
\label{tab:SAR_E1_synt}
\resizebox{0.65\linewidth}{!}{%
\begin{tabular}{llccccccc}
\toprule
\multicolumn{2}{c}{ } & \multicolumn{2}{c}{Spatial parameter} & \multicolumn{2}{c}{Intercept} & \multicolumn{3}{c}{Nonlinear components}\\
\multicolumn{2}{c}{ } & \multicolumn{2}{c}{$\rho$} & \multicolumn{2}{c}{$f_0$} & \multicolumn{1}{c}{$f_1$} & \multicolumn{1}{c}{$f_2$} & \multicolumn{1}{c}{$f_3$}\\
Estimator &   & Bias & RMSE & Bias & RMSE & RMSE & RMSE & RMSE\\
\midrule
\multicolumn{9}{l}{\textbf{$n=1000,\;SNR=0.7$}}\\\addlinespace[1ex]
\multirow{4}{*}{ BSPA\_SAR\_CFE} & 0 & -0.0018 & 0.0314 & 0.0003 & 0.0138 & 0.0275 & 0.0432 & 0.0307\\
 & 0.2 & -0.0016 & 0.0284 & 0.0003 & 0.0138 & 0.0275 & 0.0432 & 0.0307\\
 & 0.6 & -0.0012 & 0.0191 & 0.0003 & 0.0138 & 0.0275 & 0.0433 & 0.0308\\
 & 0.9 & -0.0005 & 0.0078 & 0.0002 & 0.0139 & 0.0274 & 0.0432 & 0.0308\\
\addlinespace[1ex]
\multirow{4}{*}{ GAM\_SAR\_CFE} & 0 & -0.0017 & 0.0312 & 0.0003 & 0.0138 & 0.0201 & 0.0415 & 0.0338\\
 & 0.2 & -0.0027 & 0.0282 & 0.0003 & 0.0138 & 0.0202 & 0.0415 & 0.0339\\
 & 0.6 & -0.0039 & 0.0195 & 0.0003 & 0.0139 & 0.0202 & 0.0416 & 0.0339\\
 & 0.9 & -0.0030 & 0.0094 & 0.0002 & 0.0143 & 0.0203 & 0.0416 & 0.0339\\
\addlinespace[1ex]
\multicolumn{9}{l}{\textbf{$n=1000,\;SNR=0.9$}}\\\addlinespace[1ex]
\multirow{4}{*}{ BSPA\_SAR\_CFE} & 0 & -0.0009 & 0.0199 & 0.0001 & 0.0070 & 0.0153 & 0.0267 & 0.0169\\
 & 0.2 & -0.0008 & 0.0179 & 0.0001 & 0.0070 & 0.0153 & 0.0267 & 0.0169\\
 & 0.6 & -0.0006 & 0.0121 & 0.0001 & 0.0070 & 0.0153 & 0.0267 & 0.0170\\
 & 0.9 & -0.0003 & 0.0049 & 0.0001 & 0.0071 & 0.0152 & 0.0267 & 0.0170\\
\addlinespace[1ex]
\multirow{4}{*}{ GAM\_SAR\_CFE} & 0 & -0.0009 & 0.0196 & 0.0001 & 0.0070 & 0.0103 & 0.0260 & 0.0184\\
 & 0.2 & -0.0024 & 0.0179 & 0.0001 & 0.0070 & 0.0103 & 0.0260 & 0.0184\\
 & 0.6 & -0.0041 & 0.0129 & 0.0001 & 0.0071 & 0.0104 & 0.0261 & 0.0184\\
 & 0.9 & -0.0032 & 0.0073 & 0.0001 & 0.0074 & 0.0104 & 0.0261 & 0.0184\\
\addlinespace[1ex]
\multicolumn{9}{l}{\textbf{$n=2000,\;SNR=0.7$}}\\\addlinespace[1ex]
\multirow{4}{*}{ BSPA\_SAR\_CFE} & 0 & -0.0021 & 0.0223 & 0.0002 & 0.0101 & 0.0197 & 0.0338 & 0.0222\\
 & 0.2 & -0.0019 & 0.0202 & 0.0002 & 0.0101 & 0.0197 & 0.0338 & 0.0222\\
 & 0.6 & -0.0013 & 0.0136 & 0.0001 & 0.0101 & 0.0197 & 0.0338 & 0.0222\\
 & 0.9 & -0.0005 & 0.0056 & 0.0001 & 0.0101 & 0.0197 & 0.0338 & 0.0222\\
\addlinespace[1ex]
\multirow{4}{*}{ GAM\_SAR\_CFE} & 0 & -0.0021 & 0.0223 & 0.0002 & 0.0101 & 0.0135 & 0.0321 & 0.0243\\
 & 0.2 & -0.0024 & 0.0202 & 0.0002 & 0.0101 & 0.0135 & 0.0321 & 0.0243\\
 & 0.6 & -0.0026 & 0.0138 & 0.0001 & 0.0101 & 0.0135 & 0.0321 & 0.0243\\
 & 0.9 & -0.0016 & 0.0059 & 0.0001 & 0.0102 & 0.0135 & 0.0322 & 0.0243\\
\addlinespace[1ex]
\multicolumn{9}{l}{\textbf{$n=2000,\;SNR=0.9$}}\\\addlinespace[1ex]
\multirow{4}{*}{ BSPA\_SAR\_CFE} & 0 & -0.0012 & 0.0139 & 0.0001 & 0.0051 & 0.0106 & 0.0211 & 0.0122\\
 & 0.2 & -0.0011 & 0.0125 & 0.0001 & 0.0051 & 0.0106 & 0.0211 & 0.0122\\
 & 0.6 & -0.0007 & 0.0085 & 0.0001 & 0.0051 & 0.0106 & 0.0211 & 0.0122\\
 & 0.9 & -0.0003 & 0.0035 & 0.0001 & 0.0051 & 0.0107 & 0.0211 & 0.0122\\
\addlinespace[1ex]
\multirow{4}{*}{ GAM\_SAR\_CFE} & 0 & -0.0012 & 0.0138 & 0.0001 & 0.0051 & 0.0069 & 0.0216 & 0.0132\\
 & 0.2 & -0.0019 & 0.0126 & 0.0001 & 0.0051 & 0.0069 & 0.0216 & 0.0132\\
 & 0.6 & -0.0023 & 0.0088 & 0.0001 & 0.0051 & 0.0069 & 0.0216 & 0.0132\\
 & 0.9 & -0.0015 & 0.0041 & 0.0001 & 0.0052 & 0.0070 & 0.0216 & 0.0132\\
\addlinespace[1ex]
\multicolumn{9}{l}{\textbf{$n=5000,\;SNR=0.7$}}\\\addlinespace[1ex]
\multirow{4}{*}{ BSPA\_SAR\_CFE} & 0 & -0.0003 & 0.0140 & -0.0003 & 0.0060 & 0.0135 & 0.0244 & 0.0147\\
 & 0.2 & -0.0003 & 0.0127 & -0.0003 & 0.0060 & 0.0135 & 0.0244 & 0.0147\\
 & 0.6 & -0.0002 & 0.0085 & -0.0003 & 0.0060 & 0.0135 & 0.0244 & 0.0147\\
 & 0.9 & -0.0001 & 0.0035 & -0.0003 & 0.0060 & 0.0135 & 0.0245 & 0.0148\\
\addlinespace[1ex]
\multirow{4}{*}{ GAM\_SAR\_CFE} & 0 & -0.0004 & 0.0140 & -0.0003 & 0.0060 & 0.0092 & 0.0243 & 0.0160\\
 & 0.2 & -0.0005 & 0.0127 & -0.0003 & 0.0060 & 0.0092 & 0.0243 & 0.0160\\
 & 0.6 & -0.0007 & 0.0085 & -0.0003 & 0.0060 & 0.0092 & 0.0243 & 0.0160\\
 & 0.9 & -0.0005 & 0.0035 & -0.0003 & 0.0060 & 0.0092 & 0.0243 & 0.0160\\
\addlinespace[1ex]
\multicolumn{9}{l}{\textbf{$n=5000,\;SNR=0.9$}}\\\addlinespace[1ex]
\multirow{4}{*}{ BSPA\_SAR\_CFE} & 0 & -0.0001 & 0.0086 & -0.0001 & 0.0030 & 0.0066 & 0.0172 & 0.0080\\
 & 0.2 & -0.0001 & 0.0078 & -0.0001 & 0.0030 & 0.0066 & 0.0172 & 0.0080\\
 & 0.6 & -0.0001 & 0.0053 & -0.0001 & 0.0030 & 0.0066 & 0.0172 & 0.0080\\
 & 0.9 & -0.0000 & 0.0022 & -0.0001 & 0.0031 & 0.0066 & 0.0172 & 0.0081\\
\addlinespace[1ex]
\multirow{4}{*}{ GAM\_SAR\_CFE} & 0 & -0.0002 & 0.0086 & -0.0001 & 0.0030 & 0.0047 & 0.0184 & 0.0087\\
 & 0.2 & -0.0005 & 0.0078 & -0.0001 & 0.0030 & 0.0047 & 0.0184 & 0.0087\\
 & 0.6 & -0.0007 & 0.0053 & -0.0001 & 0.0031 & 0.0047 & 0.0184 & 0.0087\\
 & 0.9 & -0.0005 & 0.0023 & -0.0001 & 0.0031 & 0.0047 & 0.0184 & 0.0087\\
\addlinespace[1ex]
\bottomrule
\end{tabular}}
\end{table}

\begin{table}[htbp]
\centering
\scriptsize
\caption{Results for experiment E1, nonlinear SEM, varying size and SNR (CFE estimators, 1000 replications; \texttt{mstop} by spatial-block cross-validation).}
\label{tab:SEM_E1_synth}
\resizebox{0.65\linewidth}{!}{%
\begin{tabular}{llccccccc}
\toprule
\multicolumn{2}{c}{ } & \multicolumn{2}{c}{Spatial parameter} & \multicolumn{2}{c}{Intercept} & \multicolumn{3}{c}{Nonlinear components}\\
\multicolumn{2}{c}{ } & \multicolumn{2}{c}{$\lambda$} & \multicolumn{2}{c}{$f_0$} & \multicolumn{1}{c}{$f_1$} & \multicolumn{1}{c}{$f_2$} & \multicolumn{1}{c}{$f_3$}\\
Estimator &   & Bias & RMSE & Bias & RMSE & RMSE & RMSE & RMSE\\
\midrule
\multicolumn{9}{l}{\textbf{$n=1000,\;SNR=0.7$}}\\\addlinespace[1ex]
\multirow{4}{*}{ BSPA\_SEM\_CFE} & 0 & -0.0025 & 0.0456 & 0.0003 & 0.0138 & 0.0275 & 0.0432 & 0.0307\\
 & 0.2 & -0.0053 & 0.0414 & 0.0004 & 0.0171 & 0.0271 & 0.0428 & 0.0303\\
 & 0.6 & -0.0096 & 0.0292 & 0.0009 & 0.0306 & 0.0227 & 0.0373 & 0.0254\\
 & 0.9 & -0.0264 & 0.0292 & 0.0025 & 0.0756 & 0.0107 & 0.0208 & 0.0121\\
\addlinespace[1ex]
\multirow{4}{*}{ GAM\_SEM\_CFE} & 0 & -0.0024 & 0.0456 & 0.0003 & 0.0139 & 0.0213 & 0.0470 & 0.0368\\
 & 0.2 & -0.0053 & 0.0414 & 0.0003 & 0.0139 & 0.0354 & 0.1019 & 0.0697\\
 & 0.6 & -0.0096 & 0.0293 & 0.0004 & 0.0126 & 0.0881 & 0.2515 & 0.1593\\
 & 0.9 & -0.0227 & 0.0267 & 0.0003 & 0.0094 & 0.1282 & 0.3355 & 0.2098\\
\addlinespace[1ex]
\multicolumn{9}{l}{\textbf{$n=1000,\;SNR=0.9$}}\\\addlinespace[1ex]
\multirow{4}{*}{ BSPA\_SEM\_CFE} & 0 & -0.0025 & 0.0456 & 0.0002 & 0.0070 & 0.0152 & 0.0267 & 0.0169\\
 & 0.2 & -0.0063 & 0.0416 & 0.0002 & 0.0089 & 0.0151 & 0.0264 & 0.0167\\
 & 0.6 & -0.0124 & 0.0304 & 0.0007 & 0.0187 & 0.0126 & 0.0230 & 0.0140\\
 & 0.9 & -0.0329 & 0.0395 & 0.0022 & 0.0606 & 0.0053 & 0.0163 & 0.0069\\
\addlinespace[1ex]
\multirow{4}{*}{ GAM\_SEM\_CFE} & 0 & -0.0025 & 0.0456 & 0.0001 & 0.0071 & 0.0123 & 0.0340 & 0.0234\\
 & 0.2 & -0.0063 & 0.0416 & 0.0002 & 0.0072 & 0.0308 & 0.0963 & 0.0633\\
 & 0.6 & -0.0121 & 0.0303 & 0.0003 & 0.0077 & 0.0865 & 0.2495 & 0.1569\\
 & 0.9 & -0.0317 & 0.0393 & 0.0003 & 0.0079 & 0.1267 & 0.3331 & 0.2081\\
\addlinespace[1ex]
\multicolumn{9}{l}{\textbf{$n=2000,\;SNR=0.7$}}\\\addlinespace[1ex]
\multirow{4}{*}{ BSPA\_SEM\_CFE} & 0 & -0.0023 & 0.0338 & 0.0002 & 0.0100 & 0.0197 & 0.0338 & 0.0222\\
 & 0.2 & -0.0038 & 0.0306 & 0.0003 & 0.0125 & 0.0194 & 0.0334 & 0.0219\\
 & 0.6 & -0.0061 & 0.0212 & 0.0007 & 0.0220 & 0.0164 & 0.0289 & 0.0185\\
 & 0.9 & -0.0149 & 0.0183 & 0.0025 & 0.0559 & 0.0071 & 0.0176 & 0.0086\\
\addlinespace[1ex]
\multirow{4}{*}{ GAM\_SEM\_CFE} & 0 & -0.0024 & 0.0338 & 0.0001 & 0.0101 & 0.0145 & 0.0361 & 0.0266\\
 & 0.2 & -0.0038 & 0.0306 & 0.0002 & 0.0100 & 0.0319 & 0.0984 & 0.0655\\
 & 0.6 & -0.0059 & 0.0212 & 0.0002 & 0.0089 & 0.0873 & 0.2522 & 0.1582\\
 & 0.9 & -0.0131 & 0.0174 & 0.0002 & 0.0064 & 0.1291 & 0.3384 & 0.2100\\
\addlinespace[1ex]
\multicolumn{9}{l}{\textbf{$n=2000,\;SNR=0.9$}}\\\addlinespace[1ex]
\multirow{4}{*}{ BSPA\_SEM\_CFE} & 0 & -0.0024 & 0.0337 & 0.0001 & 0.0051 & 0.0106 & 0.0211 & 0.0122\\
 & 0.2 & -0.0046 & 0.0307 & 0.0002 & 0.0065 & 0.0105 & 0.0209 & 0.0120\\
 & 0.6 & -0.0079 & 0.0219 & 0.0005 & 0.0133 & 0.0086 & 0.0189 & 0.0100\\
 & 0.9 & -0.0213 & 0.0299 & 0.0021 & 0.0448 & 0.0035 & 0.0154 & 0.0054\\
\addlinespace[1ex]
\multirow{4}{*}{ GAM\_SEM\_CFE} & 0 & -0.0024 & 0.0338 & 0.0001 & 0.0051 & 0.0085 & 0.0271 & 0.0170\\
 & 0.2 & -0.0046 & 0.0307 & 0.0001 & 0.0052 & 0.0297 & 0.0953 & 0.0619\\
 & 0.6 & -0.0079 & 0.0219 & 0.0002 & 0.0054 & 0.0864 & 0.2510 & 0.1568\\
 & 0.9 & -0.0221 & 0.0330 & 0.0002 & 0.0054 & 0.1277 & 0.3361 & 0.2084\\
\addlinespace[1ex]
\multicolumn{9}{l}{\textbf{$n=5000,\;SNR=0.7$}}\\\addlinespace[1ex]
\multirow{4}{*}{ BSPA\_SEM\_CFE} & 0 & -0.0007 & 0.0207 & -0.0003 & 0.0060 & 0.0135 & 0.0244 & 0.0147\\
 & 0.2 & -0.0015 & 0.0187 & -0.0003 & 0.0074 & 0.0133 & 0.0242 & 0.0145\\
 & 0.6 & -0.0028 & 0.0128 & -0.0004 & 0.0129 & 0.0110 & 0.0212 & 0.0121\\
 & 0.9 & -0.0068 & 0.0103 & 0.0000 & 0.0357 & 0.0042 & 0.0158 & 0.0060\\
\addlinespace[1ex]
\multirow{4}{*}{ GAM\_SEM\_CFE} & 0 & -0.0007 & 0.0206 & -0.0003 & 0.0060 & 0.0096 & 0.0263 & 0.0173\\
 & 0.2 & -0.0015 & 0.0186 & -0.0002 & 0.0059 & 0.0302 & 0.0968 & 0.0631\\
 & 0.6 & -0.0028 & 0.0128 & -0.0002 & 0.0052 & 0.0867 & 0.2534 & 0.1576\\
 & 0.9 & -0.0065 & 0.0102 & -0.0000 & 0.0038 & 0.1293 & 0.3406 & 0.2101\\
\addlinespace[1ex]
\multicolumn{9}{l}{\textbf{$n=5000,\;SNR=0.9$}}\\\addlinespace[1ex]
\multirow{4}{*}{ BSPA\_SEM\_CFE} & 0 & -0.0006 & 0.0207 & -0.0001 & 0.0030 & 0.0066 & 0.0172 & 0.0080\\
 & 0.2 & -0.0020 & 0.0187 & -0.0001 & 0.0038 & 0.0065 & 0.0171 & 0.0079\\
 & 0.6 & -0.0042 & 0.0132 & -0.0001 & 0.0078 & 0.0052 & 0.0163 & 0.0068\\
 & 0.9 & -0.0149 & 0.0314 & 0.0002 & 0.0297 & 0.0023 & 0.0148 & 0.0043\\
\addlinespace[1ex]
\multirow{4}{*}{ GAM\_SEM\_CFE} & 0 & -0.0006 & 0.0206 & -0.0001 & 0.0031 & 0.0055 & 0.0209 & 0.0109\\
 & 0.2 & -0.0021 & 0.0187 & -0.0001 & 0.0030 & 0.0291 & 0.0953 & 0.0613\\
 & 0.6 & -0.0046 & 0.0133 & -0.0001 & 0.0032 & 0.0862 & 0.2525 & 0.1567\\
 & 0.9 & -0.0144 & 0.0254 & 0.0001 & 0.0033 & 0.1282 & 0.3386 & 0.2088\\
\addlinespace[1ex]
\bottomrule
\end{tabular}}
\end{table}
 
\subsection{Empirical coverage rates of CFE confidence intervals}
\label{sec:coverage_rates}
Tables~\ref{tab:coverage_SAR} and~\ref{tab:coverage_SEM}
report the empirical coverage rates and variance ratios of
nominal 95\,\% confidence intervals
$\hat\rho_{CFE}\pm 1.96\,\hat\sigma(\rho)$ for the SAR and
SEM designs respectively, across all E1 configurations
(1000 replications, $\mathrm{SNR}=0.7$).
The variance ratio $\hat\sigma^2_{MC}/\overline{\hat\sigma^2(\hat\rho)}$
measures whether the analytical variance formula correctly
tracks the Monte Carlo sampling variance: values close to~1
indicate calibration, values below~1 indicate over-dispersion
of the confidence interval.

For the \textbf{SAR design} (Table~\ref{tab:coverage_SAR}),
coverage remains close to or above the nominal 95\% level across all
values of $\rho$ and sample sizes. The variance
ratio falls well below~1 for large $\rho$ (around 0.62 at
$\rho=0.6$ and 0.25--0.33 at $\rho=0.9$), reflecting systematic
over-estimation of the sampling variance in strongly autocorrelated
designs; the resulting conservative intervals explain the coverage
rates at or above 0.991 in that regime. Notably, the ratio is
stable across sample sizes at fixed $\rho$ and decreasing in
$\rho$: this is precisely the conservativeness pattern documented
for the \emph{linear} CFE by \citet{smirnov2020} (his Tables~3
and~6: 10--20\% over-estimation at moderate dependence, a factor of
2--3 under strong dependence), inherited here as discussed in
Remark~\ref{rem:conservative} of
Appendix~\ref{appendix_asymptotic_cfe}; the formula is accurately
calibrated for $\rho\leq 0.2$.

For the \textbf{SEM design} (Table~\ref{tab:coverage_SEM}),
the pattern is qualitatively similar at low to moderate
$\lambda$, with coverage rates between 0.95 and 0.99.
At $\lambda=0.9$, however, the closed-form variance itself breaks
down: the denominator $b-2\hat\lambda a$ becomes negative in a
sizeable share of replications, so that the \emph{mean} closed-form
variance can turn negative (the negative ratios in
Table~\ref{tab:coverage_SEM}) and coverage falls to 0.90--0.94 once
replications with an invalid variance are counted as non-covering.
This behaviour is directly linked to the
near-singularity of $(I-\lambda W)$ and the systematic
activation of the \texttt{cfe\_weighted\_projection} fallback
documented in Section~\ref{sec:simulations}: the fallback estimator recovers a point estimate, but the associated variance formula does not account for the induced non-smooth behaviour. Practitioners
should not rely on the closed-form CFE confidence intervals for the
SEM model when the estimated $\hat\lambda$ exceeds~0.85, and should
switch to the full ML variance there.

\begin{table}[htbp]
\centering
\small
\caption{Empirical coverage rates and variance ratios for $\hat\rho_{CFE}\pm 1.96\,\hat\sigma(\rho)$ in the nonlinear SAR design (E1, SNR$=0.7$, 1000 replications). Ratio: empirical variance of $\hat\rho$ over the mean closed-form variance; values below 1 indicate a conservative formula.}
\label{tab:coverage_SAR}
\resizebox{\linewidth}{!}{%
\begin{tabular}{lcccccccccccc}
\toprule
$\rho$ & \multicolumn{6}{c}{\texttt{BSPA\_SAR\_CFE}} & \multicolumn{6}{c}{\texttt{GAM\_SAR\_CFE}} \\
\cmidrule(lr){2-7} \cmidrule(lr){8-13}
 & \multicolumn{2}{c}{$n=1000$} & \multicolumn{2}{c}{$n=2000$} & \multicolumn{2}{c}{$n=5000$} & \multicolumn{2}{c}{$n=1000$} & \multicolumn{2}{c}{$n=2000$} & \multicolumn{2}{c}{$n=5000$} \\
 & Cov. & Ratio & Cov. & Ratio & Cov. & Ratio & Cov. & Ratio & Cov. & Ratio & Cov. & Ratio \\
\midrule
0 & 0.950 & 1.01 & 0.946 & 1.01 & 0.951 & 1.01 & 0.950 & 1.00 & 0.947 & 1.02 & 0.947 & 1.01 \\
0.2 & 0.962 & 0.91 & 0.963 & 0.92 & 0.958 & 0.91 & 0.960 & 0.91 & 0.959 & 0.92 & 0.957 & 0.91 \\
0.6 & 0.992 & 0.62 & 0.989 & 0.62 & 0.985 & 0.61 & 0.988 & 0.62 & 0.987 & 0.62 & 0.984 & 0.61 \\
0.9 & 1.000 & 0.25 & 1.000 & 0.25 & 1.000 & 0.25 & 0.991 & 0.33 & 0.999 & 0.27 & 1.000 & 0.25 \\
\bottomrule
\end{tabular}}
\end{table}

\begin{table}[htbp]
\centering
\small
\caption{Empirical coverage rates and variance ratios for $\hat\lambda_{CFE}\pm 1.96\,\hat\sigma(\lambda)$ in the nonlinear SEM design (E1, SNR$=0.7$, 1000 replications). Ratio: empirical variance of $\hat\lambda$ over the mean closed-form variance; values below 1 indicate a conservative formula.}
\label{tab:coverage_SEM}
\resizebox{\linewidth}{!}{%
\begin{tabular}{lcccccccccccc}
\toprule
$\lambda$ & \multicolumn{6}{c}{\texttt{BSPA\_SEM\_CFE}} & \multicolumn{6}{c}{\texttt{GAM\_SEM\_CFE}} \\
\cmidrule(lr){2-7} \cmidrule(lr){8-13}
 & \multicolumn{2}{c}{$n=1000$} & \multicolumn{2}{c}{$n=2000$} & \multicolumn{2}{c}{$n=5000$} & \multicolumn{2}{c}{$n=1000$} & \multicolumn{2}{c}{$n=2000$} & \multicolumn{2}{c}{$n=5000$} \\
 & Cov. & Ratio & Cov. & Ratio & Cov. & Ratio & Cov. & Ratio & Cov. & Ratio & Cov. & Ratio \\
\midrule
0 & 0.959 & 0.92 & 0.953 & 1.01 & 0.955 & 0.96 & 0.958 & 0.92 & 0.950 & 1.01 & 0.955 & 0.95 \\
0.2 & 0.970 & 0.83 & 0.958 & 0.91 & 0.963 & 0.86 & 0.968 & 0.83 & 0.957 & 0.91 & 0.965 & 0.86 \\
0.6 & 0.988 & 0.56 & 0.983 & 0.61 & 0.989 & 0.58 & 0.986 & 0.56 & 0.985 & 0.61 & 0.989 & 0.57 \\
0.9 & 0.931 & 0.66 & 0.930 & -0.68 & 0.901 & -0.24 & 0.936 & -0.87 & 0.921 & -0.39 & 0.908 & -0.27 \\
\bottomrule
\end{tabular}}
\end{table}

\end{document}